\documentclass[pdflatex,sn-nature, twocolumn]{sn-jnl}

\makeatletter
\@twosidefalse
\@mparswitchfalse
\makeatother

\usepackage{geometry}
  
\usepackage{graphicx}%
\usepackage{multirow}%
\usepackage{amsmath,amssymb,amsfonts}%
\usepackage{amsthm}%
\usepackage{mathrsfs}%
\usepackage[title]{appendix}%
\usepackage[dvipsnames]{xcolor}%
\usepackage{textcomp}%
\usepackage{manyfoot}%
\usepackage{booktabs}%
\usepackage{algorithm}%
\usepackage{algorithmicx}%
\usepackage{algpseudocode}%
\usepackage{listings}%
\usepackage[normalem]{ulem}

\usepackage{amsmath,amssymb}

\newcommand{\ket}[1]{\left|#1\right\rangle}
\newcommand{\bra}[1]{\left\langle#1\right|}

\usepackage{bbm}
\theoremstyle{thmstyleone}%
\theoremstyle{thmstyletwo}%

\theoremstyle{thmstylethree}%

\begin{document}

\title[Article Title]{Experimental realization of the minimal two-dimensional multi-orbital kagome model}



\author[1,2]{\fnm{Bing} \sur{Liu}}
\equalcont{These authors contributed equally to this work.}

\author[2,3]{\fnm{Arka} \sur{Bandyopadhyay}}
\equalcont{These authors contributed equally to this work.}

\author[2,3]{\fnm{Manish} \sur{Verma}}

\author[1,2]{\fnm{Jonas} \sur{Erhardt}}

\author[1,2]{\fnm{Tim} \sur{Wagner}}

\author[1,2]{\fnm{Jing} \sur{Qi}}

\author[1,2]{\fnm{Kilian} \sur{Strauß}}

\author[4]{\fnm{Domenico Di} \sur{Sante}}

\author[5]{\fnm{Carmine} \sur{Ortix}}

\author[1,2,6]{\fnm{Simon} \sur{Moser}}

\author[1,2]{\fnm{Jörg} \sur{Schäfer}}

\author[2,3]{\fnm{Ronny} \sur{Thomale}}

\author*[2,3]{\fnm{Giorgio} \sur{Sangiovanni}}\email{sangiovanni@physik.uni-wuerzburg.de}

\author*[1,2]{\fnm{Ralph} \sur{Claessen}}\email{claessen@physik.uni-wuerzburg.de}

\affil[1]{\orgdiv{Physikalisches Institut}, \orgname{Universit\"at W\"urzburg}, \orgaddress{\city{W\"urzburg}, \postcode{D-97074}, \country{Germany}}}

\affil[2]{\orgdiv{W\"urzburg-Dresden Cluster of Excellence ctd.qmat}, \orgname{Universit\"at W\"urzburg}, \orgaddress{\city{W\"urzburg}, \postcode{D-97074}, \country{Germany}}}

\affil[3]{\orgdiv{Institut f\"ur Theoretische Physik und Astrophysik}, \orgname{Universit\"at W\"urzburg}, \orgaddress{\city{W\"urzburg}, \postcode{D-97074}, \country{Germany}}}

\affil[4]{\orgdiv{Department of Physics and Astronomy}, \orgname{University of Bologna}, \orgaddress{\city{Bologna}, \postcode{40127}, \country{Italy}}}

\affil[5]{\orgdiv{Dipartimento di Fisica “E. R. Caianiello”}, \orgname{Università di Salerno}, \orgaddress{\city{Fisciano}, \country{Italy}}}

\affil[6]{\orgdiv{Thin Films \& Physics of Quantum Materials, Faculty of Physics}, \orgname{Bielefeld University}, \orgaddress{\city{Bielefeld}, \postcode{33615}, \country{Germany}}}

\abstract{Kagome materials have emerged as a major platform for correlated and topological quantum matter, hosting flat bands, Dirac dispersions and van Hove singularities rooted in frustrated lattice geometry. While their essential physics is epitomized by the canonical single-orbital kagome model, real kagome materials are intrinsically multi-orbital and typically chemically and structurally complex, with additional three-dimensional coupling making kagome geometry and orbital degrees of freedom difficult to disentangle. Here we realize a truly two-dimensional, elemental two-orbital kagome system in monolayer Sb on SiC(0001). Substrate-induced orbital filtering isolates the in-plane Sb \(p_x/p_y\) orbitals into a six-band kagome manifold, establishing a chemically simple, minimal platform for multi-orbital kagome physics. We demonstrate two consequences that distinguish this system from the canonical single-orbital model. First, chemically enforced half filling pins the Fermi level to a closed nodal line, whose finite density of states drives a unit-cell-conserving breathing instability and opens a giant insulating gap. Second, we uncover orbital-resolved atomic obstruction: individual orbital-derived manifolds realize obstructed atomic limits, whereas their combined half-filled valence manifold is non-obstructed. Our results establish kagome antimonene as a benchmark system for exploring orbital-driven electronic, structural and topological kagome physics in the genuine two-dimensional limit.}




\maketitle


Geometrically frustrated lattices provide a unique route to electronic states
whose properties are governed primarily by lattice geometry rather than local
chemistry. Among them, the kagome lattice, composed of corner-sharing
triangles, has emerged as one of the paradigmatic model systems of condensed
matter physics because its symmetry and connectivity give rise to a
characteristic electronic structure featuring Dirac cones, van Hove
singularities (VHSs), and a topological flat band originating from destructive
interference of electronic wave functions
\cite{disante2026kagome,kang2020dirac,slot2017experimental,liu2021screening,
jiang2019topological,yan2026engineering,hu2022rich,yin2022topological,ye2024hopping,kang2020topological}.
Together, these ingredients provide an ideal setting for correlated and
topological quantum phenomena, including charge-density
waves~\cite{teng2022discovery,tan2021charge,yong2023phonon},
unconventional
superconductivity~\cite{wu2021nature,zhao2021cascade,deng2024chiral,jiang2023kagome},
quantum spin
liquids~\cite{han2012fractionalized,lee2007quantum}, and, upon introducing a
breathing distortion that alternates the size of neighbouring triangles,
topologically non-trivial insulating phases~\cite{ezawa2018higher,kempkes2019robust,van2020topological,lu2020orbital,zhang2023realization,wang2023sub}.

\begin{figure*}[htbp]
\centering
  \includegraphics[width=1.0\textwidth]{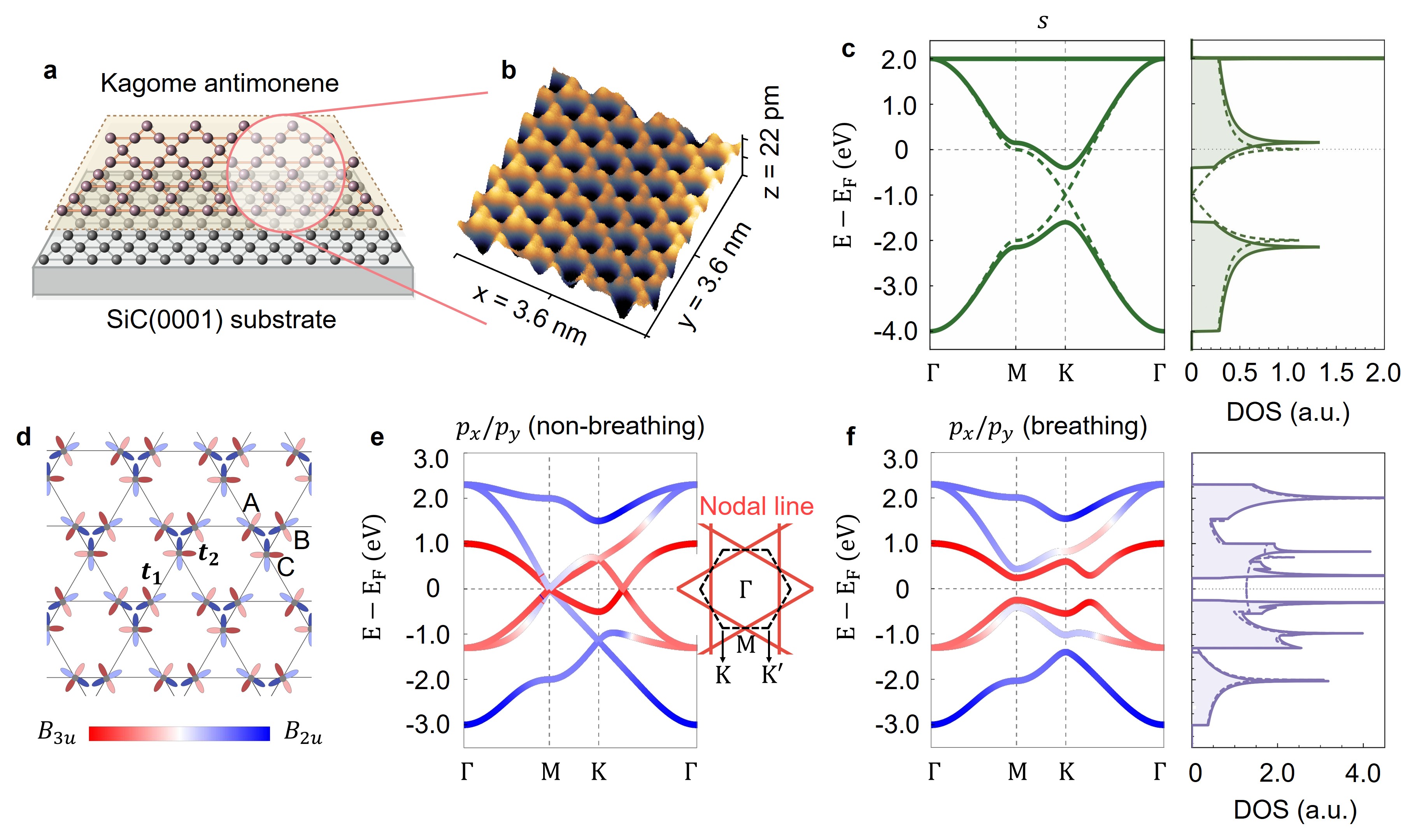}
 \hfill
    	\caption{Tight-binding model of the in-plane \textit{p}-orbital dominated 
kagome lattice. 
\textbf{a,} Sketch of a kagome monolayer supported on a substrate 
with $C_{3}$-symmetry. 
\textbf{b,} Atomically resolved STM topographic image of 
kagome antimonene grown on the SiC(0001) substrate, acquired at -3.2 V and 60 pA 
by using a Sb-functionalized tip. 
\textbf{c,} Left panel: The $s$-orbital dominated non-breathing (dashed green curves) and 
breathing (solid green curves, with hopping amplitudes $t_1 = 0.8$ and $t_2 = 1.2$) 
kagome bands. Right panel: Momentum-integrated density of states (DOS) corresponding to the non-breathing (dashed green curve) and breathing (solid green curve) $s$-orbital kagome bands shown in the left panel.
\textbf{d,} Atomic structure of the $p_{\text{x}}/p_{\text{y}}$-orbital dominated 
kagome lattice with spatial orientation of the $B_{2u}$ (blue) and $B_{3u}$ (red) 
orbitals. The atomic sites are labeled as A, B, and C. The hopping amplitudes within 
the upwards and downwards triangles are labeled as $t_1$ and $t_2$, respectively. 
\textbf{e} and \textbf{f,} Band structure along the high-symmetry path for the 
$p_{\text{x}}/p_{\text{y}}$-derived non-breathing kagome lattice 
($t_1^{\sigma} = t_2^{\sigma} = 1$, $t_1^{\pi} = t_2^{\pi} = 0$) 
(\textbf{e}) and breathing kagome lattice ($t_2^{\sigma} = 1.2$, $t_1^{\sigma}= 0.8$, 
$t_2^{\pi} = 0$, $ t_1^{\pi} = 0$) (left panel of \textbf{f}), with the orbital-projected 
contributions color-coded on the bands. The three valence bands are labeled as $\alpha$, $\beta$ 
and $\gamma$. Inset of \textbf{e}: Fermi surface of the non-breathing kagome lattice, showing a closed nodal line. The dashed black hexagon indicates the kagome Brillouin zone (BZ). Right panel of \textbf{f}: Momentum-integrated DOS corresponding to the kagome bands shown in \textbf{e} (dashed purple curve) and in the left panel of \textbf{f} (solid purple curve).
}
 \label{fgr: figure1}
\end{figure*}

Much of our present understanding of kagome physics is rooted in the canonical
tight-binding (TB) description of an ideal two-dimensional kagome lattice with
isotropic nearest-neighbour hopping between equivalent $s$ (or $p_z$) orbitals
(Fig.~\ref{fgr: figure1}). Although this single-orbital model successfully
captures the essential features of kagome band topology, experimentally
accessible kagome metals are intrinsically more complex. Prominent examples
such as Fe$_3$Sn$_2$,
CoSn, as well as the 135- and 166-families~\cite{disante2026kagome,zhao2021cascade,shi2022new,morali2019fermi,kang2020dirac,chen2023visualizing,ye2018massive}
derive their kagome bands from several transition-metal $3d$ orbitals, whose
directional bonding produces anisotropic hopping amplitudes and an intrinsically
multi-orbital electronic structure~\cite{aretz2025strong,grytsiuk2024nb3cl8}.
Additional hybridization with ligand states, generally non-integer filling of
the kagome-derived bands, and finite interlayer coupling further move these
materials away from the strictly two-dimensional limit of the minimal
Hamiltonian. While these systems have revealed a wealth of remarkable
correlated and topological phenomena, the fundamental consequences of orbital
degrees of freedom remain difficult to disentangle from chemical complexity and
material-specific details. As the minimal extension beyond the single-orbital limit, a two-orbital kagome system offers the simplest setting in which to uncover genuine multi-orbital kagome physics, yet an experimentally accessible realization has remained elusive.

A hitherto route towards such a model system is provided by
orbital filtering in epitaxial monolayers of elemental $p$-metal atoms on
semiconductor substrates. Compared with transition-metal kagome compounds,
their valence electronic structure is intrinsically much simpler, involving
only the three $p$ orbitals of the adsorbed atoms. Strong hybridization with
the substrate removes the out-of-plane $p_z$ orbital from the low-energy
electronic structure while preserving the in-plane $p_x$ and $p_y$ orbitals
\cite{reis2017bismuthene,stuhler2020tomonaga,stuhler2022effective,
bauernfeind2021design,schmitt2024achieving,erhardt2024bias}. Orbital filtering
thus reduces the active orbital manifold from three orbitals to only two,
naturally realizing the simplest multi-orbital extension of the canonical
kagome model. The resulting electronic structure is governed by anisotropic
$\sigma$- and $\pi$-bonding and provides an experimentally accessible platform
for investigating the fundamental consequences of multi-orbital kagome physics.

Here we realize this concept experimentally in an epitaxial, fully elemental Sb
monolayer on SiC(0001) (Figs.~\ref{fgr: figure1}a,b), going beyond the canonical single-orbital kagome model (Fig.~\ref{fgr: figure1}c). Orbital filtering by the
substrate isolates the in-plane Sb $p_x$ and $p_y$ orbitals into a well-defined
six-band breathing kagome manifold (Figs.~\ref{fgr: figure1}d-f), establishing an intrinsically
two-dimensional minimal multi-orbital kagome model system. We show that its
chemically preferred half filling stabilizes a pronounced breathing distortion
through a unit-cell-conserving instability of a closed nodal line at the Fermi level (inset of Fig.~\ref{fgr: figure1}e). Remarkably,
the same half-filled electronic state also gives rise to a hierarchy of atomic
limits, in which individually obstructed orbital manifolds recombine into a
non-obstructed valence manifold. These band-resolved atomic limits are
experimentally identified by energy-dependent scanning tunnelling microscopy in
combination with first-principles calculations and symmetry-adapted
tight-binding analysis. Our work establishes elemental kagome antimonene as a
model platform for investigating the fundamental consequences of multi-orbital
kagome physics.



\begin{figure*}[htbp]
 \centering
	\includegraphics[width=1.0\textwidth]{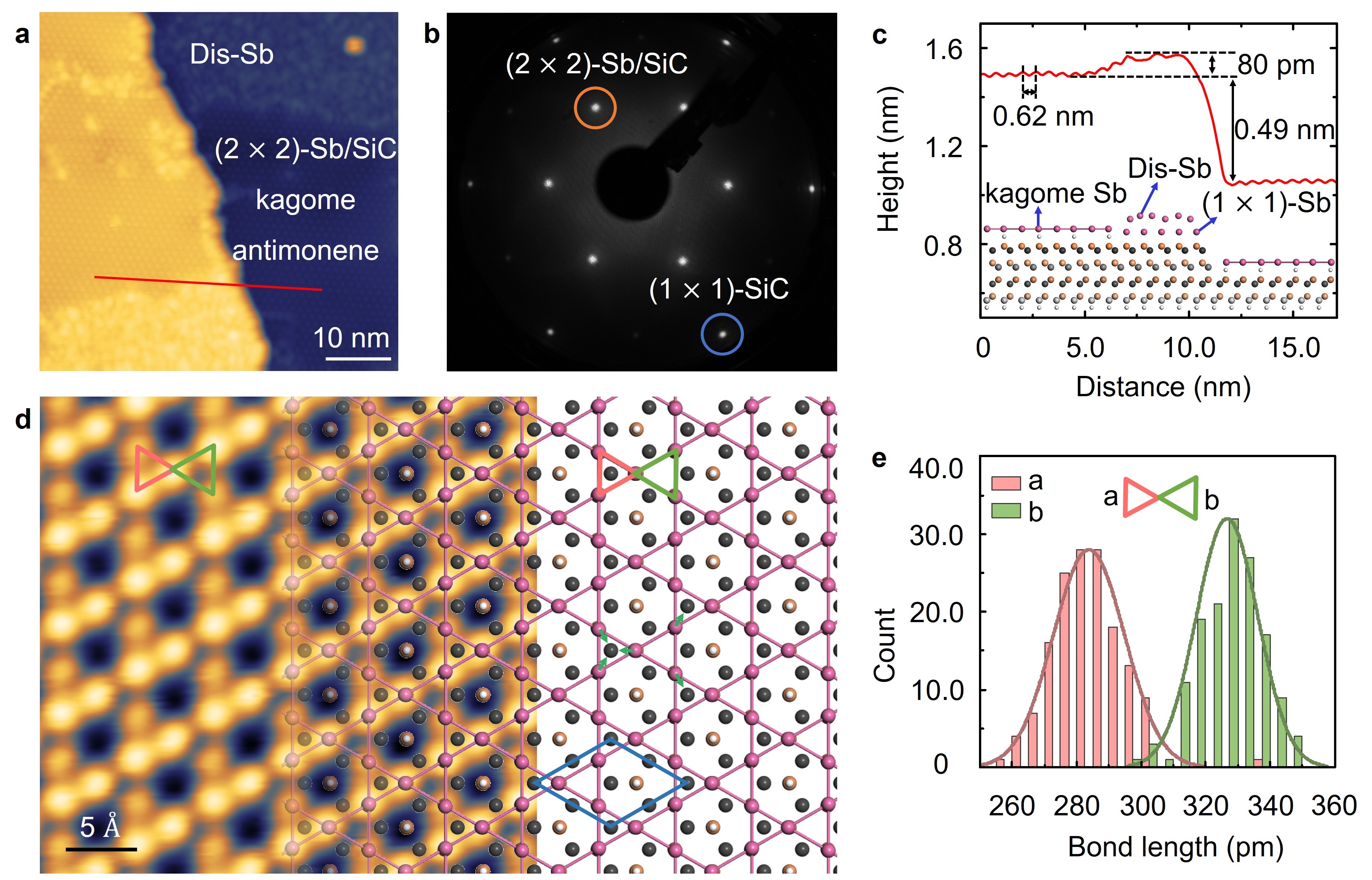}
\caption{Topography characterization of the monolayer Sb with breathing kagome
lattice on SiC(0001) substrate. 
\textbf{a,} Large-scale STM image (-3.0\,V,
10\,pA) of the surface after monolayer Sb preparation on SiC. The flat area is
the ($2\times2$)-Sb/SiC phase, i.e., breathing kagome antimonene, and around the
flat area is disordered Sb layer (labeled as Dis-Sb). 
\textbf{b}, LEED pattern
of the sample acquired at 40\,eV. Blue and orange circles indicate the $(1
\times 1)$-SiC(0001) lattice and $(2 \times 2)$-phase related to the breathing
kagome antimonene, respectively. 
\textbf{c,} Line profile along the red line in
\textbf{a}, accompanying side view of atomic structure, showing different Sb
phases on SiC. Sb: Pink spheres; Si: orange spheres; C: dark and light gray
spheres, indicating various layers. 
\textbf{d,} Zoomed-in STM image ($-3.0$ V, 50 pA) of the $(2 \times 2)$-Sb/SiC region acquired with an Sb-functionalized tip, in which an Sb cluster was adsorbed at the tip apex to enhance spatial resolution, revealing the kagome lattice. The overlaid DFT structural model shows
excellent agreement with the experimental image. For simplicity, the atomic
model includes only the Sb atoms and the first-layer Si and C atoms. The Sb
monolayer with breathing kagome lattice generates a ($2\times2$) reconstruction,
as marked by the blue rhombus, related to SiC(0001) substrate. The smaller and
larger triangles in the breathing kagome lattice are indicated by pink and green
triangular frames, respectively. Green arrows on the atoms indicate the lattice
distortion due to the breathing mode. 
\textbf{e,} Statistical analysis of
bonding length in the kagome lattice.}
	\label{fgr: figure2}
\end{figure*}

The orbital-filtering strategy introduced above is implemented here in an
epitaxial Sb monolayer on SiC(0001), where the threefold symmetry of the
substrate provides the structural template for an atomically thin kagome
lattice (Fig.~\ref{fgr: figure1}a). Following Sb deposition onto
hydrogen-etched SiC(0001) and gentle post-annealing, a well-ordered
$(2\times2)$ surface phase forms (with respect to the substrate lattice
periodicity), corresponding to kagome antimonene. Scanning tunneling microscopy (STM) topography
(Fig.~\ref{fgr: figure2}a) reveals extended kagome domains with typical lateral
dimensions of 30--50\,nm, interrupted only by small residual regions of the
underlying $(1\times1)$-Sb phase. The high structural quality is further
corroborated by sharp $(2\times2)$ diffraction spots in low-energy electron
diffraction (LEED) (Fig.~\ref{fgr: figure2}b). The line profile in
Fig.~\ref{fgr: figure2}c yields a lattice constant of
0.61\,nm---twice that of the SiC(0001) substrate~\cite{glass2016atomic}---and a
substrate step height of 0.49\,nm, both consistent with the proposed
structural model.

Atomic-resolution STM obtained with a functionalized tip
directly resolves the
corner-sharing triangular motif characteristic of the kagome lattice
(Fig.~\ref{fgr: figure2}d). Rather than forming an ideal kagome network, the Sb
atoms exhibit alternating small and large triangles, providing direct real-space
evidence for a breathing distortion. Statistical analysis of the STM images
(Fig.~\ref{fgr: figure2}e) yields average bond lengths of
$a=2.84\pm0.11$\,\AA\ and $b=3.27\pm0.10$\,\AA, corresponding to a bond-length
modulation of approximately 10\%.

To elucidate the microscopic origin of this structure, we performed
first-principles density functional
theory (DFT) calculations for a $3/4$-monolayer Sb overlayer occupying
the $T_1$ adsorption sites of SiC(0001), leaving one Si atom per substrate
hexagon unoccupied. Upon structural relaxation, the Sb atoms move slightly away
from the ideal adsorption sites and spontaneously stabilize a breathing kagome
lattice without developing any out-of-plane buckling. The relaxed structure
reproduces the experimentally observed bond lengths remarkably well, with
$a=2.900$\,\AA\ and $b=3.246$\,\AA\ (Fig.~\ref{fgr: figure2}d). The calculations further show that the different substrate geometries beneath neighbouring kagome triangles -- with a first-bilayer carbon atom directly beneath one triangle but not the other -- selectively stabilize the alternating short and long bonds (Supplementary Fig. S9), thereby
imprinting the breathing distortion while preserving the strictly
two-dimensional character of the elemental Sb monolayer.

The excellent agreement between STM and DFT establishes kagome antimonene as a
structurally well-defined, strictly two-dimensional elemental breathing kagome
monolayer. We next turn to its low-energy electronic structure and the minimal
multi-orbital kagome Hamiltonian realized by orbital filtering.

\label{tb_model}

The key consequence of orbital filtering is that the SiC substrate removes the
Sb $p_z$ states from the vicinity of the Fermi level, leaving a low-energy
electronic structure dominated by the in-plane $p_x$ and $p_y$ orbitals.
Kagome antimonene therefore realizes the minimal multi-orbital extension of the
canonical single-orbital kagome model, in which the low-energy electronic
structure is reduced to an isolated two-orbital ($p_x,p_y$) manifold.

In contrast to the single-orbital kagome model, each nearest-neighbour bond is
described by both $\sigma$- and $\pi$-type hopping between the in-plane
$p$ orbitals. The breathing distortion further introduces two inequivalent
hopping amplitudes, $t_1$ and $t_2$, associated with the long and short bonds,
respectively. Diagonalization of the resulting six-band tight-binding
Hamiltonian yields the symmetry-adapted $B_{3u}$ and $B_{2u}$ eigenstates shown
schematically in Fig.~\ref{fgr: figure1}d, which provide the natural basis for
describing the electronic structure. In the realistic $p$-orbital system, the
$\pi$ overlap is much weaker than the $\sigma$ overlap
($|t^\pi|\ll|t^\sigma|$), leaving the latter to dominate the electronic
structure. A detailed derivation of the tight-binding Hamiltonian is provided
in section 1 in the Supplementary Information.

The corresponding tight-binding band structure of the non-breathing lattice is
shown in Fig.~\ref{fgr: figure1}e. Compared with the canonical single-orbital
kagome model, the additional orbital degree of freedom qualitatively enriches
the fermiology. Besides the familiar Dirac cones at K, four bands meet at
the M point, while additional band crossings appear along the
K--$\Gamma$ direction. These crossings are not isolated point degeneracies but
belong to triangular nodal rings surrounding the K points (Supplementary
Fig.~S3). Throughout the following discussion, the bands are labelled according
to their dominant $B_{3u}$ or $B_{2u}$ orbital character.

Introducing a breathing distortion lifts the symmetry-protected band degeneracies through inversion
symmetry breaking. As shown in Fig.~\ref{fgr: figure1}f, sizeable gaps open both
at the Fermi level and at the Dirac crossings near K, reflecting the absence of
two-dimensional irreducible representations away from the $\Gamma$ point in the
reduced space group ($p3m1$). Remarkably, the breathing-induced gap naturally
separates the occupied bands into a $2/3$-filled $B_{3u}$ manifold and a
$1/3$-filled $B_{2u}$ manifold.

\begin{figure*}[htbp]
  \centering
	\includegraphics[width=1\textwidth]{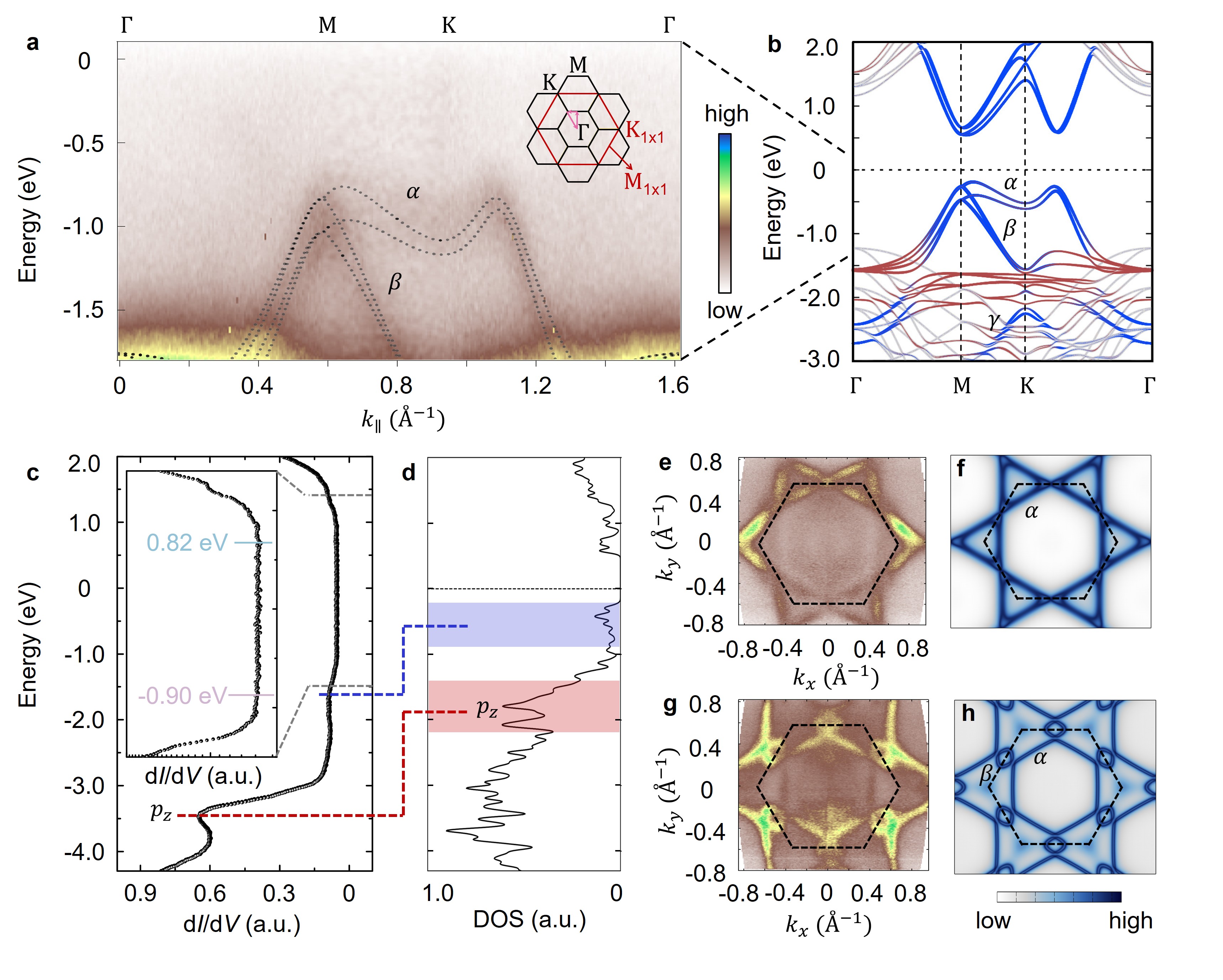}
 \hfill
	\caption{
Electronic structure of the breathing kagome antimonene. 
\textbf{a}, ARPES band
dispersion measured along the path indicated by the pink line in the BZ sketch
inset, revealing a large gap at the Fermi level, with the valence band maximum
located at approximately -0.75 eV. Inset: Sketch of the BZ of the ($1\times1$)
(red) and ($2\times2$) (black) reconstructions on SiC(0001). To distinguish
between the kagome BZ and the ($1\times1$)-SiC(0001) BZ, we use the labels
\(\Gamma\), K and M without subscripts for the kagome lattice, and the same
labels with the subscript "($1\times1$)" especially for the (1 × 1) BZ. The
DFT-calculated band structure is overlaid on the ARPES data as dashed curves.
\textbf{b}, DFT calculated band structures
of the breathing kagome antimonene lattice on SiC(0001) substrate with atomic
SOC included. The Sb $p_{\text{x}}$ and $p_{\text{y}}$ orbital-dominated bands
are plotted in blue, while Sb $p_{\text{z}}$ bands are in red, and SiC substrate
bands are in gray. 
\textbf{c}, Differential conductivity d\textit{I}/d\textit{V}
spectroscopy detected on the breathing kagome antimonene, revealing a large gap
between -0.90 eV and 0.82 eV. The inset shows zoom-in spectrum plotted with a
logarithmic $y$-axis. 
\textbf{d}, DFT calculated density of states based on the
band structure in \textbf{b}. 
\textbf{e} and \textbf{g,} The CECs acquired from
ARPES measurements at -0.78 eV, and -1.30 eV, respectively, indicating remnant nodal-line-character at the VBM (\textbf{e}). The dashed black
hexagon indicates the kagome BZ. 
\textbf{f} and \textbf{h,} Theoretically
calculated CECs at the energies 0.05 eV and 0.51 eV below the VBM, for comparison with the experimental results in \textbf{e} and
\textbf{g}, respectively.
     }
	\label{fgr: figure3}
\end{figure*}

\label{electronic_structure}

We now demonstrate that the electronic structure of kagome antimonene closely
follows the predictions of the minimal multi-orbital kagome Hamiltonian.
Angle-resolved photoemission spectroscopy (ARPES) was performed on samples
dominated by the kagome antimonene phase [see Supplementary Information for a
comparison with the competing $(1\times1)$-Sb phase]. The measured band
structure, shown in Fig.~\ref{fgr: figure3}a, exhibits two prominent
kagome-derived bands. The $\beta$ band forms a hole-like parabola with its
maximum at the M point, whereas the $\alpha$ band is relatively flat between M
and K, reaches its maximum close to K, and subsequently disperses rapidly
towards $\Gamma$. Most notably, ARPES directly reveals a pronounced insulating
gap, with the valence-band maximum located approximately 0.75\,eV below the
Fermi level.

First-principles DFT calculations reproduce the measured electronic structure
remarkably well (Fig.~\ref{fgr: figure3}b), including the dispersions of the
$\alpha$ and $\beta$ bands. An additional in-plane Sb $p_x/p_y$-derived band,
labelled $\gamma$, is predicted at higher binding energies. Experimentally,
however, this band overlaps with the intense and spectrally complex SiC valence
bands and therefore cannot be unambiguously resolved by ARPES. The following
analysis consequently focuses on the low-binding-energy electronic structure.

The calculations further clarify the microscopic origin of the observed bands.
By tracking the evolution of the electronic structure upon sequential inclusion
of the SiC substrate and the breathing distortion (Supplementary Information Fig. S7),
we find that orbital filtering shifts the Sb $p_z$-derived bands well away from
the Fermi level, leaving the low-energy states dominated by the in-plane
$p_x$ and $p_y$ orbitals. Their dispersions are quantitatively captured by the
minimal multi-orbital tight-binding Hamiltonian introduced above. Comparison of
the ARPES and DFT results with the tight-binding bands in
Fig.~\ref{fgr: figure1}f identifies the $\alpha$ and $\beta$ bands as
predominantly $B_{3u}$-derived, whereas the $\gamma$ band is dominated by
$B_{2u}$ character. This orbital assignment provides the key for understanding
the two emergent phenomena discussed below: the breathing instability at half
filling and the orbital-resolved atomic obstruction.

The insulating state is independently confirmed by scanning tunnelling
spectroscopy (STS). The differential-conductance spectra ($dI/dV$), which
reflect the local density of states (LDOS), display a pronounced gap extending
from $-0.90$ to $0.82$\,eV (Fig.~\ref{fgr: figure3}c), with a valence-band
maximum consistent with that determined by ARPES. Comparison with the
calculated density of states (Fig.~\ref{fgr: figure3}d) assigns the broad STS
feature near $-1.8$\,eV to the $\alpha$ and $\beta$ bands, whereas the
pronounced peak near $-3.5$\,eV originates predominantly from Sb $p_z$ states.
Both spectral features are reproduced by the calculated density of states. The
remaining difference between the calculated and measured gaps is consistent
with the well-known tendency of conventional DFT to underestimate band gaps.

The momentum-space structure of the occupied states is further resolved by
ARPES constant-energy contours (CECs). At the valence-band maximum, a characteristic
kagome-shaped contour is observed (Fig.~\ref{fgr: figure3}e). At higher binding
energies (Fig.~\ref{fgr: figure3}g), the triangular contours expand and an
additional elliptical pocket develops around the M point, originating from the
$\beta$ band. The corresponding DFT constant-energy contours
(Figs.~\ref{fgr: figure3}f,h) closely reproduce the experimental observations
and are in excellent agreement with the predictions of the minimal
multi-orbital tight-binding Hamiltonian. Notably, the kagome-shaped contour at
the valence-band maximum directly traces the momentum-space position of the nodal line before the breathing distortion opens the gap. It therefore
not only provides a fingerprint of the underlying multi-orbital band topology,
but also constitutes the manifold of initial states for direct optical
transitions across the breathing-induced gap.

Taken together, the excellent agreement among ARPES, STS, DFT and the minimal
multi-orbital tight-binding Hamiltonian demonstrates that kagome antimonene
constitutes a faithful experimental realization of the predicted minimal
multi-orbital kagome model. Having established the correspondence between
structure, Hamiltonian and electronic structure, we next turn to the new
physics that emerges from this orbital-filtered kagome system.

\label{breathing}

We first address the origin of the pronounced breathing instability, one of
the defining consequences of the two-orbital kagome Hamiltonian.
It resembles the canonical single-orbital kagome model, where a
unit-cell-conserving breathing distortion alternates the bond lengths of the
two inequivalent kagome triangles and opens a gap at the Dirac point at K.
However, the electronic driving force is expected to be weak in that case,
because the Dirac point coincides with a vanishing density of states (Fig.~\ref{fgr: figure1}c).
Furthermore, the required one-third filling of the kagome bands (two electrons
per primitive unit cell) cannot be realized in an elemental kagome lattice
without external charge transfer.

At half filling, the two-orbital kagome Hamiltonian realized in kagome
antimonene behaves fundamentally differently. Orbital filtering removes one of
the three Sb $5p$ electrons from the kagome manifold by covalent bonding to the
SiC substrate, leaving exactly two in-plane $p$ electrons per Sb atom. With
three Sb atoms per primitive unit cell, this yields precisely six electrons,
corresponding to half filling of the six-band $p_x/p_y$ manifold. Remarkably,
this filling is not externally tuned but follows directly from the chemistry of
the orbital-filtered Sb monolayer, thereby pinning the Fermi level to the closed
nodal line of the undistorted lattice. Unlike the isolated Dirac point of the
single-orbital model, the nodal line forms an extended Fermi contour (inset of Fig.~\ref{fgr: figure1}e) with a
finite density of states (right panels in Fig.~\ref{fgr: figure1}c and Fig.~\ref{fgr: figure1}f), thereby providing a much
stronger electronic driving force for the breathing distortion.

The nodal-line Fermi contour exhibits a characteristic spanning vector along $\Gamma$-M, 
reminiscent of the nesting vector associated with the van Hove singularity of the single-orbital 
kagome model and the charge-density-wave order found in several transition-metal kagome materials 
\cite{Ortiz2020CsV3Sb5,zhao2021cascade,tan2021charge,czn6-ndzm}, suggesting a possible finite-$q$ instability. 
However, DFT phonon calculations for the undistorted kagome lattice on the SiC substrate reveal the strongest instability 
at $\Gamma$ rather than M, signaled by an imaginary phonon frequency whose eigenvector corresponds 
to the Sb breathing distortion (see Section 5 of the Supplementary Information). The resulting $q=0$ 
mode preserves the unit cell and arises from the interplay of chemically enforced half filling, the 
closed nodal-line electronic structure, and electron–phonon coupling, allowing the electronic system 
to lower its energy through the breathing distortion. These results identify the latter as an intrinsic 
instability of the two-orbital kagome electronic structure realized in antimonene.

While this instability favors breathing, it does not by itself determine the orientation of the distortion. 
By spontaneously breaking the inversion symmetry of the ideal kagome lattice, the breathing distortion 
gives rise to two symmetry-related domains that are degenerate within the two-orbital kagome Hamiltonian. 
First-principles calculations for Sb/SiC(0001), however, reveal that this degeneracy is weakly lifted by 
the substrate. Carbon atoms in the second layer of the SiC substrate break the symmetry between the two 
domains \cite{bauernfeind2021design}, slightly lowering the energy of one relative to the other 
(Section 5 of the Supplementary Information). The substrate thus plays a dual role: orbital filtering 
establishes the two-orbital kagome electronic structure susceptible to breathing, while its weak 
symmetry-breaking potential selects a preferred domain, consistent with the single-domain 
breathing phase observed by STM.

\begin{figure*}[htbp]
\centering
\includegraphics[width=0.8\textwidth]{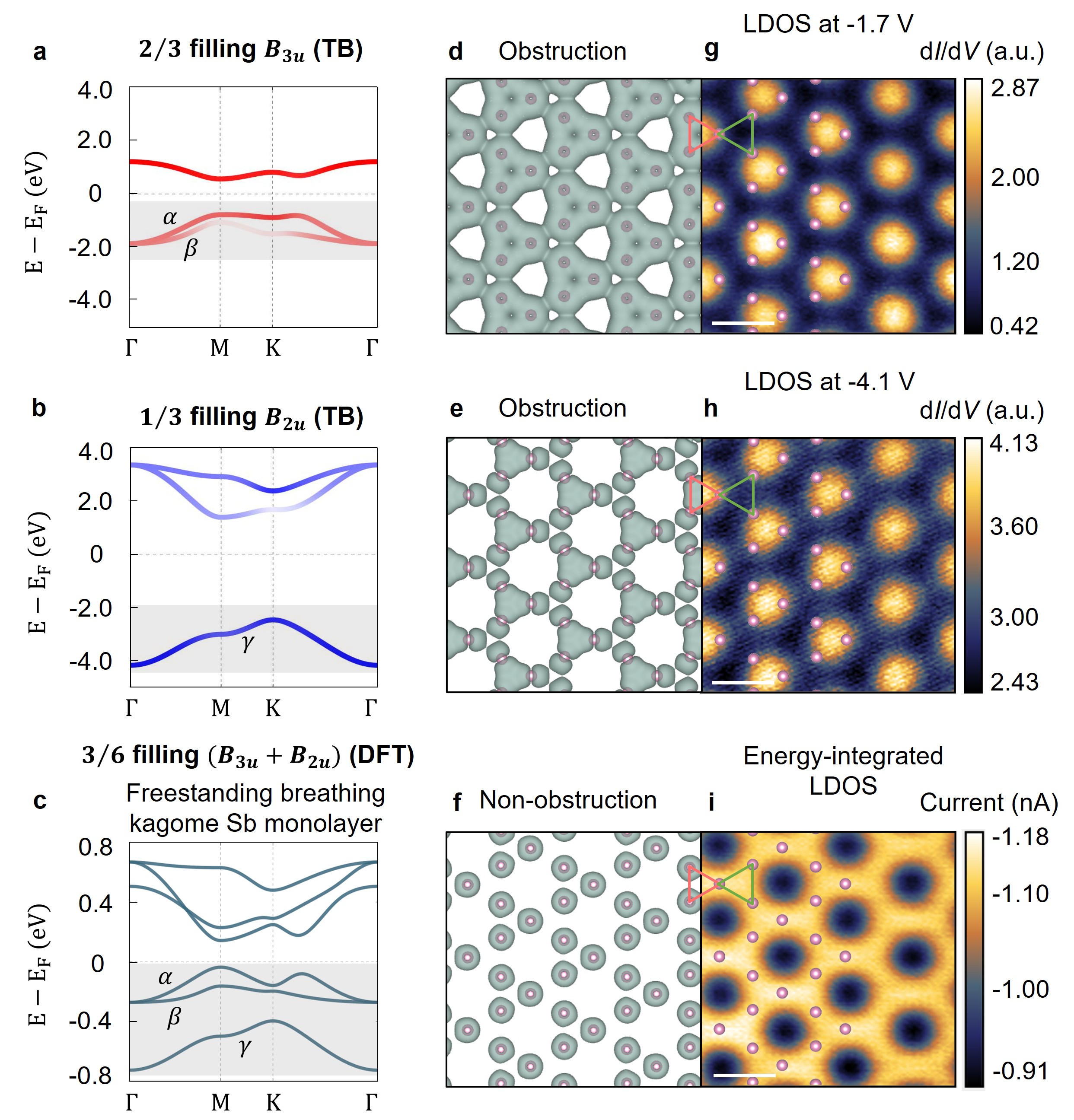}
\caption{Band-resolved obstructed atomic limits in the two-orbital breathing
kagome model. 
\textbf{a,b}, Tight-binding-calculated breathing kagome band
structures dominated by the $B_{3u}$ (red) and $B_{2u}$ (blue) orbitals, at
$2/3$ and $1/3$ filling, respectively. 
\textbf{c}, DFT band structure of a
simplified model system -- freestanding hydrogenated breathing kagome Sb
monolayer, where the similar occupied $\alpha$, $\beta$, and $\gamma$ bands are
associated with the $B_{3u}$ and $B_{2u}$ orbitals, respectively. 
\textbf{d--f},
DFT-calculated LDOS maps based on the band structure in \textbf{c}, obtained by
integrating over the $\alpha$ and $\beta$ bands (\textbf{d}), the $\gamma$ band
only (\textbf{e}), and all three valence bands (\textbf{f}). Sb atoms are shown
in pink. 
\textbf{g,h}, Experimentally measured STS maps acquired in
constant-height mode at bias voltages of $-1.7$\,V and $-4.1$\,V, respectively.
\textbf{i}, Constant-height STM current image acquired at $-3.8$\,V, reflecting
the energy-integrated LDOS. The smaller and larger triangles are indicated by
pink and green frames, respectively, and are overlaid on the STM/STS images
together with the Sb atomic lattice. The scale bar applies to all STM/STS images
and corresponds to $0.5$\,nm. }
\label{fgr:figure4}
\end{figure*}

\label{atomic_obstruction}

The same half filling that stabilizes the breathing instability also proves to
be special from a topological point of view. In the canonical single-orbital
breathing kagome model, theory predicts that opening the Dirac gap invariably produces an obstructed
atomic insulator, whose Wannier centres are displaced away from the kagome sites to
the centres of the breathing triangles \cite{ezawa2018higher}. This is analogous to obstructed phases recently identified in triangular-lattice transition-metal dichalcogenides \cite{cualuguaru2026observation,holbrook2026real}. Within this single-orbital picture, the atomic limit 
is therefore uniquely determined once the insulating gap is opened. The minimal
two-orbital kagome Hamiltonian realized here, however, exhibits a
qualitatively richer behaviour. Because the breathing distortion opens two
independent gaps in the six-band spectrum
(Fig.~\ref{fgr: figure1}f), the occupied subspaces at $1/6$ filling and at half
filling ($3/6$) can be classified independently. Remarkably, these insulating
phases possess different atomic limits: the isolated one-band and two-band
sectors are each obstructed, whereas their combined three-band valence manifold
at half filling realizes a non-obstructed atomic insulator. The same
half-filled insulating state that underlies the breathing instability therefore
also marks the transition from obstructed to non-obstructed atomic topology.

This hierarchy of atomic limits follows from a combined analysis of
symmetry-adapted Wannier functions, Wilson loops and symmetry indicators (see section 6 in the 
Supplementary Information for details). Both the isolated
$B_{2u}$-derived $\gamma$ band and the connected $B_{3u}$-derived
$\alpha+\beta$ manifold realize obstructed atomic limits, with Wannier centres
located at the triangle-centred Wyckoff positions and the corresponding
fractional corner charges. At half filling, these two obstructed sectors
recombine into a symmetry-allowed Wannier representation centred on the Sb
kagome sites. The complete three-band valence manifold therefore realizes a
non-obstructed atomic limit, demonstrating that
the atomic obstruction is a property of the occupied band projector rather than
of the crystal itself.

This hierarchy of atomic limits has direct experimental consequences, because
the obstructed and non-obstructed sectors are associated with distinct
real-space charge distributions that can be visualized by energy-resolved
local-density-of-states maps.

We now connect this theoretical picture to breathing kagome antimonene. To
isolate the intrinsic kagome-layer physics, we first consider a freestanding
hydrogenated Sb monolayer, which preserves the relevant symmetry and orbital
character while eliminating the SiC substrate. The DFT-calculated band
structure and energy-resolved LDOS
(Figs.~\ref{fgr:figure4}c--f) closely follow the Wannier analysis. The occupied
$\alpha+\beta$ manifold exhibits spectral weight concentrated at the centres of
the breathing triangles (Figs.~\ref{fgr:figure4}a and d), consistent with its obstructed atomic limit. The
isolated $\gamma$ band displays a complementary orbital texture but likewise
remains centred on the same triangle-centred Wyckoff position (Figs.~\ref{fgr:figure4}b and e), confirming it as
a second, independent obstructed sector. Only when all three in-plane valence
bands are considered together does the integrated LDOS evolve towards a
kagome-site-centred charge distribution, consistent with the predicted
non-obstructed atomic limit at half filling.

Scanning tunnelling microscopy and spectroscopy (STM/STS) provide direct
experimental support for this picture. Constant-height differential-conductance
($dI/dV$) maps acquired at $-1.7$\,V probe the $\alpha+\beta$ manifold and
reveal spectral weight concentrated near the centres of the breathing
triangles, in excellent agreement with the calculated LDOS (Figs.~\ref{fgr:figure4}d and g). At $-4.1$\,V,
where the $\gamma$ band contributes, the spectral weight again remains centred
on the breathing triangles (Figs.~\ref{fgr:figure4}e and h), confirming its obstructed character. By contrast,
constant-current STM images integrate all occupied states from the Fermi level
down to the measurement bias. As the integration window reaches the deeper
$\gamma$ band, the charge distribution progressively extends towards the Sb
sites (Figs.~\ref{fgr:figure4}f and i), consistent with the non-obstructed atomic limit of the complete
three-band valence manifold.

In the realistic material, the weakly dispersive Sb $p_z$ bands lie
energetically between the $\alpha+\beta$ and $\gamma$ manifolds
(Figs.~\ref{fgr: figure3}b and S7c). While these states modify the energetic
ordering of the bands, they neither alter the orbital character of the
in-plane kagome states nor their symmetry-preserving Wannier representation.
Taken together, the symmetry analysis, first-principles calculations and
STM/STS measurements establish orbital-resolved atomic obstruction in breathing
kagome antimonene. Although the underlying topology is defined by the occupied
band projector, the distinct $B_{2u}$ and $B_{3u}$ orbital character of the
corresponding manifolds enables its direct experimental identification. The
same elemental two-dimensional kagome crystal therefore hosts both obstructed
and non-obstructed atomic limits within a single electronic structure,
depending solely on the occupied band manifold.

\label{conclusion}

In conclusion, breathing kagome antimonene constitutes the first experimental
realization of the two-orbital extension of the canonical kagome
model.
Looking beyond the present work, the breathing kagome lattice provides a
natural platform for investigating higher-order topology and its connection to
corner states. Equally intriguing is the unusual electronic structure at the
fundamental direct band gap, where the lowest-energy optical transitions occur
between the gapped remnants of the nodal lines rather than isolated band
extrema. The resulting extended momentum-space phase space for direct optical
transitions may enable unconventional excitonic states and many-body optical
phenomena unique to multi-orbital kagome systems.

\section*{Methods}\label{sec11}

\subsection*{Sample preparation and STM/ARPES measurements}

Commercially available 4H-SiC(0001) wafers, n-type doped with a resistivity of
0.01--0.03~$\Omega$~cm, were treated by a hydrogen dry-etching process to obtain
a clean and well-defined $(1 \times 1)$ surface, with the top-layer Si atoms
passivated by hydrogen.
After etching, the samples were transferred \textit{in situ} into an
ultrahigh-vacuum (UHV) chamber for epitaxial growth, and the $(1 \times 1)$
surface was verified by low-energy electron diffraction (LEED).
In the preparation chamber, with a base pressure below $3 \times 10^{-10}$~mbar,
high-purity antimony (99.9999\%) was evaporated from a commercial cracker cell
(Dr. Eberl MBE-Komponenten GmbH).
The crucible, tube, and cracker temperatures were set to
460~$^{\circ}\mathrm{C}$, 620~$^{\circ}\mathrm{C}$, and
820~$^{\circ}\mathrm{C}$, respectively, ensuring efficient cracking of Sb$_4$
molecules into atomic Sb.
During Sb evaporation, the SiC sample was first heated to
600~$^{\circ}\mathrm{C}$ to desorb surface hydrogen and was then cooled to room
temperature to accumulate a large amount of Sb.
After Sb evaporation, the sample was annealed at about 400~$^{\circ}\mathrm{C}$
to form the $(2 \times 2)$-Sb/SiC phase, i.e., kagome antimonene.

After preparation, the sample was transferred \textit{in situ} into the UHV STM
and ARPES chambers for characterization. STM measurements were performed using a
commercial low-temperature STM system (Omicron LT-STM) operated at 4.7 K under a
base pressure below $5\times10^{-11}$ mbar. Tungsten tips were used and checked
on an Ag(111) surface for both topographic imaging and spectroscopic
measurements. ARPES measurements were carried out using a commercial
photoemission system (SPECS), equipped with a hemispherical analyzer (PHOIBOS
100) and a He-VUV discharge lamp (UVS 300), providing photons with an energy of
21.2 eV.

\subsection*{Density functional theory and lattice dynamics calculations}

First-principles calculations for kagome monolayer of antimony (Sb) on a
SiC(0001) substrate were performed within the framework of density functional
theory as implemented in the Vienna ab-intio simulation package
(VASP)~\cite{Kresse1, Kresse2} with the projector augmented wave (PAW)
basis~\cite{Kresse3}. The generalized gradient approximation was used for the
exchange correlation functional in the implementation of Perdew, Burke, and
Ernzerhof~\cite{Perdew}, by expanding the Kohn–Sham wave functions into
plane-waves up to an energy cut-off of 500 eV. To sample the Brillouin zone, we
have used a $12 \times 12 \times 1$ Monkhorst-Pack k-mesh~\cite{Monkhorst} and
employed spin-orbit coupling self-consistently. We consider a $(2 \times 2)$
reconstruction of triangular Sb on four layers of Si-terminated SiC(0001) with
an in-plane lattice constant of 3.07 \AA. The equilibrium structures were
obtained by relaxing all atoms until forces were converged below 0.001 eV/\AA.
To disentangle the electronic states of both surfaces, a vacuum distance of at
least 20 \AA~between periodic replicas in \textit{z}-direction is assumed and
the dangling bonds of the substrate terminated surface are saturated by
hydrogen. Structural models were visualized with VESTA~\cite{Momma:db5098}. To
ascertain the lattice dynamical stability of breathing kagome monolayer of
antimony (Sb) on a SiC(0001) substrate, lattice dynamics calculation was
performed using the supercell finite-differences approach implemented in the
Phonopy package~\cite{Togo_2023-A, Togo_2023-B}. The second-order (harmonic)
force constants were calculated using the $2 \times 2 \times 1$ supercell
containing 160 atoms, leading to 60 displacements. Note that the $2 \times 2
\times 1$ supercell is with respect to the $(2 \times 2)$ unit cell of the
kagome Sb monolayer on SiC(0001) substrate having 40 atoms.

\backmatter

\bmhead{Supplementary information}

The online version contains supplementary material is
available, including in-plane \textit{p}-orbital tight-binding model for the pristine and breathing kagome lattice, symmetry-enforced and accidental band degeneracies,
Sb atomic layer phases on SiC(0001), DFT picture of the Sb breathing kagome phase on SiC(0001), substrate stabilization of the breathing distortion, projector, Wilson-loop and Wannier diagnosis of the
orbital-resolved obstruction, and more evidence for the orbital-resolved atomic obstruction.

\bmhead{Acknowledgments}

B.L. gratefully acknowledges valuable discussions with X. Cai.
We are grateful for financial support by the Deutsche Forschungsgemeinschaft
(DFG, German Research Foundation) through the Würzburg-Dresden Cluster of
Excellence ctd.qmat – Complexity, Topology and Dynamics in Quantum Matter (EXC
2147, project-id 390858490) as well as through the Collaborative Research
Center SFB 1170 ToCoTronics (Project ID 258499086). B.L. gratefully acknowledges
support by the Alexander von Humboldt Foundation (Bonn, Germany). D.D.S acknowledges MUR funding within the FIS2 (n. 1236, 01-08-2023) Project no. FIS-2023-00144 (CUP J53C25001880001). We gratefully
acknowledge the Gauss Centre for Supercomputing e.V.
(https://www.gauss-centre.eu) for funding this project by providing computing
time on the GCS Supercomputer SuperMUC-NG at Leibniz Super-computing Centre
(https://www.lrz.de).


\section*{Competing interests}
The authors declare no competing interests.
\section*{Data availability} 
The data that support the findings of this study are available from the corresponding authors upon reasonable request.
\section*{Author contribution} 
B.L. grew the monolayer material, and acquired STM/STS data, with support from T.W. and J.Q.. B.L. performed ARPES measurement with support from J.E. and S.M.. B.L. analyzed experimental data with support from K.S. and J.S.. A.B. performed the theoretical calculations with support from D.D.S. and C.O.. A.B. and M.V. performed the DFT calculation and analysis. B.L. and A.B. prepared the paper with substantial contributions from all authors. J.S., R.T., G.S. and R.C. supervised the project.

\bigskip


\bibliography{Sb_kagome_SiC}

@article{disante2026kagome,
  author  = {Di Sante, Domenico and Titus Neupert and Giorgio Sangiovanni and
             Ronny Thomale and Riccardo Comin and
             Joseph G. Checkelsky and Ilija Zeljkovic and
             Stephen D. Wilson},
  title   = {Kagome metals},
  journal = {Rev. Mod. Phys.},
  volume  = {98},
  number  = {1},
  pages   = {015002},
  year    = {2026},
  doi     = {10.1103/RevModPhys.98.015002},
  publisher = {American Physical Society}
}

@article{bauernfeind2021design,
  title={Design and realization of topological Dirac fermions on a triangular lattice},
  author={Bauernfeind, Maximilian and others},
  journal={Nat. Commun.},
  volume={12},
  number={1},
  pages={5396},
  year={2021},
  publisher={Nature Publishing Group UK London}
}

@article{schmitt2024achieving,
  title={Achieving environmental stability in an atomically thin quantum spin Hall insulator via graphene intercalation},
  author={Schmitt, Cedric and others},
  journal={Nat. Commun.},
  volume={15},
  number={1},
  pages={1486},
  year={2024},
  publisher={Nature Publishing Group UK London}
}

@article{erhardt2024bias,
  title={Bias-free access to orbital angular momentum in two-dimensional quantum materials},
  author={Erhardt, Jonas and others},
  journal={Phys. Rev. Lett.},
  volume={132},
  number={19},
  pages={196401},
  year={2024},
  publisher={APS}
}

@article{stuhler2020tomonaga,
  title={Tomonaga--Luttinger liquid in the edge channels of a quantum spin Hall insulator},
  author={St{\"u}hler, R and others},
  journal={Nat. Phys.},
  volume={16},
  number={1},
  pages={47--51},
  year={2020},
  publisher={Nature Publishing Group UK London}
}

@article{stuhler2022effective,
  title={Effective lifting of the topological protection of quantum spin Hall edge states by edge coupling},
  author={St{\"u}hler, R and others},
  journal={Nat. Commun.},
  volume={13},
  number={1},
  pages={3480},
  year={2022},
  publisher={Nature Publishing Group UK London}
}

@article{reis2017bismuthene,
  title={Bismuthene on a SiC substrate: A candidate for a high-temperature quantum spin Hall material},
  author={Reis, F and others},
  journal={Science},
  volume={357},
  number={6348},
  pages={287--290},
  year={2017},
  publisher={American Association for the Advancement of Science}
}

@article{kang2020dirac,
  title={Dirac fermions and flat bands in the ideal kagome metal {FeSn}},
  author={Kang, Mingu and others},
  journal={Nat. Mater.},
  volume={19},
  number={2},
  pages={163--169},
  year={2020},
  publisher={Nature Publishing Group UK London}
}

@article{slot2017experimental,
  title={Experimental realization and characterization of an electronic Lieb lattice},
  author={Slot, Marlou R and others},
  journal={Nat. Phys.},
  volume={13},
  number={7},
  pages={672--676},
  year={2017},
  publisher={Nature Publishing Group UK London}
}

@article{zhang2023realization,
  title={Realization of photonic \textit{p}-orbital higher-order topological insulators},
  author={Zhang, Yahui and others},
  journal={eLight},
  volume={3},
  number={1},
  pages={5},
  year={2023},
  publisher={Springer}
}

@article{lu2020orbital,
  title={Orbital corner states on breathing kagome lattices},
  author={Lu, Xiancong and Chen, Ying and Chen, Huanyang},
  journal={Phys. Rev. B},
  volume={101},
  number={19},
  pages={195143},
  year={2020},
  publisher={APS}
}

@article{liu2021screening,
  title={Screening two-dimensional materials with topological flat bands},
  author={Liu, Hang and Meng, Sheng and Liu, Feng},
  journal={Phys. Rev. Mater.},
  volume={5},
  number={8},
  pages={084203},
  year={2021},
  publisher={APS}
}

@article{jiang2019topological,
  title={Topological band evolution between Lieb and kagome lattices},
  author={Jiang, Wei and others},
  journal={Phys. Rev. B},
  volume={99},
  number={12},
  pages={125131},
  year={2019},
  publisher={APS}
}

@article{hu2022rich,
  title={Rich nature of Van Hove singularities in Kagome superconductor $\rm{CsV_{3}Sb_{5}}$},
  author={Hu, Yong and others},
  journal={Nat. Commun.},
  volume={13},
  number={1},
  pages={2220},
  year={2022},
  publisher={Nature Publishing Group UK London}
}

@article{wu2021nature,
  title={Nature of unconventional pairing in the kagome superconductors $\rm{AV_{3}Sb_{5} ~(A= K, Rb, Cs)}$},
  author={Wu, Xianxin and others},
  journal={Phys. Rev. Lett.},
  volume={127},
  number={17},
  pages={177001},
  year={2021},
  publisher={APS}
}

@article{yin2022topological,
  title={Topological kagome magnets and superconductors},
  author={Yin, Jia-Xin and Lian, Biao and Hasan, M Zahid},
  journal={Nature},
  volume={612},
  number={7941},
  pages={647--657},
  year={2022},
  publisher={Nature Publishing Group UK London}
}

@article{ye2024hopping,
  title={Hopping frustration-induced flat band and strange metallicity in a kagome metal},
  author={Ye, Linda and others},
  journal={Nat. Phys.},
  volume={20},
  number={4},
  pages={610--614},
  year={2024},
  publisher={Nature Publishing Group UK London}
}

@article{zhao2021cascade,
  title={Cascade of correlated electron states in the kagome superconductor $\rm{CsV_{3}Sb_{5}}$},
  author={Zhao, He and others},
  journal={Nature},
  volume={599},
  number={7884},
  pages={216--221},
  year={2021},
  publisher={Nature Publishing Group UK London}
}

@article{teng2022discovery,
  title={Discovery of charge density wave in a kagome lattice antiferromagnet},
  author={Teng, Xiaokun and others},
  journal={Nature},
  volume={609},
  number={7927},
  pages={490--495},
  year={2022},
  publisher={Nature Publishing Group UK London}
}

@article{tan2021charge,
  title={Charge density waves and electronic properties of superconducting kagome metals},
  author={Tan, Hengxin and Liu, Yizhou and Wang, Ziqiang and Yan, Binghai},
  journal={Phys. Rev. Lett.},
  volume={127},
  number={4},
  pages={046401},
  year={2021},
  publisher={APS}
}

@article{Ortiz2020CsV3Sb5,
  author       = {Brenden R. Ortiz and Samuel M. L. Teicher and Yong Hu and
                  Julia L. Zuo and Paul M. Sarte and Emily C. Schueller and
                  A. M. Milinda Abeykoon and Matthew J. Krogstad and
                  Stefan Rosenkranz and Raymond Osborn and
                  Ram Seshadri and Leon Balents and
                  Junfeng He and Stephen D. Wilson},
  title        = {CsV$_3$Sb$_5$: A {$\mathbb{Z}_2$} Topological Kagome Metal with a Superconducting Ground State},
  journal      = {Physical Review Letters},
  volume       = {125},
  pages        = {247002},
  year         = {2020},
  doi          = {10.1103/PhysRevLett.125.247002}
}

@article{yong2023phonon,
  title={Phonon promoted charge density wave in topological kagome metal $\rm{ScV_{6}Sn_{6}}$},
  author={Hu, Yong and others},
  journal={Nat. Commun.},
  volume={15},
  number={1},
  pages={1658},
  year={2024},
  publisher={Nature Publishing Group UK London}
}

@article{deng2024chiral,
  title={Chiral kagome superconductivity modulations with residual Fermi arcs},
  author={Deng, Hanbin and others},
  journal={Nature},
  volume={632},
  number={8026},
  pages={775--781},
  year={2024},
  publisher={Nature Publishing Group UK London}
}

@article{jiang2023kagome,
  title={Kagome superconductors $\rm{AV_{3}Sb_{5}~(A=K, Rb, Cs)}$},
  author={Jiang, Kun and others},
  journal={Natl. Sci. Rev.},
  volume={10},
  number={2},
  pages={nwac199},
  year={2023},
  publisher={Oxford University Press}
}

@article{han2012fractionalized,
  title={Fractionalized excitations in the spin-liquid state of a kagome-lattice antiferromagnet},
  author={Han, Tian-Heng and others},
  journal={Nature},
  volume={492},
  number={7429},
  pages={406--410},
  year={2012},
  publisher={Nature Publishing Group UK London}
}

@article{lee2007quantum,
  title={Quantum-spin-liquid states in the two-dimensional kagome antiferromagnets $\rm{Zn_{x}Cu_{4- x} (OD)_{6}Cl_{2}}$},
  author={Lee, S-H and others},
  journal={Nat. Mater.},
  volume={6},
  number={11},
  pages={853--857},
  year={2007},
  publisher={Nature Publishing Group UK London}
}

@article{ezawa2018higher,
  title={Higher-order topological insulators and semimetals on the breathing kagome and pyrochlore lattices},
  author={Ezawa, Motohiko},
  journal={Phys. Rev. Lett.},
  volume={120},
  number={2},
  pages={026801},
  year={2018},
  publisher={APS}
}

@article{kempkes2019robust,
  title={Robust zero-energy modes in an electronic higher-order topological insulator},
  author={Kempkes, SN and others},
  journal={Nat. Mater.},
  volume={18},
  number={12},
  pages={1292--1297},
  year={2019},
  publisher={Nature Publishing Group UK London}
}

@article{wang2023sub,
  title={Sub-symmetry-protected topological states},
  author={Wang, Ziteng and others},
  journal={Nat. Phys.},
  volume={19},
  number={7},
  pages={992--998},
  year={2023},
  publisher={Nature Publishing Group UK London}
}

@article{shi2022new,
  title={A new class of bilayer kagome lattice compounds with Dirac nodal lines and pressure-induced superconductivity},
  author={Shi, Mengzhu and others},
  journal={Nat. Commun.},
  volume={13},
  number={1},
  pages={2773},
  year={2022},
  publisher={Nature Publishing Group UK London}
}

@article{morali2019fermi,
  title={Fermi-arc diversity on surface terminations of the magnetic Weyl semimetal $\rm{Co_{3}Sn_{2}S_{2}}$},
  author={Morali, Noam and others},
  journal={Science},
  volume={365},
  number={6459},
  pages={1286--1291},
  year={2019},
  publisher={American Association for the Advancement of Science}
}

@article{kang2020topological,
  title={Topological flat bands in frustrated kagome lattice CoSn},
  author={Kang, Mingu and others},
  journal={Nat. Commun.},
  volume={11},
  number={1},
  pages={4004},
  year={2020},
  publisher={Nature Publishing Group UK London}
}

@article{slater1954simplified,
  title={Simplified LCAO method for the periodic potential problem},
  author={Slater, John C and Koster, George F},
  journal={Phys. Rev.},
  volume={94},
  number={6},
  pages={1498},
  year={1954},
  publisher={APS}
}

@article{van2020topological,
  title={On the topological immunity of corner states in two-dimensional crystalline insulators},
  author={van Miert, Guido and Ortix, Carmine},
  journal={npj Quantum Mater.},
  volume={5},
  number={1},
  pages={63},
  year={2020},
  publisher={Nature Publishing Group UK London}
}

@article{bradlyn2017topological,
  title={Topological quantum chemistry},
  author={Bradlyn, Barry and others},
  journal={Nature},
  volume={547},
  number={7663},
  pages={298--305},
  year={2017},
  publisher={Nature Publishing Group UK London}
}

@article{henkelman2000climbing,
  title={A climbing image nudged elastic band method for finding saddle points and minimum energy paths},
  author={Henkelman, Graeme and Uberuaga, Blas P and J{\'o}nsson, Hannes},
  journal={J. Chem. Phys.},
  volume={113},
  number={22},
  pages={9901--9904},
  year={2000},
  publisher={American Institute of Physics}
}

@article{Kresse1,
  title = {Ab initio molecular dynamics for liquid metals},
  author = {Kresse, G. and Hafner, J.},
  journal = {Phys. Rev. B},
  volume = {47},
  issue = {1},
  pages = {558--561},
  numpages = {0},
  year = {1993},
  month = {Jan},
  publisher = {American Physical Society}
}

@article{Kresse2,
  title = {Efficient iterative schemes for ab initio total-energy calculations using a plane-wave basis set},
  author = {Kresse, G. and Furthm\"uller, J.},
  journal = {Phys. Rev. B},
  volume = {54},
  issue = {16},
  pages = {11169--11186},
  numpages = {0},
  year = {1996},
  month = {Oct},
  publisher = {American Physical Society}
}

@article{Kresse3,
  title = {From ultrasoft pseudopotentials to the projector augmented-wave method},
  author = {Kresse, G. and Joubert, D.},
  journal = {Phys. Rev. B},
  volume = {59},
  issue = {3},
  pages = {1758--1775},
  numpages = {0},
  year = {1999},
  month = {Jan}
}

@article{Perdew,
  title = {Generalized Gradient Approximation Made Simple},
  author = {Perdew, John P. and Burke, Kieron and Ernzerhof, Matthias},
  journal = {Phys. Rev. Lett.},
  volume = {77},
  issue = {18},
  pages = {3865--3868},
  numpages = {0},
  year = {1996},
  month = {Oct},
  publisher = {American Physical Society}
}

@article{Monkhorst,
  title = {Special points for Brillouin-zone integrations},
  author = {Monkhorst, Hendrik J. and Pack, James D.},
  journal = {Phys. Rev. B},
  volume = {13},
  issue = {12},
  pages = {5188--5192},
  numpages = {0},
  year = {1976},
  month = {Jun},
  publisher = {American Physical Society}
}

@article{Momma:db5098,
author = "Momma, Koichi and Izumi, Fujio",
title = "{{\it VESTA3} for three-dimensional visualization of crystal, volumetric and morphology data}",
journal = "J. Appl. Crystallogr.",
year = "2011",
volume = "44",
number = "6",
pages = "1272--1276",
month = "Dec"
}

@article{Togo_2023-A,
year = {2023},
month = {jun},
publisher = {IOP Publishing},
volume = {35},
number = {35},
pages = {353001},
author = {Togo, Atsushi and Chaput, Laurent and Tadano, Terumasa and Tanaka, Isao},
title = {Implementation strategies in phonopy and phono3py},
journal = {J. Phys. Condens. Matter}
}

@article{Togo_2023-B,
author = {Togo ,Atsushi},
title = {First-principles Phonon Calculations with Phonopy and Phono3py},
journal = {J. Phys. Soc. Jpn.},
volume = {92},
number = {1},
pages = {012001},
year = {2023},
}

@article{yan2026engineering,
  title={Engineering phase-frustration-induced flat bands in an aza-triangulene covalent kagome lattice},
  author={Yan, Yuyi and others},
  journal={Nat. Mat.},
  volume={25},
  pages={982--989},
  year={2026},
  publisher={Nature Publishing Group UK London}
}

@article{cualuguaru2026observation,
  title={Observation of an obstructed atomic band in a transition metal dichalcogenide},
  author={C{\u{a}}lug{\u{a}}ru, Dumitru and others},
  journal={Nat. Phys.},
  volume={22},
  pages={686--691},
  year={2026},
  publisher={Nature Publishing Group UK London}
}

@article{holbrook2026real,
  title={Real-space imaging of the band topology of transition metal dichalcogenides},
  author={Holbrook, Madisen and Ingham, Julian and Kaplan, Daniel and others},
  journal={Nat. Phys.},
  volume={22},
  pages={680--685},
  year={2026},
  publisher={Nature Publishing Group UK London}
}

@article{glass2016atomic,
  title={Atomic-scale mapping of layer-by-layer hydrogen etching and passivation of SiC (0001) substrates},
  author={Glass, Stefan and Reis, Felix and Bauernfeind, Maximilian and Aulbach, Julian and Scholz, Markus R and Adler, Florian and Dudy, Lenart and Li, Gang and Claessen, Ralph and Schäfer, Jörg},
  journal={J. Phys. Chem. C},
  volume={120},
  number={19},
  pages={10361--10367},
  year={2016},
  publisher={ACS Publications}
}

@article{aretz2025strong,
  title={From strong to weak correlations in breathing-mode kagome van der waals materials: $\rm{Nb_{3}(F, Cl, Br, I)_{8}}$ as a robust and versatile platform for many-body engineering},
  author={Aretz, Joost and Grytsiuk, Sergii and Liu, Xiaojing and Feraco, Giovanna and Knekna, Chrystalla and Waseem, Muhammad and Dan, Zhiying and Bianchi, Marco and Hofmann, Philip and Ali, Mazhar N and others},
  journal={Phys. Rev. X},
  volume={15},
  number={4},
  pages={041042},
  year={2025},
  publisher={APS}
}

@article{grytsiuk2024nb3cl8,
  title={$\rm{Nb_3Cl_8}$: a prototypical layered Mott-Hubbard insulator},
  author={Grytsiuk, Sergii and Katsnelson, Mikhail I and Loon, Erik GCP van and R{\"o}sner, Malte},
  journal={npj Quantum Mater.},
  volume={9},
  number={1},
  pages={8},
  year={2024},
  publisher={Nature Publishing Group UK London}
}

@article{chen2023visualizing,
  title={Visualizing the localized electrons of a kagome flat band},
  author={Chen, Caiyun and Zheng, Jiangchang and Yu, Ruopeng and Sankar, Soumya and Law, Kam Tuen and Po, Hoi Chun and J{\"a}ck, Berthold},
  journal={Physical Review Research},
  volume={5},
  number={4},
  pages={043269},
  year={2023},
  publisher={APS}
}

@article{ye2018massive,
  title={Massive Dirac fermions in a ferromagnetic kagome metal},
  author={Ye, Linda and Kang, Mingu and Liu, Junwei and Von Cube, Felix and Wicker, Christina R and Suzuki, Takehito and Jozwiak, Chris and Bostwick, Aaron and Rotenberg, Eli and Bell, David C and others},
  journal={Nature},
  volume={555},
  number={7698},
  pages={638--642},
  year={2018},
  publisher={Nature Publishing Group UK London}
}

@article{czn6-ndzm,
  title = {Phonon fluctuation diagnostics: Origin of charge order in $\rm{AV_3Sb_5}$ kagome metals},
  author = {Enzner, Stefan and Berges, Jan and Schobert, Arne and Oh, Dongjin and Kang, Mingu and Comin, Riccardo and Thomale, Ronny and Wehling, Tim and Di Sante, Domenico and Sangiovanni, Giorgio},
  journal = {Phys. Rev. B},
  volume = {114},
  issue = {14},
  pages = {144304},
  numpages = {8},
  year = {2026},
  month = {Sep},
  publisher = {American Physical Society},
  doi = {10.1103/czn6-ndzm}
}

\end{document}


\title[Supplementary Information]{Supplementary Information: Experimental realization of the minimal two-dimensional multi-orbital kagome model}

\author[1,2]{\fnm{Bing} \sur{Liu}}
\equalcont{These authors contributed equally to this work.}

\author[2,3]{\fnm{Arka} \sur{Bandyopadhyay}}
\equalcont{These authors contributed equally to this work.}

\author[2,3]{\fnm{Manish} \sur{Verma}}

\author[1,2]{\fnm{Jonas} \sur{Erhardt}}

\author[1,2]{\fnm{Tim} \sur{Wagner}}

\author[1,2]{\fnm{Jing} \sur{Qi}}

\author[1,2]{\fnm{Kilian} \sur{Strauß}}

\author[4]{\fnm{Domenico Di} \sur{Sante}}

\author[5]{\fnm{Carmine} \sur{Ortix}}

\author[1,2,6]{\fnm{Simon} \sur{Moser}}

\author[1,2]{\fnm{Jörg} \sur{Schäfer}}

\author[2,3]{\fnm{Ronny} \sur{Thomale}}

\author*[2,3]{\fnm{Giorgio} \sur{Sangiovanni}}\email{sangiovanni@physik.uni-wuerzburg.de}

\author*[1,2]{\fnm{Ralph} \sur{Claessen}}\email{claessen@physik.uni-wuerzburg.de}

\affil[1]{\orgdiv{Physikalisches Institut}, \orgname{Universit\"at W\"urzburg}, \orgaddress{\city{W\"urzburg}, \postcode{D-97074}, \country{Germany}}}

\affil[2]{\orgdiv{W\"urzburg-Dresden Cluster of Excellence ctd.qmat}, \orgname{Universit\"at W\"urzburg}, \orgaddress{\city{W\"urzburg}, \postcode{D-97074}, \country{Germany}}}

\affil[3]{\orgdiv{Institut f\"ur Theoretische Physik und Astrophysik}, \orgname{Universit\"at W\"urzburg}, \orgaddress{\city{W\"urzburg}, \postcode{D-97074}, \country{Germany}}}

\affil[4]{\orgdiv{Department of Physics and Astronomy}, \orgname{University of Bologna}, \orgaddress{\city{Bologna}, \postcode{40127}, \country{Italy}}}

\affil[5]{\orgdiv{Dipartimento di Fisica “E. R. Caianiello”}, \orgname{Università di Salerno}, \orgaddress{\city{Fisciano}, \country{Italy}}}

\affil[6]{\orgdiv{Thin Films \& Physics of Quantum Materials, Faculty of Physics}, \orgname{Bielefeld University}, \orgaddress{\city{Bielefeld}, \postcode{33615}, \country{Germany}}}

\maketitle

\clearpage
\tableofcontents
\newpage

\section{In-plane \textit{p}-orbital tight-binding model for the pristine and breathing kagome lattice}
\label{sec:tb}

In this section we define a minimal multi-orbital kagome Hamiltonian built from in-plane
$p_x$ and $p_y$ orbitals and follow its band structure as a function of the two control
parameters of the model: the $\pi/\sigma$ hopping ratio and the breathing amplitude. The
interpretation of the resulting spectra in terms of symmetry and of an emergent nodal
contour is given in Sec.~\ref{sec:symmetry}.

\subsection*{Lattice, unit cell and bond convention}

The kagome lattice is a triangular Bravais lattice with a three-site basis forming
corner-sharing triangles. We use the primitive vectors
\begin{equation}
\mathbf{a}_1=a\left(1,0\right),\qquad
\mathbf{a}_2=a\left(-1/2,\sqrt3/2\right),
\qquad \mathbf{a}_1\cdot\mathbf{a}_2=-a^2/2,
\label{eq:lattice_vectors}
\end{equation}
with reciprocal vectors defined by $\mathbf{a}_i\cdot\mathbf{b}_j=2\pi\delta_{ij}$. The
high-symmetry momenta used throughout are $\Gamma=\mathbf{0}$,
$K=(\mathbf{b}_1+\mathbf{b}_2)/3$, $K'=-K$ and $M=(\mathbf{b}_1-\mathbf{b}_2)/2$, with
$|K|=4\pi/(3a)$ and $|M|=2\pi/(\sqrt3a)$. We place
the origin at the centre of a kagome hexagon (Wyckoff position $1a$). The three sublattices
$A$, $B$ and $C$ occupy the Wyckoff position $3f$. Among the equivalent representatives we
choose those that form a single \emph{small} triangle, which we take as the unit cell:
$\boldsymbol{\tau}_A=\mathbf{a}_1/2$, $\boldsymbol{\tau}_B=(\mathbf{a}_1+\mathbf{a}_2)/2$
and $\boldsymbol{\tau}_C=\mathbf{a}_1+\mathbf{a}_2/2$, i.e.\ $(1/2,0)$, $(1/2,1/2)$ and $(1,1/2)$ in fractional coordinates.
The centroids of the small and large triangles are then
$\boldsymbol{\tau}_{t_2}=(2/3,1/3)$ and $\boldsymbol{\tau}_{t_1}=(1/3,2/3)$
(together forming the Wyckoff position $2c$ of $p6/mmm$, which splits into two
inequivalent one-fold positions once $t_1\neq t_2$). Shifting the origin to site $A$ gives
$\boldsymbol{\tau}_A=(0,0)$, $\boldsymbol{\tau}_B=(0,1/2)$,
$\boldsymbol{\tau}_C=(1/2,1/2)$ and $\boldsymbol{\tau}_{t_2}=(1/6,1/3)$,
which is the embedding used in Sec.~\ref{sec:wilson}.

The three bond orientations are
\begin{equation}
\boldsymbol{\delta}_1=\mathbf{a}_2=a\left(-1/2,\sqrt3/2\right),
\qquad
\boldsymbol{\delta}_2=\mathbf{a}_1=a\left(1,0\right),
\qquad
\boldsymbol{\delta}_3=-(\mathbf{a}_1+\mathbf{a}_2)=a\left(-1/2,-\sqrt3/2\right),
\label{eq:delta_vectors}
\end{equation}
with $\boldsymbol{\delta}_1+\boldsymbol{\delta}_2+\boldsymbol{\delta}_3=0$; bond family
$i=1,2,3$ connects the sublattice pairs $A$--$B$, $B$--$C$ and $C$--$A$, respectively. Each
$\boldsymbol{\delta}_i$ is a lattice translation and, up to a factor $1/2$, the displacement
of a nearest-neighbour bond. Every bond family contains one bond per triangle.
Throughout this Supplement $t_2$ denotes the hopping on the short bonds (small triangles) and
$t_1$ that on the long bonds (large triangles), so that $|t_2|>|t_1|$ in the breathing
phase and $t_1=t_2$ in the pristine lattice; this is the convention of Fig.~1d--f of the
main text and of Sec.~\ref{sec:wilson}. With this unit cell, the nearest-neighbour bonds, written as
matrix elements $\bra{\alpha,\mathbf{0}}\hat H\ket{\beta,\mathbf{R}}$ between orbitals in
the home cell and in cell $\mathbf{R}$, are the intra-cell (small-triangle, $t_2$) bonds
$(B,\mathbf{0})$--$(A,\mathbf{0})$, $(C,\mathbf{0})$--$(B,\mathbf{0})$ and
$(A,\mathbf{0})$--$(C,\mathbf{0})$, and the inter-cell (large-triangle, $t_1$) bonds
$(B,\mathbf{0})$--$(A,\boldsymbol{\delta}_1)$, $(C,\mathbf{0})$--$(B,\boldsymbol{\delta}_2)$ and
$(A,\mathbf{0})$--$(C,\boldsymbol{\delta}_3)$, together with their Hermitian conjugates.

\subsection*{Orbital basis and local frames}

At every site we retain the two in-plane orbitals $p_x$ and $p_y$. We distinguish the
\emph{global} Cartesian orbital basis, in which the Hamiltonian is most
compactly written, from the \emph{local} site-adapted basis in which the orbitals carry
their symmetry labels. The site-symmetry group of $3f$ in the pristine lattice is $mmm$
($D_{2h}$, order 8), whose two in-plane twofold axes are defined geometrically (Fig.~1d of the main text). The
triangle-radial axis $\hat{\mathbf{y}}_\alpha$ points from site $\alpha$ to the centroid of the
small triangle to which it belongs; the same line passes through the centroid of the large
triangle on the opposite side, and $\hat{\mathbf{y}}_\alpha$ bisects the two bonds of the
\emph{same} triangle. The hexagon-radial axis
$\hat{\mathbf{e}}_\alpha=\hat{\mathbf{y}}_\alpha\times\hat{\mathbf{z}}$ points from the centre of
the adjacent hexagon to site $\alpha$ and bisects two bonds belonging to \emph{different}
triangles.
For site $A$ at $\boldsymbol{\tau}_A=\mathbf{a}_1/2$, $\hat{\mathbf{e}}_A=\hat{\mathbf{x}}$ and
$\hat{\mathbf{y}}_A=\hat{\mathbf{y}}$; the frames of $C$ and $B$ follow by $C_3$ rotations about the triangle centroid by
$120^\circ$ and $240^\circ$, respectively. The standard $D_{2h}$
labelling in the local frame gives
\begin{equation}
p_{\parallel}^{(\alpha)}\equiv \hat{\mathbf{e}}_\alpha\cdot\mathbf{p}\;\rightarrow\;B_{3u},
\qquad
p_{\perp}^{(\alpha)}\equiv \hat{\mathbf{y}}_\alpha\cdot\mathbf{p}\;\rightarrow\;B_{2u}.
\label{eq:orbital_labels}
\end{equation}
In the language of an isolated trimer, the $B_{2u}$ orbital is therefore \emph{radial}
(pointing towards the trimer centre) and the $B_{3u}$ orbital \emph{tangential}. These are
the orbitals sketched in Fig.~1d of the main text, and they are recovered independently by
the Wannier construction of Sec.~\ref{sec:dft}.

\subsection*{Spinless Hamiltonian}

Because the $p$ orbitals are directional, the hopping amplitudes depend on the relative
orientation of the orbital lobes and the bond. This is captured by the Slater--Koster
formalism~\cite{slater1954simplified}, which decomposes each bond into a $\sigma$
(bond-parallel) and a $\pi$ (bond-perpendicular) channel. With
$\hat{\mathbf{n}}_i=\boldsymbol{\delta}_i/|\boldsymbol{\delta}_i|$ we define the orthogonal
projectors in the $\{p_x,p_y\}$ subspace,
\begin{equation}
P_i^{\sigma}=\hat{\mathbf{n}}_i\hat{\mathbf{n}}_i^{\top},\qquad
P_i^{\pi}=\mathbb{I}_2-P_i^{\sigma},\qquad
P_i^{\sigma}P_i^{\pi}=0,
\label{eq:P_proj}
\end{equation}
and the breathing-resolved Bloch form factors
\begin{equation}
f_i^{\sigma}(\mathbf{k})=V_2^{\sigma}+V_1^{\sigma}e^{\,i\mathbf{k}\cdot\boldsymbol{\delta}_i},
\qquad
f_i^{\pi}(\mathbf{k})=V_2^{\pi}+V_1^{\pi}e^{\,i\mathbf{k}\cdot\boldsymbol{\delta}_i},
\label{eq:f_sigma_pi}
\end{equation}
where $V^{\sigma}$ and $V^{\pi}$ are the Slater--Koster integrals of the global
$(p_x,p_y)$ basis, the subscripts $1$ and $2$ refer to the long and short bonds, and the
phase-free term is the intra-cell (small-triangle) bond. The
$2\times2$ hopping block along bond orientation $i$ has the standard Slater--Koster form
\begin{equation}
T_i(\mathbf{k})=f_i^{\sigma}(\mathbf{k})\,P_i^{\sigma}+f_i^{\pi}(\mathbf{k})\,P_i^{\pi}.
\label{eq:T_i}
\end{equation}
In the six-dimensional basis
$\Psi_{\mathbf{k}}=(a_{\mathbf{k},x},a_{\mathbf{k},y},b_{\mathbf{k},x},b_{\mathbf{k},y},
c_{\mathbf{k},x},c_{\mathbf{k},y})^{\top}$, where $a,b,c$ label the sublattices and $x,y$
the global orbital components, the spinless Bloch Hamiltonian is
\begin{equation}
\mathcal{H}(\mathbf{k})=
\begin{pmatrix}
0 & T_1^{\dagger}(\mathbf{k}) & T_3(\mathbf{k})\\[2pt]
T_1(\mathbf{k}) & 0 & T_2^{\dagger}(\mathbf{k})\\[2pt]
T_3^{\dagger}(\mathbf{k}) & T_2(\mathbf{k}) & 0
\end{pmatrix}.
\label{eq:H_k}
\end{equation}
It is written in the periodic (basis) gauge,
$\mathcal{H}_{\alpha\beta}(\mathbf{k})=\sum_{\mathbf{R}}
\bra{\alpha,\mathbf{0}}\hat H\ket{\beta,\mathbf{R}}\,e^{i\mathbf{k}\cdot\mathbf{R}}$, so that
$\mathcal{H}(\mathbf{k}+\mathbf{G})=\mathcal{H}(\mathbf{k})$, and follows directly from
the bond list given above. The sublattice embedding
$\boldsymbol{\tau}_\alpha$ does not enter $\mathcal{H}(\mathbf{k})$; it enters through the
phases $e^{-i\mathbf{k}\cdot\boldsymbol{\tau}_\alpha}$ in the symmetry operators of
Sec.~\ref{sec:symmetry} and in the Wilson loops of Sec.~\ref{sec:wilson}. Since all hopping
amplitudes are real, $\mathcal{H}(\mathbf{k})$ obeys spinless time-reversal symmetry,
$\mathcal{H}^{*}(\mathbf{k})=\mathcal{H}(-\mathbf{k})$.

The Hamiltonian of Eq.~\eqref{eq:H_k} is constructed in the global $(p_x,p_y)$ basis, in
which the Slater--Koster blocks take their simplest form. Because the three sublattices are
related by $C_3$, the natural basis for discussing the results is the locally rotated one
of Eq.~\eqref{eq:orbital_labels}. The two bases are related by the site-diagonal,
momentum-independent unitary $U=\mathrm{diag}(R_A,R_B,R_C)$, where $R_\alpha$ rotates
$(\hat{\mathbf{x}},\hat{\mathbf{y}})$ onto the right-handed local frame
$(\hat{\mathbf{e}}_\alpha,\hat{\mathbf{y}}_\alpha)$, and
$\mathcal{H}_{\mathrm{loc}}(\mathbf{k})=U^{\dagger}\mathcal{H}(\mathbf{k})U$. The band structures are identical in both bases, and the orbital character shown
in Figs.~\ref{fig:S1}--\ref{fig:S4} is the weight on the rotated $B_{2u}$ and $B_{3u}$
orbitals. In the rotated basis we parametrize each nearest-neighbour bond by
\begin{equation}
t^{\sigma}=V^{\sigma},\qquad t^{\pi}=-V^{\pi},
\label{eq:t_from_V}
\end{equation}
so that the block coupling $(B_{3u},B_{2u})$ on site $\alpha$ to $(B_{3u},B_{2u})$ on site
$\beta$, with $(\alpha,\beta)=(A,B),(B,C),(C,A)$, reads
\begin{equation}
R_\alpha^{\top}\,T\,R_\beta=
\begin{pmatrix}
(t^{\sigma}+3t^{\pi})/4 & \sqrt3\,(t^{\sigma}-t^{\pi})/4\\[2pt]
-\sqrt3\,(t^{\sigma}-t^{\pi})/4 & -(3t^{\sigma}+t^{\pi})/4
\end{pmatrix},
\label{eq:local_block}
\end{equation}
the same for all three bond orientations, as required by $C_3$. All hopping parameters quoted
in this Supplement, in the figures and in the main text are the rotated-basis parameters
$t^{\sigma}$ and $t^{\pi}$ of Eq.~\eqref{eq:local_block}; the global integrals follow from
Eq.~\eqref{eq:t_from_V}. In all parameter sets used here $t^{\sigma}>0$ and $t^{\pi}\le0$, i.e.\ $V^{\sigma}>0$ and
$V^{\pi}\ge0$.

Two limits illustrate the role of the $\pi$ channel. For $t^{\pi}=-t^{\sigma}$
($V^{\pi}=V^{\sigma}$) the two projectors in Eq.~\eqref{eq:T_i} combine to the identity,
$T_i=f_i\,\mathbb{I}_2$, and the model reduces to two identical, degenerate copies of the
single-orbital kagome model. For $t^{\pi}=0$ the blocks $T_i$ are rank one and only the
bond-parallel orbital component participates in hopping. Kagome antimonene is close to the
second limit, $|t^{\pi}|\ll|t^{\sigma}|$.

\subsection*{Spin--orbit coupling and symmetry-allowed on-site terms}

We set $\hbar=1$ and write the spin as $\mathbf{s}=\boldsymbol{\sigma}/2$. Within the
$\{p_x,p_y\}$ manifold the orbital angular momentum operator is
\begin{equation}
L_z=\begin{pmatrix}0&-i\\ i&0\end{pmatrix}=\tau_y ,
\label{eq:Lz_pxpy}
\end{equation}
with $\boldsymbol{\tau}$ the Pauli matrices in orbital space. Projecting the atomic
spin--orbit coupling $\lambda_{\mathrm{SO}}\,\mathbf{L}\cdot\mathbf{s}$ onto this manifold leaves only
the $L_zs_z$ component,
\begin{equation}
\mathcal{H}^{\mathrm{at}}_{\mathrm{SOC}}=\lambda_{\mathrm{SO}}\,L_z s_z
=(\lambda_{\mathrm{SO}}/2)\,\tau_y\otimes\sigma_z
=\frac{\lambda_{\mathrm{SO}}}{2}
\begin{pmatrix}
0&0&-i&0\\ 0&0&0&i\\ i&0&0&0\\ 0&-i&0&0
\end{pmatrix},
\label{eq:H_SOC_atomic}
\end{equation}
where the matrix is written in the basis
$(p_x\!\uparrow,p_x\!\downarrow,p_y\!\uparrow,p_y\!\downarrow)^{\top}$.  Because $L_z$ is invariant
under all in-plane rotations, Eq.~\eqref{eq:H_SOC_atomic} takes the same form in the global
and local frames. In the $p_z$-extended model described below, the full $\lambda_{\mathrm{SO}}\,\mathbf{L}\cdot\mathbf{s}$ is
used, which in addition couples $p_z$ to $p_{x,y}$ through $L_x$ and $L_y$.

The freestanding planar layer possesses the horizontal mirror $\sigma_h$. The SiC
substrate removes it and thereby permits additional spin--orbit terms, whose form is
strongly constrained. Within the $\{p_x,p_y\}$ manifold the only time-reversal-odd
orbital operator is $L_z=\tau_y$, so any time-reversal-even on-site spin--orbit term must
have the structure $\tau_y\otimes(\boldsymbol{\sigma}\cdot\hat{\mathbf{m}})$. In the
substrate-supported breathing structure the site-symmetry group of the kagome sites is
$C_s=\{E,\sigma_v^{(\alpha)}\}$, where $\sigma_v^{(\alpha)}$ is the vertical mirror that
contains $\hat{\mathbf{y}}_\alpha$ (and hence both triangle centroids) and has normal
$\hat{\mathbf{e}}_\alpha$. The other vertical site mirror of $D_{2h}$, which contains
$\hat{\mathbf{e}}_\alpha$, interchanges the small and the large triangle and is lost upon
breathing. Under $\sigma_v^{(\alpha)}$, $L_z\to-L_z$, while the spin components lying in
the mirror plane change sign and the component along its normal does not. Invariance
therefore requires $\hat{\mathbf{m}}$ to lie in the mirror plane,
$\hat{\mathbf{m}}\in\mathrm{span}\{\hat{\mathbf{y}}_\alpha,\hat{\mathbf{z}}\}$. The $\hat{\mathbf{z}}$ component reproduces the atomic term, so the only new term is a
substrate-induced, on-site Rashba-type spin--orbit coupling,
\begin{equation}
\mathcal{H}_{\mathrm{R}}=\lambda_{R}\sum_{\alpha=A,B,C}
\left(L_z\right)_{\alpha}\otimes
\left(\boldsymbol{\sigma}\cdot\hat{\mathbf{y}}_{\alpha}\right).
\label{eq:H_Rashba}
\end{equation}
Equation~\eqref{eq:H_Rashba} is Hermitian and time-reversal even. It is invariant under
$C_3$ and the three vertical mirrors of $p3m1$, and odd under $\sigma_h$, under $C_{2z}$
and under the site mirror containing $\hat{\mathbf{e}}_\alpha$. It is therefore allowed only if $\sigma_h$ is broken, which the substrate does, and if the two triangles meeting at each site are inequivalent, which breaks $C_{2z}$. The latter
is provided by the breathing distortion and, already for an undistorted Sb net, by the
registry with the substrate, whose inequivalent carbon atoms beneath adjacent triangles
reduce the symmetry to $C_{3v}$ (Sec.~\ref{sec:dft}). The analogous term with $\hat{\mathbf{e}}_\alpha$ in place of $\hat{\mathbf{y}}_\alpha$
is odd under $\sigma_v^{(\alpha)}$ and hence forbidden.  Once $\sigma_h$ is broken, a spin-dependent inter-site term of the conventional Rashba form
is also allowed,
\begin{equation}
\mathcal{H}'_{\mathrm{R}}=i\lambda'_{R}\sum_{\langle ij\rangle}\sum_{a=x,y}
c_{ia}^{\dagger}\,\bigl(\hat{\mathbf{d}}_{ij}\times\boldsymbol{\sigma}\bigr)_z\,c_{ja},
\label{eq:H_Rashba_inter}
\end{equation}
where $\hat{\mathbf{d}}_{ij}$ is the unit vector from site $i$ to its nearest neighbour $j$
and $c_{ia}$ is a spinor. Equation~\eqref{eq:H_Rashba_inter} is Hermitian, time-reversal
even and invariant under the operations of $p3m1$.

One further on-site term is symmetry allowed and will be needed below. Since $B_{2u}$ and
$B_{3u}$ are distinct one-dimensional representations of the site-symmetry group,
nothing forbids an energy splitting between them,
\begin{equation}
\mathcal{H}_{\Delta}=\Delta\sum_{\alpha}
\left(\ket{B_{3u}^{(\alpha)}}\bra{B_{3u}^{(\alpha)}}
-\ket{B_{2u}^{(\alpha)}}\bra{B_{2u}^{(\alpha)}}\right)
=\Delta\sum_{\alpha}
\left(2\,\hat{\mathbf{e}}_{\alpha}\hat{\mathbf{e}}_{\alpha}^{\top}-\mathbb{I}_2\right)_{\alpha}.
\label{eq:H_Delta}
\end{equation}
It depends only on the axis $\hat{\mathbf{e}}_\alpha\hat{\mathbf{e}}_\alpha^{\top}$ and not
on its orientation. It is invariant under the full space group of the pristine lattice,
and hence also under those of the breathing structures. It is used twice below: it removes the accidental degeneracy at $M$
(Sec.~\ref{sec:symmetry}), and it provides the adiabatic path that
connects the half-filled manifold to a site-centred atomic limit
(Sec.~\ref{sec:wilson}). Collecting all contributions, the full spinful Hamiltonian in
the basis (sublattice)$\otimes$(orbital)$\otimes$(spin) reads
\begin{equation}
\mathcal{H}_{\mathrm{full}}(\mathbf{k})=
\mathcal{H}(\mathbf{k})\otimes\mathbb{I}_2
+\mathbb{I}_3\otimes\mathcal{H}^{\mathrm{at}}_{\mathrm{SOC}}
+\mathcal{H}_{\mathrm{R}}
+\mathcal{H}'_{\mathrm{R}}(\mathbf{k})
+\mathcal{H}_{\Delta}\otimes\mathbb{I}_2 .
\label{eq:H_full}
\end{equation}

\subsection*{Extension to the out-of-plane channel}

For a quantitative comparison with density functional theory (DFT) the model must be
extended by the out-of-plane channel. Orbital filtering pushes the Sb $p_z$ states away
from the Fermi level but does not eliminate them. In the realistic material three weakly
dispersive $p_z$-derived bands lie energetically between the $\alpha,\beta$ and $\gamma$
in-plane bands (Fig.~3b of the main text; see also Sec.~\ref{sec:dft}). Adding $p_z$
introduces an on-site energy $\varepsilon_z$, a nearest-neighbour $p_z$--$p_z$ integral
$V_z$ of pure $\pi$ character and, once $\sigma_h$ is broken by the substrate, a
$p_z$--$p_{x,y}$ integral $V_{z,x/y}$. The latter follows the bond direction: for a bond from
site $\alpha$ to site $\beta$ with unit vector $\hat{\mathbf{n}}$,
$\bra{p_z,\alpha}\hat H\ket{p_a,\beta}=V_{z,x/y}\,\hat n_a=-\bra{p_a,\alpha}\hat H\ket{p_z,\beta}$
($a=x,y$), so that it changes sign under bond reversal. The in-plane block is unchanged,
the full atomic $\lambda_{\mathrm{SO}}\,\mathbf{L}\cdot\mathbf{s}$ is used, and the same spatial symmetries and symmetry constraints on the in-plane subspace are preserved. With the parameters of Table~\ref{tab:params}, this model reproduces the DFT band
structure near the Fermi level in the presence of the substrate and spin--orbit coupling. The purely in-plane parameter sets used for the schematic band
structures of Figs.~\ref{fig:S1}--\ref{fig:S4} are quoted in the corresponding captions and
are given in eV.

\begin{table*}[t!]
\centering
\caption{Tight-binding parameters (in eV) reproducing the DFT band structure of
breathing kagome antimonene in the presence of the substrate and spin--orbit coupling. $t_1$ ($t_2$) refers to the long (short)
bond, i.e.\ the large (small) triangle. The in-plane hoppings are the rotated-basis parameters
of Eq.~\eqref{eq:local_block}; $\Delta$ is the coefficient of the on-site splitting term of Eq.~\eqref{eq:H_Delta}, so that the bare $B_{3u}$--$B_{2u}$ level separation is $2\Delta$; $\varepsilon_p$ and $\varepsilon_z$ are the on-site energies of the
in-plane and $p_z$ orbitals relative to $E_F$; $V_z$ and $V_{z,x/y}$ are the global $p_z$
integrals; $\lambda_{\mathrm{SO}}$ is the atomic spin--orbit coupling, and $\lambda_R$ and $\lambda'_R$
are the substrate-induced on-site and inter-site terms of Eqs.~\eqref{eq:H_Rashba} and
\eqref{eq:H_Rashba_inter}.}
\label{tab:params}
\rowcolors{2}{gray!10}{white}
\begin{tabular}{l *{12}{c}}
\toprule
Parameter & $t_1^{\sigma}$ & $t_2^{\sigma}$ & $t_1^{\pi}$ & $t_2^{\pi}$ & $\Delta$ &
$\varepsilon_p$ & $\varepsilon_z$ & $V_z$ & $V_{z,x/y}$ & $\lambda_{\mathrm{SO}}$ & $\lambda_{R}$ & $\lambda'_{R}$ \\
\midrule
Value & $1.39$ & $2.12$ & $-0.16$ & $0.00$ & $0.08$ & $0.11$ & $-1.70$ & $0.02$ & $0.14$ &
$0.44$ & $-0.05$ & $-0.10$ \\
\bottomrule
\end{tabular}
\end{table*}

\subsection*{Evolution with the $\pi/\sigma$ ratio}

\begin{figure}[hbt]
\centering
  \includegraphics[width=1.0\columnwidth]{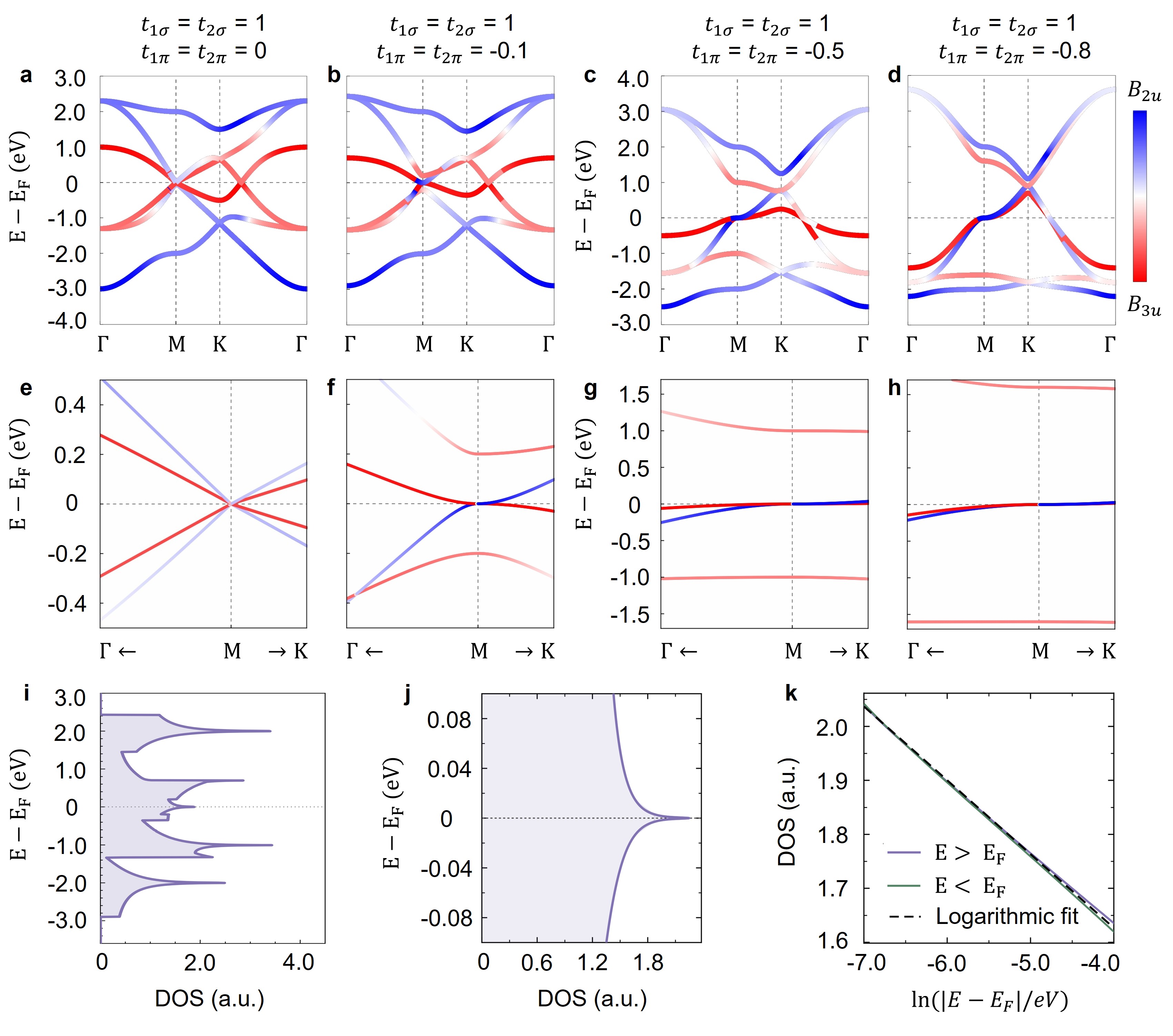}
  \caption{\textbf{Impact of $\pi$ bonding on the in-plane $p$-orbital kagome band
structure.} Tight-binding band structures with the projected $B_{2u}$ (blue) and $B_{3u}$
(red) orbital weight colour-coded on the bands.
\textbf{a--d,} Dispersions along $\Gamma$--$M$--$K$--$\Gamma$ for the pristine lattice
$t_1^{\sigma}=t_2^{\sigma}=1$ with $t_1^{\pi}=t_2^{\pi}$ varied from $0$ (\textbf{a}) to
$-0.1$ (\textbf{b}), $-0.5$ (\textbf{c}) and $-0.8$ (\textbf{d}).
Panel \textbf{a} corresponds to Fig.~1e of the main text, and panel \textbf{b} to the other physically relevant regime $|t^{\pi}|\ll|t^{\sigma}|$ of kagome antimonene, whereas panel \textbf{d} approaches the limit in which the spectrum reduces to two identical, degenerate copies of the single-orbital kagome model.
\textbf{e--h,} Enlargements around the $M$ point corresponding to \textbf{a--d}. The
fourfold crossing at $t^{\pi}=0$ (\textbf{e}) splits into a twofold crossing plus two
singlets as soon as $t^{\pi}\neq0$ (\textbf{f--h}); both are accidental degeneracies of the
nearest-neighbour model, as analysed in Sec.~\ref{sec:symmetry}.
\textbf{i,} $k$-integrated density of states (DOS) for the band structure of \textbf{b}.
\textbf{j,} Enlargement of the DOS around $E_F$, showing that it remains finite and is
logarithmically enhanced at the Fermi level.
\textbf{k,} Logarithmic fit $N(E)=N_0-c\ln|E-E_F|$ to the DOS of \textbf{j}, evaluated
separately above and below $E_F$.}
  \label{fig:S1}
\end{figure}

Figure~\ref{fig:S1} isolates the role of $\pi$ bonding at fixed
$t_1^{\sigma}=t_2^{\sigma}$. As $t^{\pi}$ approaches $-t^{\sigma}$ the six bands evolve towards two degenerate copies of
the single-orbital kagome spectrum (Fig.~\ref{fig:S1}d). Reducing $|t^{\pi}|$ hybridizes the two copies and generates
the characteristic multi-orbital fermiology discussed in the main text. Four bands meet at
$M$ at $E=0$ for $t^{\pi}=0$, two for $t^{\pi}\neq0$, and additional band
touchings appear along $K$-$\Gamma$ (Figs.~\ref{fig:S1}a,b, with enlargements in
Figs.~\ref{fig:S1}e,f).

Figures~\ref{fig:S1}i--k show the corresponding $k$-integrated density of states (DOS).
At half filling the DOS at the Fermi level is finite. This contrasts with the single-orbital kagome
model filled up to its Dirac point ($1/3$ or $2/3$ filling, depending on the sign of the
hopping), where $N(E_F)=0$. On top of
this finite background we resolve a weak logarithmic van Hove enhancement,
$N(E)\simeq N_0-c\ln|E-E_F|$ (Fig.~\ref{fig:S1}k).
Section~\ref{sec:symmetry} shows that the finite background originates from a contour of
band touching pinned at $E_F$, and that the logarithmic correction comes from the $M$
points, where two branches of that contour intersect.

\subsection*{Evolution with the breathing amplitude}

\begin{figure}[hbt]
\centering
  \includegraphics[width=1.0\columnwidth]{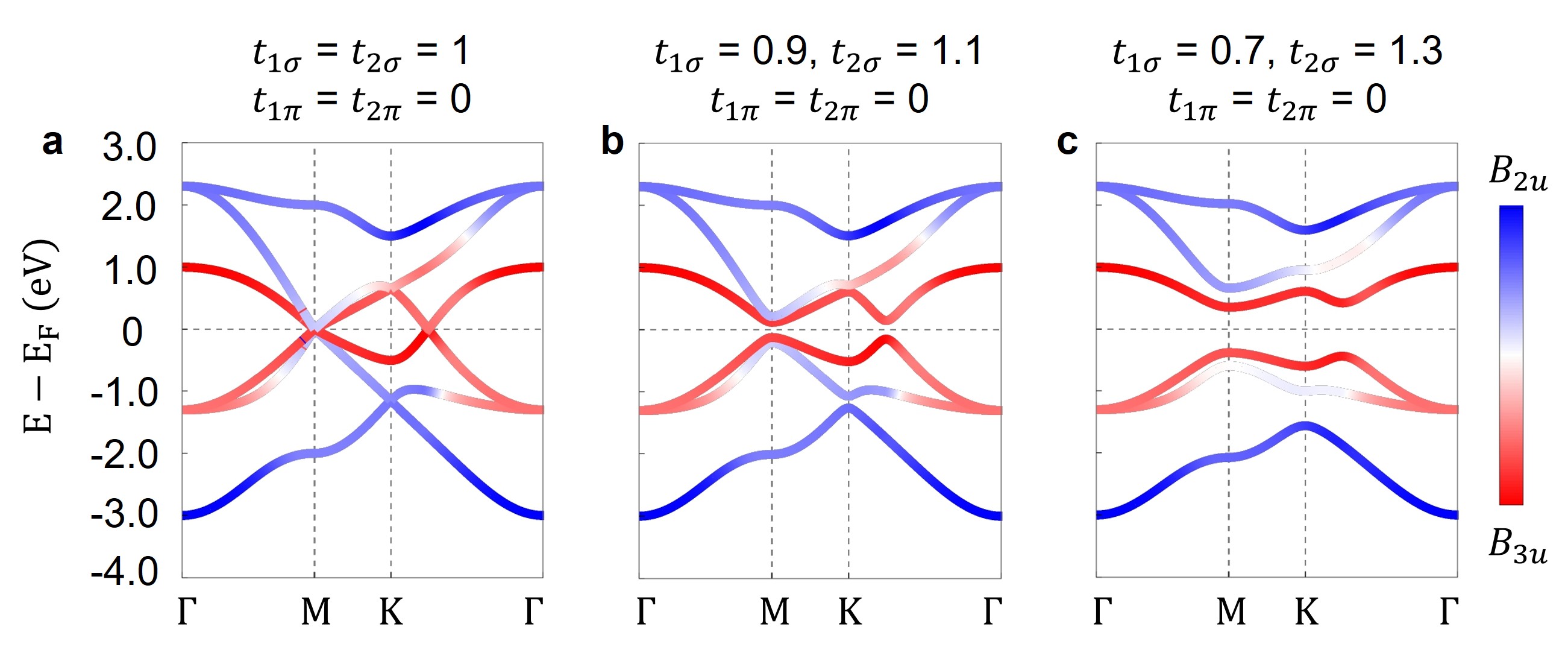}
  \caption{\textbf{Impact of the breathing amplitude on the in-plane $p$-orbital kagome band structure.} Projected $B_{2u}$ (blue) and $B_{3u}$ (red) orbital weight as in Fig.~\ref{fig:S1}.
\textbf{a--c,} Dispersions at $t_1^{\pi}=t_2^{\pi}=0$ for
$t_1^{\sigma}/t_2^{\sigma}=1/1$ (\textbf{a}), $0.9/1.1$ (\textbf{b}) and $0.7/1.3$
(\textbf{c}). The gaps opened at the crossings of the pristine lattice grow linearly with
the breathing amplitude (Table~\ref{tab:gaplaw}); unlike the single-orbital model, where only isolated Dirac
points are gapped, the distortion here gaps an entire nodal contour (Sec.~\ref{sec:symmetry}).}
  \label{fig:S2}
\end{figure}

Figure~\ref{fig:S2} shows the complementary evolution with breathing amplitude at
$t^{\pi}=0$. No degeneracy away from $\Gamma$ remains symmetry-enforced
(Sec.~\ref{sec:symmetry}), and for the parameters shown all degeneracies away from $\Gamma$ are lifted. The gap at the Fermi level opens
linearly with the breathing amplitude,
\begin{equation}
E_{\mathrm{gap}}\simeq\sqrt{3/2}\,\bigl|t_2^{\sigma}-t_1^{\sigma}\bigr|,
\label{eq:gap_law}
\end{equation}
to leading order and independently of the mean hopping (Table~\ref{tab:gaplaw}). The
microscopic origin of this
law is explained in Sec.~\ref{sec:symmetry}. The parameter set
$t_1^{\sigma}=0.8$, $t_2^{\sigma}=1.2$, $t_1^{\pi}=0.00$, $t_2^{\pi}=0.00$, used for Fig.~1f of the main text and in Sec.~\ref{sec:wilson}, has a breathing ratio close to that of Table~\ref{tab:params}.

\begin{table}[h]
\centering
\caption{Indirect gap at half filling of the six-band $p_x/p_y$ kagome model as a function
of breathing amplitude, for $t^{\pi}=0$ and $(t_1^{\sigma}+t_2^{\sigma})/2=1$.}
\label{tab:gaplaw}
\begin{tabular}{lccccc}
\hline
$|t_2^{\sigma}-t_1^{\sigma}|$ & $\to0$ & 0.10 & 0.20 & 0.40 & 0.80 \\
\hline
$E_{\mathrm{gap}}$ & --- & 0.1225 & 0.2448 & 0.4888 & 0.9707 \\
$E_{\mathrm{gap}}/|t_2^{\sigma}-t_1^{\sigma}|$ & 1.2247 & 1.2246 & 1.2240 & 1.2219 & 1.2133 \\
\hline
\end{tabular}
\end{table}

\begin{figure}[hbt]
\centering
  \includegraphics[width=1.0\columnwidth]{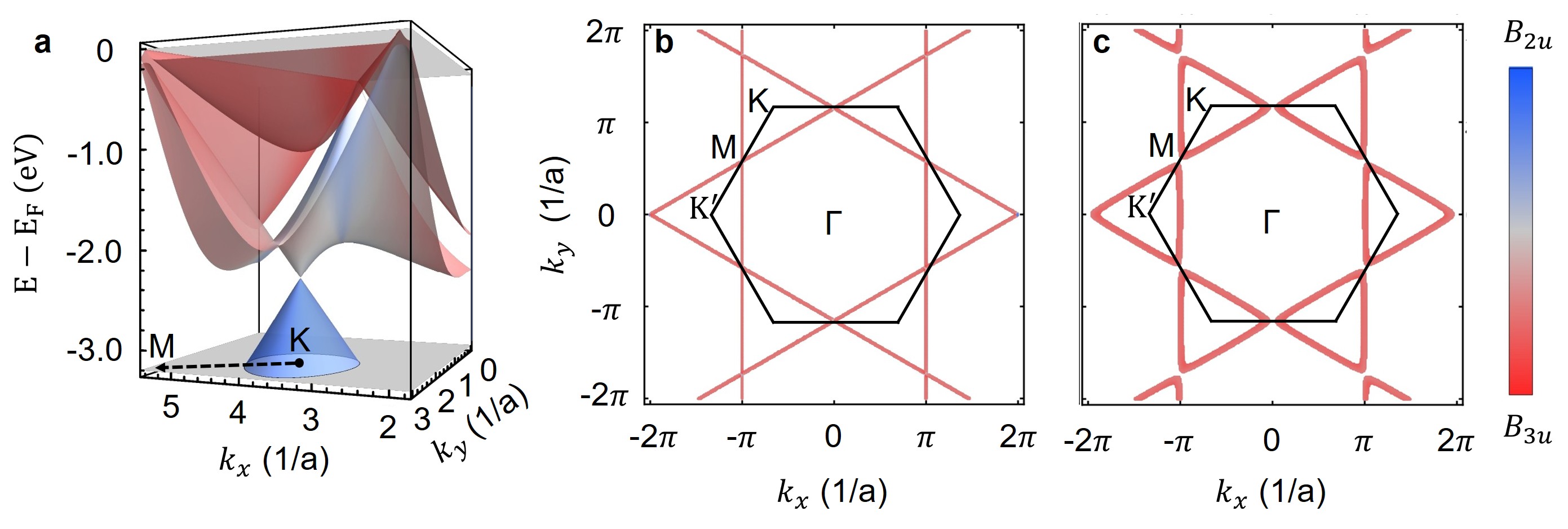}
  \caption{\textbf{Nodal-contour character of the states at the Fermi level.}
\textbf{a,} Three-dimensional band structure $E(k_x,k_y)$ of the pristine $p_x/p_y$ kagome
model near $K$, with the orbital projection colour-coded. The $B_{3u}$-derived bands touch
at $E_F$ along triangular loops around $K$ rather than at an isolated cone; the Dirac cone
visible below $E_F$ at $K$ is predominantly of $B_{2u}$ character.
\textbf{b,c,} Constant-energy contours at $E_F$ for the non-breathing lattice of Fig.~1e
(\textbf{b}) and close to the valence-band maximum for the breathing lattice of Fig.~1f
(\textbf{c}). The contours are dominated by $B_{3u}$ weight and consist of straight
segments forming triangular loops around $K$ and $K'$ that touch at $M$, i.e.\ the
kagome-shaped contour of Eq.~\eqref{eq:nodal_lines}. The pattern survives as a remnant in
the gapped breathing lattice (\textbf{c}), where the loops no longer touch at $M$. Black
lines mark the Brillouin zone, which is common to the pristine and breathing lattices.}
  \label{fig:S3}
\end{figure}

Figure~\ref{fig:S3} resolves the momentum-space structure of the states at the Fermi
level. The three-dimensional dispersion around $K$ (Fig.~\ref{fig:S3}a) shows that the
touchings at $M$ and along $K$--$\Gamma$ in Fig.~1e are not isolated Dirac points. Instead, the $B_{3u}$-derived bands touch at $E_F$ along closed triangular loops encircling $K$ and
$K'$, with neighbouring loops meeting at $M$, which produces the kagome-shaped Fermi
contour of Fig.~\ref{fig:S3}b. A genuine Dirac cone does occur, but below $E_F$ at $K$,
and is predominantly of $B_{2u}$ character. The kagome-shaped contour survives the breathing distortion as a remnant. As shown in Fig.~\ref{fig:S3}c, the
constant-energy contour taken close to the valence-band maximum of the gapped spectrum of
Fig.~1f still consists of triangular loops around $K$ and $K'$ that almost touch at $M$.
It matches the contour observed by ARPES near the valence-band maximum in Fig.~3e of
the main text.

\begin{figure}[hbt]
\centering
  \includegraphics[width=1.0\columnwidth]{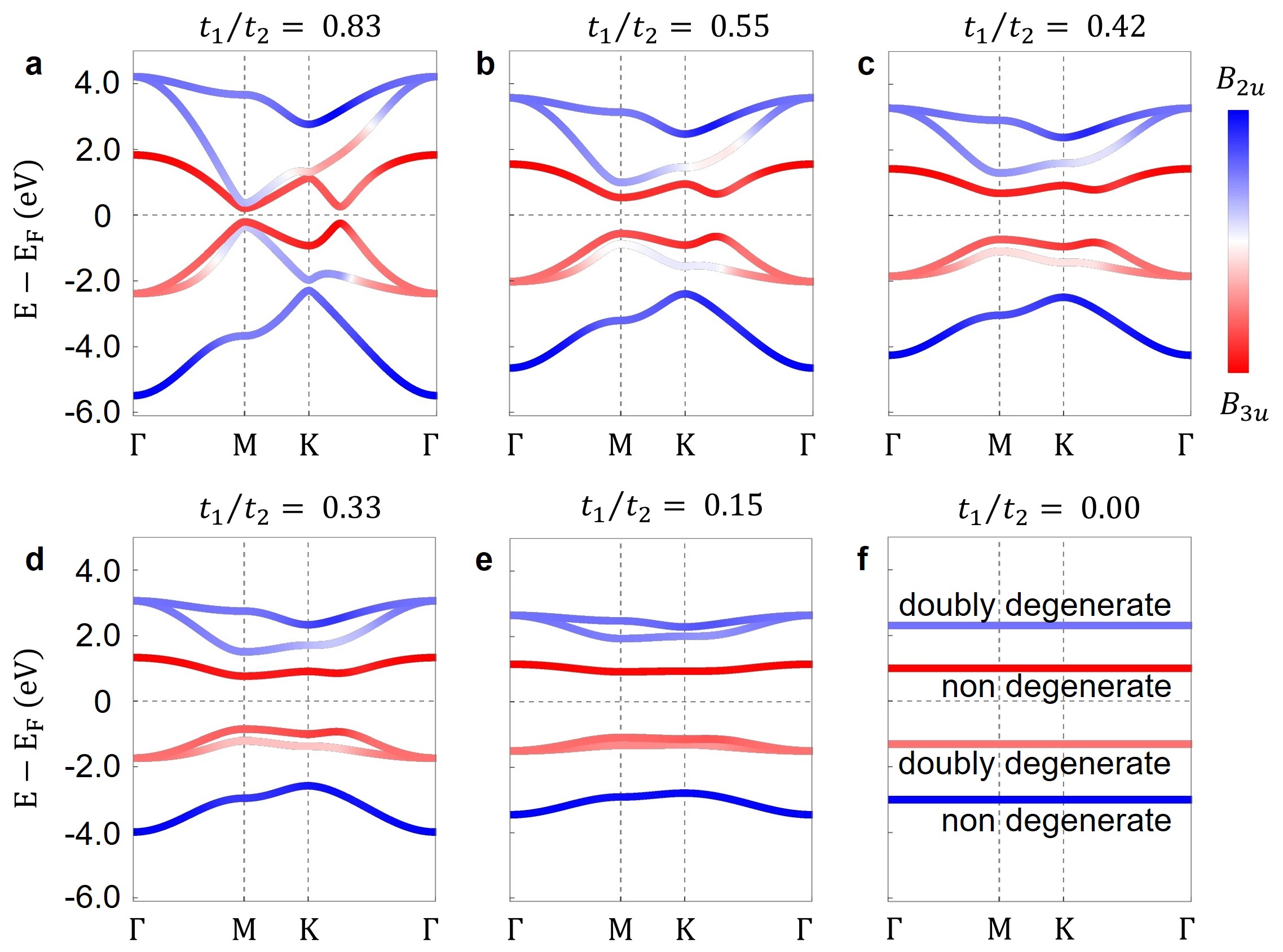}
  \caption{\textbf{Adiabatic connection to the decoupled-trimer limit.} Evolution of the
in-plane $p$-orbital breathing kagome bands as the ratio $t_1/t_2$ is reduced from $0.83$
to $0$ at $t_2^{\sigma}=2$~eV and $t_1^{\pi}=t_2^{\pi}=0$. Neither the $1/6$-
nor the $1/2$-filling gap closes
along this path. At $t_1/t_2=0$ (\textbf{f}) the spectrum consists of the four molecular levels of an
isolated $t_2$ trimer.}
  \label{fig:S4}
\end{figure}

Finally, Fig.~\ref{fig:S4} follows the six bands from $t_1/t_2=0.83$ (the ratio of Table~\ref{tab:params} is $t_1^{\sigma}/t_2^{\sigma}\approx0.66$) down to the decoupled-trimer
limit $t_1/t_2\to0$ at $t^{\pi}=0$. Neither the $1/6$- nor the $1/2$-filling gap closes along
this path. In the limit $t_1=0$ the spectrum collapses onto the four molecular levels of an isolated
$t_2$ trimer, two non-degenerate and two doubly degenerate (Fig.~\ref{fig:S4}f). The two
singlets have pure orbital character, $B_{2u}$ for the lowest level (the $\gamma$ band) and
$B_{3u}$ for the other. The two doublets mix the two local orbitals,
the lower one predominantly $B_{3u}$ (the $\alpha,\beta$ bands) and the upper one
predominantly $B_{2u}$; the mixing is allowed because both doublets transform as the same
two-dimensional irrep of the trimer point group. This adiabatic connection justifies
labelling the bands of the breathing lattice by their dominant orbital character, and it
underlies the trimer picture used in Sec.~\ref{sec:wilson}.

\clearpage

\section{Symmetry-enforced and accidental band degeneracies}
\label{sec:symmetry}

\subsection*{Band representation and enforced degeneracies}

At a given momentum $\mathbf{k}$ the Bloch states form a six-dimensional representation of
the little co-group $G_{\mathbf{k}}$, the subgroup of point-group operations that leave
$\mathbf{k}$ invariant modulo a reciprocal lattice vector. All layer groups considered
here are symmorphic, so the small representations are ordinary irreducible
representations (irreps) of $G_{\mathbf{k}}$. Whenever the representation contains a
two-dimensional irrep, a twofold degeneracy is enforced. Whenever it decomposes entirely
into one-dimensional irreps, no degeneracy is required and any observed degeneracy is
accidental. In the spinless case time-reversal symmetry does not pair any of the irreps
encountered below, since all irreps at $\Gamma$ and $M$ are real and $K$ is mapped onto
$K'$. We apply this criterion at $\Gamma$, $K$ and $M$; the band representation below can be
compared with the tabulated elementary band representations of space group
191~\cite{bradlyn2017topological}.

The freestanding, strictly planar pristine kagome monolayer has the point group $D_{6h}$
(24 elements), corresponding to the layer group $p6/mmm$ and, in the three-dimensional
setting, to space group 191 ($P6/mmm$). Its three sites occupy the Wyckoff position $3f$,
whose site-symmetry group is $mmm$ ($D_{2h}$, order 8, multiplicity $24/8=3$). We therefore
analyse the band representation induced from the two in-plane orbitals of
Eq.~\eqref{eq:orbital_labels},
\begin{equation}
\rho=\left(B_{2u}\oplus B_{3u}\right)_{3f}\uparrow G ,
\qquad \dim\rho=3\times2=6 .
\label{eq:bandrep}
\end{equation}
Because $p$ orbitals transform as a vector, the orbital part of every operation
$g=\{R_g|\mathbf{0}\}$ is the in-plane block of $R_g$. Only sublattices mapped onto
themselves modulo a lattice vector, $R_g\boldsymbol{\tau}_\alpha=\boldsymbol{\tau}_\alpha+\mathbf{R}_\alpha$,
contribute to the character, which with the origin at the hexagon centre reads
\begin{equation}
\chi^{\mathbf{k}}(g)=
\sum_{\alpha\,:\,R_g\boldsymbol{\tau}_\alpha=\boldsymbol{\tau}_\alpha+\mathbf{R}_\alpha}
e^{-i\mathbf{k}\cdot\mathbf{R}_\alpha}\;
\mathrm{tr}_{\parallel}\!\left(R_g\right),
\label{eq:characters}
\end{equation}
where $\mathrm{tr}_{\parallel}$ denotes the trace of the in-plane $2\times2$ block. 

At $\Gamma$ the little co-group is the full $D_{6h}$ (Table~\ref{tab:D6h}; the $C_2'$ axes
pass through the kagome sites). Evaluating Eq.~\eqref{eq:characters} gives the
non-vanishing characters $\chi^{\Gamma}(E)=6$, $\chi^{\Gamma}(C_2^{z})=\chi^{\Gamma}(i)=-6$
and $\chi^{\Gamma}(\sigma_h)=6$, whence
\begin{equation}
\rho^{\Gamma}\cong B_{1u}\oplus B_{2u}\oplus 2E_{1u} .
\label{eq:gamma_decomposition}
\end{equation}
All constituents are even under $\sigma_h$, as they must be for a manifold built from
in-plane orbitals, and the two copies of the two-dimensional $E_{1u}$ enforce two twofold
degeneracies at $\Gamma$, as seen in the band structure.

At the zone corners the little co-group is $D_{3h}$ (Table~\ref{tab:D3h}). Since the
in-plane orbitals lie inside the mirror plane $\sigma_h$, only the mirror-even (singly
primed) irreps can appear, and
\begin{equation}
\rho^{K}\cong A_1'\oplus A_2'\oplus 2E' .
\label{eq:K_decomposition}
\end{equation}
The two copies of $E'$ enforce two twofold degeneracies at $K$. Whether a given $E'$
doublet disperses linearly, that is whether it forms a Dirac point, is not fixed by the
irrep content but by the local $\mathbf{k}\cdot\mathbf{p}$ expansion. In the present
model the lower $E'$ doublet is a Dirac point of predominantly $B_{2u}$ character
(Fig.~\ref{fig:S3}a), whereas the upper one, at positive energy, is predominantly of
$B_{3u}$ character.

At $M$ the little co-group is $D_{2h}$ (Table~\ref{tab:D2h}), which is abelian, so all of
its irreps are one-dimensional. We take $\mathbf{k}_M=(\mathbf{b}_1-\mathbf{b}_2)/2$ and label the $D_{2h}$ operations with
the local $x$ axis along $\mathbf{k}_M$ and $z$ along the surface normal.
With the origin at the hexagon centre, Eq.~\eqref{eq:characters} gives $\chi^{M}(E)=6$,
$\chi^{M}(C_2^{z})=\chi^{M}(i)=2$ and $\chi^{M}(\sigma_h)=6$, with zero for the two
in-plane twofold axes and the two vertical mirrors, so that
\begin{equation}
\rho^{M}\cong 2A_g\oplus 2B_{1g}\oplus B_{2u}\oplus B_{3u} .
\label{eq:M_decomposition}
\end{equation}
The gerade and ungerade labels depend on the choice of inversion centre, because the Bloch
phase multiplies the local orbital character; the absence of two-dimensional irreps does
not. No degeneracy is enforced at $M$ by the crystal symmetry of the
pristine kagome lattice, and the degeneracies seen there in Fig.~1e and
Figs.~\ref{fig:S1}a,e are accidental in the group-theoretical sense.

\subsection*{Accidental degeneracy at $M$}

The origin of these degeneracies follows once the Bloch form factors are
evaluated at $\mathbf{k}_M$, for which $\mathbf{k}_M\cdot\boldsymbol{\delta}_1=-\pi$,
$\mathbf{k}_M\cdot\boldsymbol{\delta}_2=+\pi$ and
$\mathbf{k}_M\cdot\boldsymbol{\delta}_3=0$. For the pristine lattice,
$V_1^{\sigma,\pi}=V_2^{\sigma,\pi}\equiv V^{\sigma,\pi}$, this gives
\begin{equation}
f_1^{\sigma,\pi}(\mathbf{k}_M)=f_2^{\sigma,\pi}(\mathbf{k}_M)=0,
\qquad
f_3^{\sigma,\pi}(\mathbf{k}_M)=2\,V^{\sigma,\pi} .
\label{eq:f_at_M}
\end{equation}
Two of the three bond families are switched off at $M$. Sublattice $B$, which
participates only in families 1 and 2, decouples completely, and the Hamiltonian collapses
to a single $A$--$C$ block whose spectrum is $\pm$ the singular values of
$T_3(\mathbf{k}_M)=2V^{\sigma}P_3^{\sigma}+2V^{\pi}P_3^{\pi}$ (with $|V^{\sigma,\pi}|=|t^{\sigma,\pi}|$). Hence
\begin{equation}
E(\mathbf{k}_M)=\left\{\pm2|t^{\sigma}|,\;\pm2|t^{\pi}|,\;0,\;0\right\},
\label{eq:M_spectrum}
\end{equation}
where the two exact zeros are the decoupled $B$ orbitals. At $t^{\pi}=0$ the block $T_3$ is
rank one and contributes two further zeros, producing the fourfold degeneracy of
Fig.~\ref{fig:S1}e. Any finite $t^{\pi}$ restores full rank and splits these two states
symmetrically to $\pm2|t^{\pi}|$, leaving the twofold degeneracy of
Fig.~\ref{fig:S1}f--h. Because the remaining twofold zero originates from a
decoupled sublattice rather than from a symmetry, it can be removed by a symmetry-allowed perturbation. The on-site orbital splitting
$\mathcal{H}_\Delta$ of Eq.~\eqref{eq:H_Delta}, which preserves the entire space group,
splits the two decoupled $B$ orbitals to $\pm\Delta$ and removes every degeneracy at $M$.

\subsection*{Nodal contour of the nearest-neighbour model}

The touchings along $K$--$\Gamma$ in Fig.~1e are likewise not isolated points, and their
analysis is central to the breathing instability. For the pristine lattice the form factors
of Eq.~\eqref{eq:f_sigma_pi} reduce to
$f_i(\mathbf{k})=V\,(1+e^{i\mathbf{k}\cdot\boldsymbol{\delta}_i})$ and vanish identically on
the straight lines
\begin{equation}
\mathcal{L}_i:\qquad \mathbf{k}\cdot\boldsymbol{\delta}_i=\pi \;\;(\mathrm{mod}\;2\pi),
\qquad i=1,2,3 .
\label{eq:nodal_lines}
\end{equation}
Since $f_i^{\sigma}$ and $f_i^{\pi}$ vanish together, the entire bond family $i$ is
switched off on $\mathcal{L}_i$ irrespective of the $\pi/\sigma$ ratio. On such a line the
kagome graph loses one of its three bond orientations. The two surviving families connect
one distinguished sublattice to the other two, and no closed triangle remains. The
Hamiltonian is then bipartite, with two orbitals on the minority and four on the majority
sublattice, and therefore possesses a chiral (sublattice) symmetry. A chiral Hamiltonian
has a spectrum symmetric about $E=0$ and at least $\nu$ zero modes, where $\nu$ is the
sublattice imbalance; here $\nu=|4-2|=2$, so that
\begin{equation}
E(\mathbf{k})=\left\{\pm\epsilon_1,\;\pm\epsilon_2,\;0,\;0\right\}
\qquad\text{for all }\mathbf{k}\in\mathcal{L}_1\cup\mathcal{L}_2\cup\mathcal{L}_3 .
\label{eq:nodal_spectrum}
\end{equation}
For $t^{\pi}\ge-t^{\sigma}/3$ the third and fourth bands are separated away from these lines, with
$E_3(\mathbf{k})\le0\le E_4(\mathbf{k})$ throughout the Brillouin zone, so that they touch
along an extended contour and do not overlap; for $t^{\pi}<-t^{\sigma}/3$ they overlap (Figs.~\ref{fig:S1}c,d). At $M$, which lies on two of the lines, the chiral argument still yields two
zero modes. The additional twofold zero found there for $t^{\pi}=0$ is not a consequence of
the line crossing but of the rank deficiency of $T_3(\mathbf{k}_M)$ discussed above.

The three families $\mathcal{L}_i$ are sets of parallel straight lines perpendicular to
the three bond orientations. Their union is itself a kagome pattern of triangular loops
centred on $K$ and $K'$, with adjacent loops touching pairwise at the $M$ points. This is the contour plotted in Fig.~\ref{fig:S3}b and, as a gapped remnant, in
Fig.~\ref{fig:S3}c. The three bands below $E=0$ accommodate six electrons per primitive
cell, i.e.\ two in-plane $p$ electrons per Sb atom. In the
DFT band structures of the substrate-supported layer (Figs.~\ref{fig:S7}b and
\ref{fig:S8}a) $E_F$ lies between the third and fourth bands of the in-plane manifold, so
that this manifold is half filled and $E_F$ is pinned onto the nodal contour without any
external tuning (up to the small substrate-induced gap discussed in Sec.~\ref{sec:dft}). Two consequences
follow. First, the Fermi surface degenerates into a contour of band touching rather than
enclosing electron or hole pockets, while the DOS at $E_F$ remains finite,
as seen in Fig.~\ref{fig:S1}j. Second, at
the $M$ points two branches of the contour intersect, producing saddle points that
superimpose a logarithmic van Hove enhancement on that finite background. Both features
are visible in Figs.~\ref{fig:S1}i--k.

The vanishing of the splitting
on $\mathcal{L}_i$ is an exact property of the nearest-neighbour model for any $\pi/\sigma$
ratio. It is protected by the emergent chiral symmetry on these lines, not by the space
group, and the pinning of $E_F$ onto the contour at half filling additionally requires
$t^{\pi}\ge-t^{\sigma}/3$. Symmetry-allowed perturbations beyond that
model, such as second-neighbour hopping or the on-site splitting $\mathcal{H}_{\Delta}$,
lift the exact degeneracy, but only weakly and without opening a gap. For a
$\sigma$-type second-neighbour amplitude of $\pm0.05|t^{\sigma}|$, for example, the third
and fourth bands overlap by at most $\approx0.02|t^{\sigma}|$, and the half-filled system
becomes a compensated semimetal whose small electron and hole pockets closely track the
lines $\mathcal{L}_i$. The contour-like Fermi surface and the associated finite DOS are therefore robust over a
finite range of symmetry-allowed perturbations, although no longer exactly protected.

\subsection*{Gapping of the contour by the breathing distortion}

The breathing distortion removes the contour outright. For the same-sign hopping
amplitudes relevant here, breathing gives $|V_1|\neq|V_2|$, and
\begin{equation}
\min_{\mathbf{k}}\left|f_i^{\sigma}(\mathbf{k})\right|
=\bigl|\,|V_2^{\sigma}|-|V_1^{\sigma}|\,\bigr|>0 ,
\label{eq:fmin}
\end{equation}
and likewise for $f_i^{\pi}$ when $|V_1^{\pi}|\neq|V_2^{\pi}|$. No momentum in the Brillouin
zone can then sever a bond family, the chiral structure that protected the zero modes is
destroyed everywhere at once, and the contour is gapped along its entire length. Since the
perturbation is linear in $t_2-t_1$ and acts on a pair of degenerate states along the whole
contour, the gap opens linearly, with the slope $\sqrt{3/2}$ of Eq.~\eqref{eq:gap_law} for $t^{\pi}=0$.
Unlike the single-orbital model, where only isolated Dirac points at $K$ and $K'$ are gapped, the
energy gain is collected from a one-dimensional manifold of states at $E_F$ with finite
DOS. By the standard Peierls-type estimate, a gap $E_{\mathrm{gap}}\propto u$ opened on
states of finite DOS $N(E_F)$ lowers the electronic energy by an amount of order
$N(E_F)\,E_{\mathrm{gap}}^2\ln(W/E_{\mathrm{gap}})$, with $W$ the bandwidth. The electronic susceptibility
towards the breathing channel is therefore logarithmically enhanced, and for small distortion amplitudes $u$ the electronic energy gain can exceed the
elastic cost $\propto u^2$. This points to a multi-orbital electronic mechanism for the
unit-cell-conserving breathing distortion of kagome antimonene, consistent with the
zone-centre soft phonon found in Sec.~\ref{sec:breathing_origin}; a quantitative statement
for the real material would require the electronic energy as a function of the breathing
coordinate from DFT. Although the straight segments
of the contour are nested, the instability occurs at $\mathbf{q}=0$, because the breathing
mode modulates the intra-cell and inter-cell bonds within the same unit cell and therefore
couples states at the same $\mathbf{k}$ all along the contour. The distortion shifts the contour in energy but leaves its momentum-space geometry
essentially unchanged, so a remnant of the nodal contour remains visible in the
constant-energy contours of the gapped phase (Fig.~\ref{fig:S3}c and Fig.~3e,f of the
main text).

\subsection*{Symmetry reduction by the breathing distortion and the substrate}

The breathing distortion also lowers the symmetry, and Table~\ref{tab:degeneracies}
summarizes the consequences. We determined the symmetry group of each structure directly
from the lattice, by requiring that an operation permute the sublattices and map short
bonds onto short bonds. The freestanding planar breathing layer retains the 12 operations
$\{E,2C_3,3C_2',\sigma_h,2S_3,3\sigma_v\}$ of the point group $D_{3h}$ (layer group
$p\bar{6}m2$), where the retained vertical mirrors are those containing
$\hat{\mathbf{y}}_\alpha$. The SiC substrate additionally removes $\sigma_h$ and all
in-plane twofold axes, leaving the six operations $\{E,2C_3,3\sigma_v\}$ of the polar point
group $C_{3v}$ and the plane group $p3m1$ quoted in the main text. In both reduced groups
two-dimensional irreps occur only at $\Gamma$: the little co-groups at $K$ ($C_{3h}$ and
$C_3$) and at $M$ ($C_{2v}$ and $C_s$) admit one-dimensional irreps only. Consequently,
the only symmetry-enforced degeneracies that survive are the two doublets at $\Gamma$; in
particular the two $E'$ doublets at $K$ are split. This is the pattern of Fig.~1f of the
main text and of Fig.~\ref{fig:S2}.

Two mechanisms should be distinguished here. The splitting of the
doublets at $K$ follows from symmetry, because the two-dimensional irreps that enforced them are
absent in the reduced group. The gapping of the nodal contour is not, because that contour
is not symmetry-enforced. It is removed by the breathing distortion
through the mechanism of Eq.~\eqref{eq:fmin}. The remark in the main text that the gaps
open because two-dimensional irreps are absent away from $\Gamma$ refers to the first
mechanism.

\subsection*{Spin--orbit coupling}

With spin--orbit coupling the relevant representations are the double-valued
representations of the little co-groups. The breathing distortion and the substrate break inversion, and the double-group analysis
of the substrate-supported structure ($p3m1$) gives the following pattern. At $\Gamma$
($C_{3v}$), each spinless one-dimensional irrep $A_{1,2}$ becomes the two-dimensional spinor
irrep $\bar{E}_{1/2}$, and each spinless doublet $E$ splits according to
$E\otimes\bar{E}_{1/2}=\bar{E}_{1/2}\oplus({}^1\bar{E}_{3/2}\oplus{}^2\bar{E}_{3/2})$, whose
two one-dimensional components form a Kramers pair. All states at $\Gamma$ therefore remain
twofold degenerate, and each spinless doublet splits into two Kramers doublets. At $M$
($C_s$), a time-reversal-invariant momentum, all states form Kramers doublets. At $K$
($C_3$), which is not time-reversal invariant, all states are non-degenerate, and
time-reversal symmetry relates the spectrum at $K$ to that at $K'$ with reversed spin. At
generic momenta the bands are spin split.
The substrate breaks $\sigma_h$ and, through $\lambda_R$ and $\lambda'_R$ of
Eqs.~\eqref{eq:H_Rashba} and \eqref{eq:H_Rashba_inter}, admixes in-plane spin components. The corresponding DFT band structures with and without
spin--orbit coupling are compared in Sec.~\ref{sec:dft}.

\begin{table}[h]
\centering
\caption{Character table of the point group $D_{6h}$, with the characters of the six-band
kagome $(p_x,p_y)$ representation at $\Gamma$ added in the last row. The $C_2'$ axes pass
through the kagome sites.}
\label{tab:D6h}
\begin{tabular}{c|cccccccccccc|c}
\hline
Irrep & $E$ & $2C_6$ & $2C_3$ & $C_2$ & $3C_2'$ & $3C_2''$ & $i$ & $2S_3$ & $2S_6$ & $\sigma_h$ & $3\sigma_d$ & $3\sigma_v$ & Dim. \\
\hline
$A_{1g}$ & 1 & 1 & 1 & 1 & 1 & 1 & 1 & 1 & 1 & 1 & 1 & 1 & 1 \\
$A_{2g}$ & 1 & 1 & 1 & 1 & $-1$ & $-1$ & 1 & 1 & 1 & 1 & $-1$ & $-1$ & 1 \\
$B_{1g}$ & 1 & $-1$ & 1 & $-1$ & 1 & $-1$ & 1 & $-1$ & 1 & $-1$ & 1 & $-1$ & 1 \\
$B_{2g}$ & 1 & $-1$ & 1 & $-1$ & $-1$ & 1 & 1 & $-1$ & 1 & $-1$ & $-1$ & 1 & 1 \\
$E_{1g}$ & 2 & 1 & $-1$ & $-2$ & 0 & 0 & 2 & 1 & $-1$ & $-2$ & 0 & 0 & 2 \\
$E_{2g}$ & 2 & $-1$ & $-1$ & 2 & 0 & 0 & 2 & $-1$ & $-1$ & 2 & 0 & 0 & 2 \\
$A_{1u}$ & 1 & 1 & 1 & 1 & 1 & 1 & $-1$ & $-1$ & $-1$ & $-1$ & $-1$ & $-1$ & 1 \\
$A_{2u}$ & 1 & 1 & 1 & 1 & $-1$ & $-1$ & $-1$ & $-1$ & $-1$ & $-1$ & 1 & 1 & 1 \\
$B_{1u}$ & 1 & $-1$ & 1 & $-1$ & 1 & $-1$ & $-1$ & 1 & $-1$ & 1 & $-1$ & 1 & 1 \\
$B_{2u}$ & 1 & $-1$ & 1 & $-1$ & $-1$ & 1 & $-1$ & 1 & $-1$ & 1 & 1 & $-1$ & 1 \\
$E_{1u}$ & 2 & 1 & $-1$ & $-2$ & 0 & 0 & $-2$ & $-1$ & 1 & 2 & 0 & 0 & 2 \\
$E_{2u}$ & 2 & $-1$ & $-1$ & 2 & 0 & 0 & $-2$ & 1 & 1 & $-2$ & 0 & 0 & 2 \\
\hline
$\rho^{\Gamma}$ & 6 & 0 & 0 & $-6$ & 0 & 0 & $-6$ & 0 & 0 & 6 & 0 & 0 & 6 \\
\hline
\end{tabular}
\end{table}

\begin{table}[h]
\centering
\caption{Character table of the point group $D_{3h}$, the little co-group at $K$ for the
pristine lattice and at $\Gamma$ for the freestanding breathing lattice. The last row gives
the characters of the six-band kagome $(p_x,p_y)$ representation at $K$.}
\label{tab:D3h}
\begin{tabular}{c|cccccc|c}
\hline
Irrep & $E$ & $2C_3$ & $3C_2'$ & $\sigma_h$ & $2S_3$ & $3\sigma_v$ & Dim. \\
\hline
$A_1'$   & 1 & 1 & 1 & 1 & 1 & 1 & 1 \\
$A_2'$   & 1 & 1 & $-1$ & 1 & 1 & $-1$ & 1 \\
$E'$     & 2 & $-1$ & 0 & 2 & $-1$ & 0 & 2 \\
$A_1''$  & 1 & 1 & 1 & $-1$ & $-1$ & $-1$ & 1 \\
$A_2''$  & 1 & 1 & $-1$ & $-1$ & $-1$ & 1 & 1 \\
$E''$    & 2 & $-1$ & 0 & $-2$ & 1 & 0 & 2 \\
\hline
$\rho^{K}$ & 6 & 0 & 0 & 6 & 0 & 0 & 6 \\
\hline
\end{tabular}
\end{table}

\begin{table}[h]
\centering
\caption{Character table of the point group $D_{2h}$, which is isomorphic both to the
site-symmetry group of the Wyckoff position $3f$ and to the little co-group at $M$ of the
pristine lattice. In the local site frame
$(\hat{\mathbf{e}}_\alpha,\hat{\mathbf{y}}_\alpha,\hat{\mathbf{z}})$ the orbitals
$p_\parallel$, $p_\perp$ and $p_z$ transform as $B_{3u}$, $B_{2u}$ and $B_{1u}$,
respectively. The last row gives the characters of the six-band kagome $(p_x,p_y)$
representation at $M$, with the origin at the hexagon centre and the local $x$ axis of this table along
$\mathbf{k}_M$. Since $D_{2h}$ is abelian, all its irreps are
one-dimensional and no degeneracy can be enforced at $M$.}
\label{tab:D2h}
\begin{tabular}{c|cccccccc|c}
\hline
Irrep & $E$ & $C_{2}(z)$ & $C_{2}(y)$ & $C_{2}(x)$ & $i$ & $\sigma_{xy}$ & $\sigma_{xz}$ & $\sigma_{yz}$ & Dim. \\
\hline
$A_g$    & 1 &  1 &  1 &  1 &  1 &  1 &  1 &  1 & 1 \\
$B_{1g}$ & 1 &  1 & $-1$ & $-1$ &  1 &  1 & $-1$ & $-1$ & 1 \\
$B_{2g}$ & 1 & $-1$ &  1 & $-1$ &  1 & $-1$ &  1 & $-1$ & 1 \\
$B_{3g}$ & 1 & $-1$ & $-1$ &  1 &  1 & $-1$ & $-1$ &  1 & 1 \\
$A_u$    & 1 &  1 &  1 &  1 & $-1$ & $-1$ & $-1$ & $-1$ & 1 \\
$B_{1u}$ & 1 &  1 & $-1$ & $-1$ & $-1$ & $-1$ &  1 &  1 & 1 \\
$B_{2u}$ & 1 & $-1$ &  1 & $-1$ & $-1$ &  1 & $-1$ &  1 & 1 \\
$B_{3u}$ & 1 & $-1$ & $-1$ &  1 & $-1$ &  1 &  1 & $-1$ & 1 \\
\hline
$\rho^{M}$ & 6 & 2 & 0 & 0 & 2 & 6 & 0 & 0 & 6 \\
\hline
\end{tabular}
\end{table}

\begin{table}[h]
\centering
\caption{Little co-groups, decomposition of the six-band $(p_x,p_y)$ kagome representation
(spinless), and the degeneracies enforced by the irrep content. At $M$ in the pristine
lattice the spectrum shows an additional twofold degeneracy, which is accidental and
traceable to the decoupling of one sublattice, Eq.~\eqref{eq:M_spectrum}. The decomposition at $M$ refers to the
origin at the hexagon centre.}
\label{tab:degeneracies}
\begin{tabular}{lllll}
\hline
Structure & $\mathbf{k}$ & Little co-group & Decomposition & Enforced degeneracies \\
\hline
Pristine, freestanding   & $\Gamma$ & $D_{6h}$ & $B_{1u}\oplus B_{2u}\oplus 2E_{1u}$ & $1,1,2,2$ \\
($p6/mmm$, 24 elements)  & $K$      & $D_{3h}$ & $A_1'\oplus A_2'\oplus 2E'$          & $1,1,2,2$ \\
                         & $M$      & $D_{2h}$ & $2A_g\oplus2B_{1g}\oplus B_{2u}\oplus B_{3u}$ & $1,1,1,1,1,1$ \\
\hline
Breathing, freestanding  & $\Gamma$ & $D_{3h}$ & $A_1'\oplus A_2'\oplus 2E'$ & $1,1,2,2$ \\
($p\bar{6}m2$, 12 elements) & $K$   & $C_{3h}$ & six 1D irreps               & all 1 \\
                         & $M$      & $C_{2v}$ & six 1D irreps               & all 1 \\
\hline
Breathing on SiC         & $\Gamma$ & $C_{3v}$ & $A_1\oplus A_2\oplus 2E$    & $1,1,2,2$ \\
($p3m1$, 6 elements)     & $K$      & $C_{3}$  & six 1D irreps               & all 1 \\
                         & $M$      & $C_{s}$  & six 1D irreps               & all 1 \\
\hline
\end{tabular}
\end{table}

\clearpage

\section{Sb atomic layer phases on SiC(0001)}
\label{sec:phases}

\begin{figure}[hbt]
\centering
  \includegraphics[width=1.0\columnwidth]{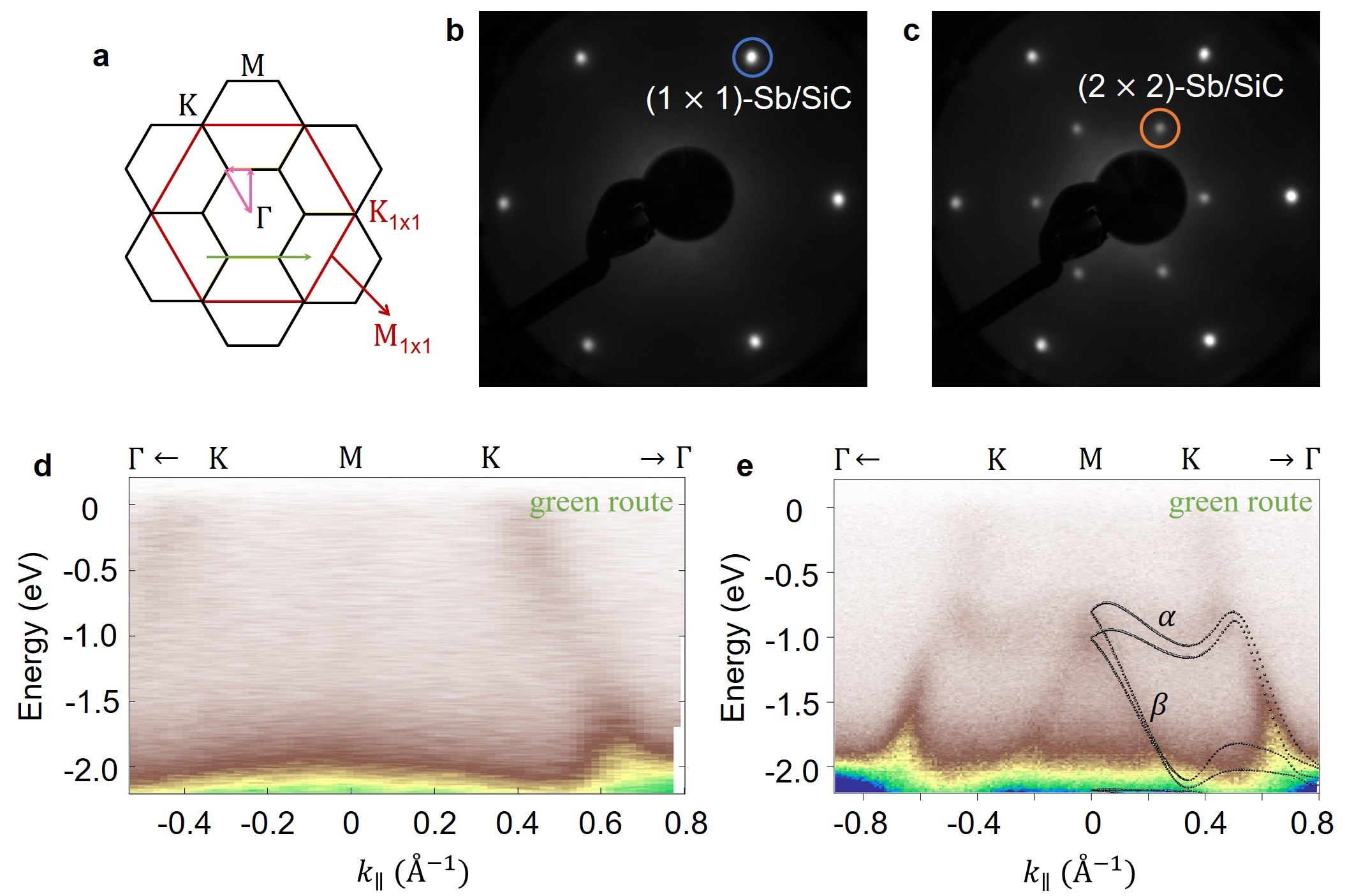}
  \caption{\textbf{Comparison of LEED and band dispersion of the $(1\times1)$-Sb/SiC and
breathing kagome phases.}
\textbf{a,} Brillouin-zone sketch, as in the inset of Fig.~3a of the main text. The green
arrow marks the momentum route along which the spectra in \textbf{d} and \textbf{e} were
acquired.
\textbf{b,c,} LEED patterns (45~eV) measured on a sample before (\textbf{b}) and after
(\textbf{c}) annealing at 420\,$^{\circ}$C, showing that a $(2\times2)$ reconstruction
appears after annealing.
\textbf{d,} Band dispersion measured along the green route in \textbf{a} on the pure
$(1\times1)$-Sb/SiC sample.
\textbf{e,} ARPES band dispersion along the same route, acquired on a sample containing
both the kagome and the $(1\times1)$ phases. Besides the linearly dispersing bands of
\textbf{d}, the dispersion from K to M is clearly resolved and agrees with the theoretical
prediction for the breathing kagome phase (overlay).}
  \label{fig:S5}
\end{figure}

Since the sample shown in Fig.~2a of the main text contains both the $(1\times1)$-Sb/SiC
and the kagome phase, it is essential to compare the band structure before and after
annealing in order to confirm the origin of the bands measured in Fig.~3. To avoid
ambiguity we use the labels $\Gamma$, K and M without subscripts for the kagome Brillouin
zone, corresponding to the $(2\times2)$ reconstruction of the SiC(0001) lattice, and the
same labels with the subscript $(1\times1)$ for the $(1\times1)$-Sb/SiC phase; both zones
are sketched in Fig.~\ref{fig:S5}a.

\begin{figure}[hbt]
\centering
  \includegraphics[width=1.0\columnwidth]{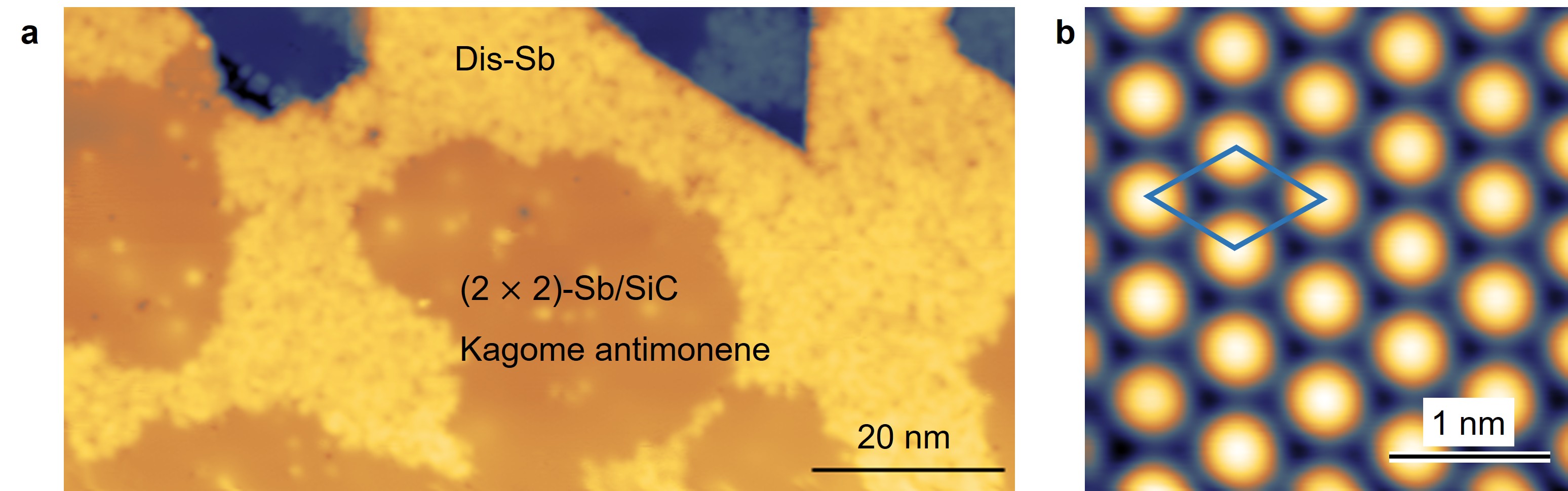}
  \caption{\textbf{Topography of kagome antimonene on SiC(0001).}
\textbf{a,} Large-scale scanning tunnelling microscopy (STM) image of the sample topography
measured at 3.0~V, 10~pA.
\textbf{b,} Zoomed-in STM image ($-2.0$~V, 50~pA) of the $(2\times2)$-Sb/SiC region,
unveiling the $(2\times2)$ reconstructed unit cell indicated by the blue rhombus.}
  \label{fig:S6}
\end{figure}

Low-energy electron diffraction (LEED) patterns acquired before and after annealing are
shown in Figs.~\ref{fig:S5}b and \ref{fig:S5}c. Before annealing only the $(1\times1)$
diffraction pattern is observed, corresponding to the $(1\times1)$-Sb/SiC phase. After
annealing at 420\,$^{\circ}$C additional $(2\times2)$ spots emerge, consistent with the formation of
the breathing kagome phase. Figure~\ref{fig:S5}d displays the band dispersion measured
along the green route of Fig.~\ref{fig:S5}a for the sample before annealing, where the
full-coverage $(1\times1)$ phase is buried beneath a rough layer; only a near-linear
dispersion between K and $\Gamma$ is observed. As shown in Fig.~\ref{fig:S5}e, after
annealing to the kagome phase the $\alpha$ and $\beta$ bands discussed in Fig.~3 emerge and
align well with the DFT-calculated bands, while the intensity of the near-linearly
dispersive bands within the gap, which originates from the small amount of coexisting
$(1\times1)$ phase, weakens significantly. This comparison shows that the
$\alpha$ and $\beta$ bands originate from the kagome antimonene phase.
Figure~\ref{fig:S6} presents additional topographic details of breathing kagome antimonene
on the SiC(0001) substrate.

\clearpage

\section{DFT picture of the Sb breathing kagome phase on SiC(0001): orbital filtering, gap opening, SOC and Wannierization}
\label{sec:dft}

To understand the band structure we build it up in three steps: freestanding undistorted
lattice, addition of the SiC substrate, and finally the breathing distortion.

\begin{figure}[hbt]
\centering
  \includegraphics[width=1.0\columnwidth]{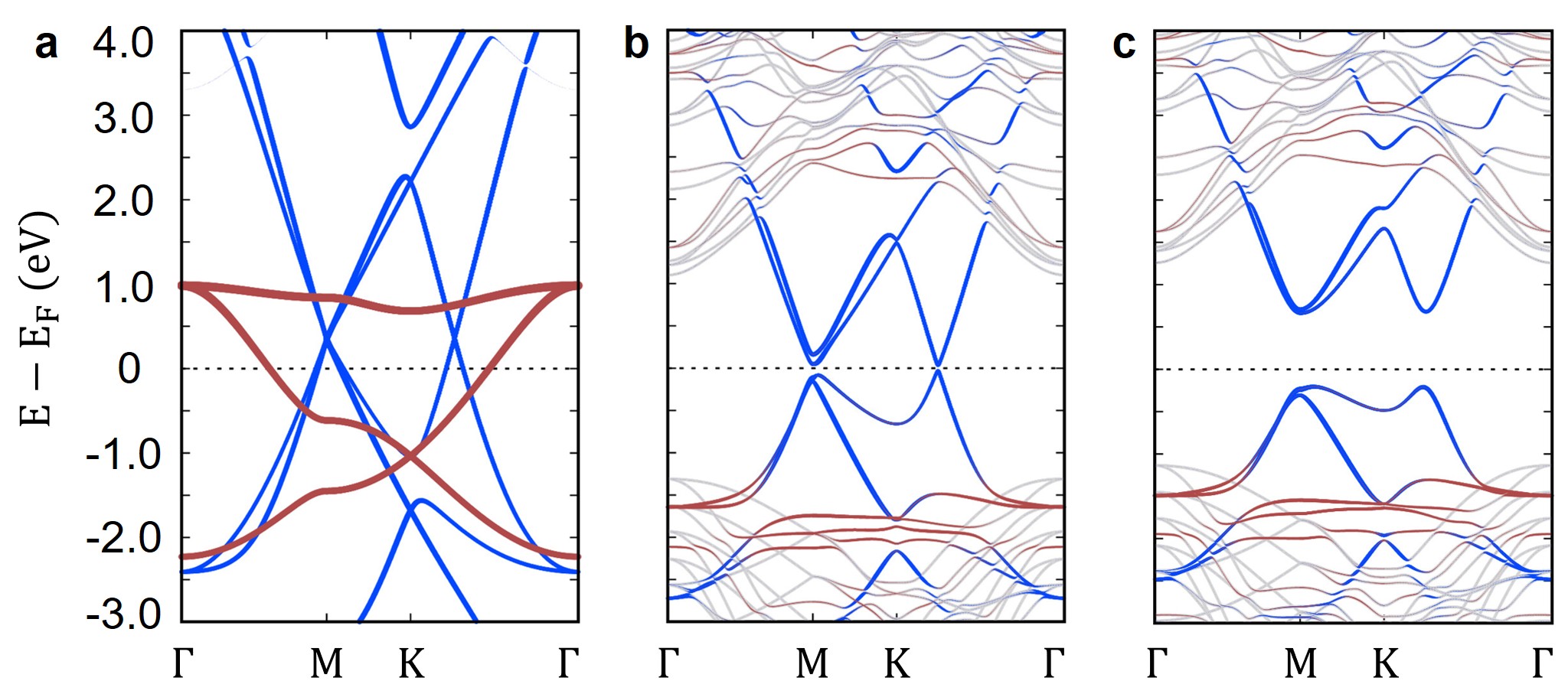}
  \caption{\textbf{Build-up of the electronic structure of kagome antimonene from DFT
(without SOC).}
\textbf{a,} Freestanding Sb kagome lattice without breathing distortion.
\textbf{b,} The same lattice with the SiC(0001) substrate included, showing the removal of
the $p_z$ states from the Fermi level by orbital filtering.
\textbf{c,} Fully relaxed, substrate-supported lattice including the breathing distortion,
which opens a 0.86~eV gap. Bands dominated by the Sb $p_x$ and $p_y$ orbitals are plotted
in blue, $p_z$-dominated bands in red, and SiC substrate bands in grey.}
  \label{fig:S7}
\end{figure}

We start from a freestanding kagome antimonene layer without any distortion
(nearest-neighbour Sb--Sb distance $a/2\approx3.05$~\AA). As shown in Fig.~\ref{fig:S7}a, the $p_z$ orbital exhibits a
dispersion similar to the classical $s$-orbital kagome system, with a flat band, two van
Hove singularities near the M point, and a Dirac cone at K. In contrast, the $p_x$ and
$p_y$ orbitals reproduce the main features of the tight-binding model of Fig.~1e, featuring four
bands that become degenerate at M near the Fermi level, two Dirac cones at K around
$\pm2.0$~eV, and two additional bands crossing between K and $\Gamma$ near the Fermi
energy. As shown in Sec.~\ref{sec:symmetry}, the fourfold degeneracy at M is not protected
by the crystal symmetry group, since it is readily lifted by an on-site energy offset
between the $B_{2u}$ and $B_{3u}$ orbitals, which respects all symmetry elements of the
group; the crossings between K and $\Gamma$, on the other hand, are part of the nodal line
derived there.

Introducing the SiC substrate (Fig.~\ref{fig:S7}b) leads to strong hybridization with the
Sb $p_z$ orbital, pushing it away from the Fermi level, similar to the orbital filtering
observed in bismuthene~\cite{reis2017bismuthene}. The in-plane Sb $p_x$ and $p_y$ orbitals
are much less affected and retain their contributions to the band structure almost
unchanged, leaving behind the isolated six-band $(p_x,p_y)$ kagome manifold
analysed throughout this work. When the Sb layer on the substrate is relaxed only along the
out-of-plane direction, that is without any breathing-mode distortion, a small gap already
opens at the Fermi level, similar to the SiC-supported indenene
lattice~\cite{bauernfeind2021design}. This gap originates from the breaking of the twofold rotation $C_{2z}$ caused by the inequivalent carbon atoms beneath adjacent triangles of the kagome
lattice, although the triangular geometry of the Sb lattice remains identical to that of
the freestanding case.

A full DFT relaxation including the in-plane degrees of freedom drives the SiC-supported
kagome antimonene toward the anticipated breathing-mode distortion, shown in the atomic
model of Fig.~2d of the main text. This substantially modifies the electronic band
structure, lifting the band degeneracies away from the $\Gamma$ point in accordance
with Table~\ref{tab:degeneracies}, and opening a wide band gap of 0.86~eV at the Fermi
level (Fig.~\ref{fig:S7}c). Consistent with Sec.~\ref{sec:symmetry}, the gap is not
confined to isolated crossing points but extends along the entire former nodal contour.

\begin{figure}[hbt]
\centering
  \includegraphics[width=1.0\columnwidth]{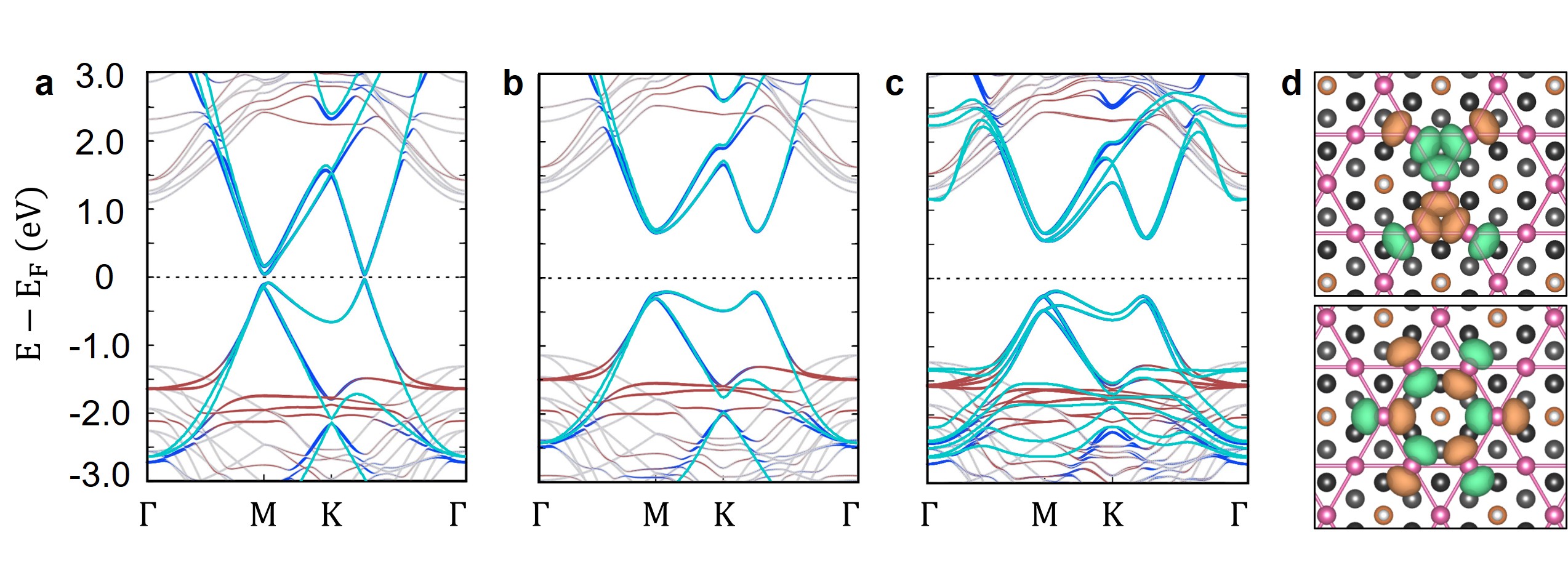}
  \caption{\textbf{Wannier representation of the in-plane kagome manifold.} DFT band
structures (blue: projected on Sb $p_x$ and $p_y$; red: projected on Sb $p_z$) overlaid
with the Wannier-interpolated bands (light blue) for the SiC(0001)-supported non-breathing
Sb kagome lattice (\textbf{a}), the breathing lattice (\textbf{b}), and the breathing
lattice including SOC (\textbf{c}).
\textbf{d,} Spatial orientation of the $B_{2u}$ (top panel) and $B_{3u}$ (bottom panel)
orbitals obtained from the Wannier functions of the non-breathing lattice, confirming the
local-frame assignment of Eq.~\eqref{eq:orbital_labels}.}
  \label{fig:S8}
\end{figure}

Figure~\ref{fig:S8} finally compares the DFT bands with a Wannier interpolation of the
in-plane kagome manifold, for the non-breathing lattice (Fig.~\ref{fig:S8}a), the breathing
lattice (Fig.~\ref{fig:S8}b) and the breathing lattice including SOC
(Fig.~\ref{fig:S8}c). The Wannier bands reproduce the \textit{ab initio} dispersion of the
$p_x/p_y$-derived manifold throughout, which validates the reduction to the six-band model
of Sec.~\ref{sec:tb}. Figure~\ref{fig:S8}d shows the resulting real-space Wannier orbitals:
they are the local $B_{2u}$ and $B_{3u}$ states of Eq.~\eqref{eq:orbital_labels}, with the
$B_{3u}$ lobes oriented along the hexagon-radial axis $\hat{\mathbf{e}}_\alpha$ and the
$B_{2u}$ lobes along the triangle-radial axis $\hat{\mathbf{y}}_\alpha$, confirming the local-frame convention adopted in Sec.~\ref{sec:tb}.

\clearpage

\section{Substrate stabilization of the breathing distortion}
\label{sec:breathing_origin}

Section~\ref{sec:symmetry} established that an orbital-filtered kagome lattice at half
filling is unstable against the breathing distortion, because the distortion removes an
entire nodal line at $E_F$. That argument is symmetric under exchange of the two triangles
and on its own predicts two energetically degenerate breathing configurations. We now show
that the substrate lifts this degeneracy and selects one of them.

\begin{figure*}[htbp]
\centering
  \includegraphics[width=1.0\textwidth]{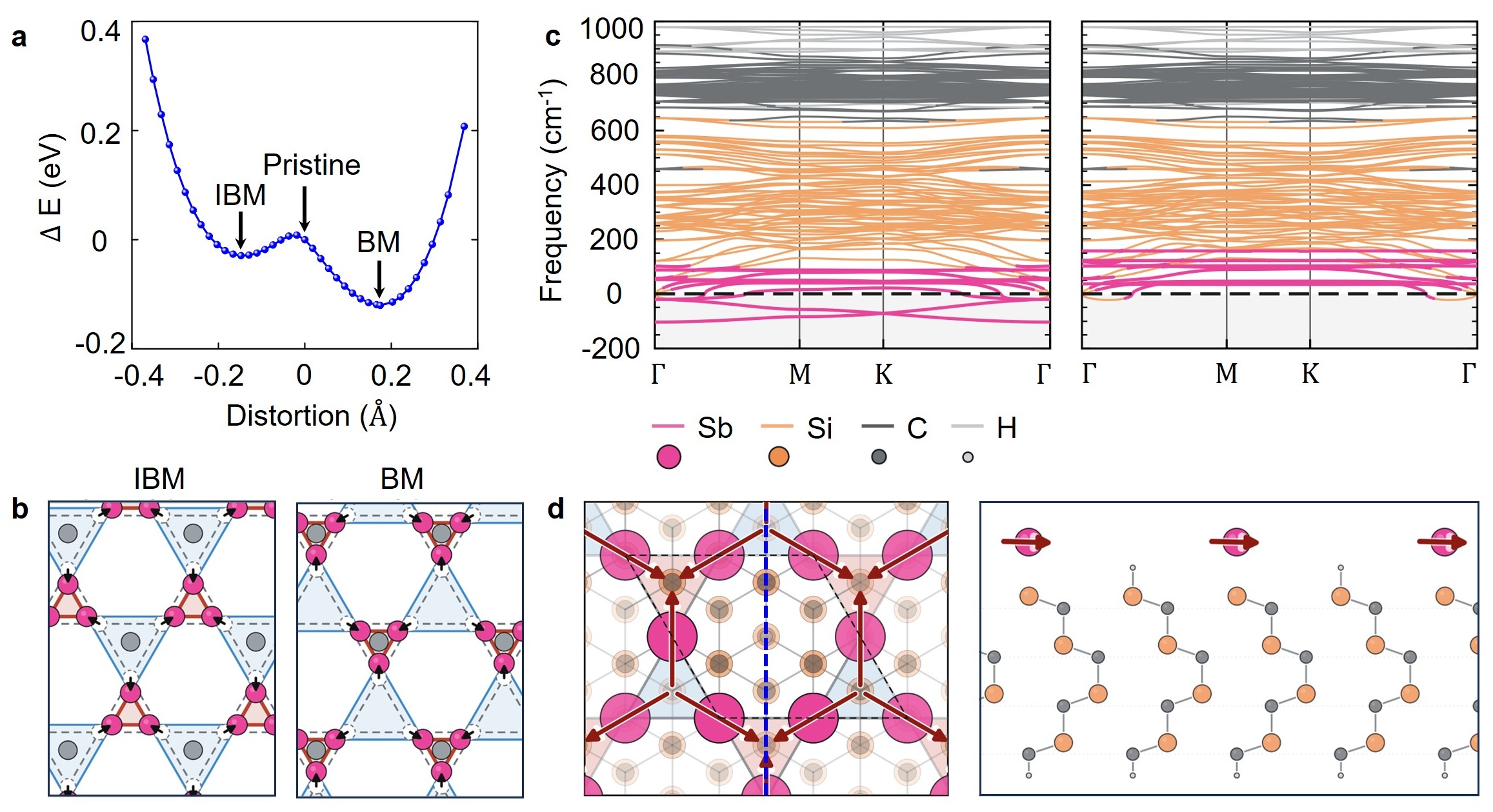}
  \caption{\textbf{Dynamical stability of the Sb overlayer on the SiC substrate.}
  \textbf{a},~Total-energy profile along the distortion coordinate connecting
  the inverse-breathing (IBM) and breathing (BM) minima through the
  undistorted (non-breathing) structure; energies are referred to the pristine
  structure.
  \textbf{b},~Schematic of the undistorted $\rightarrow$ breathing distortion
  of the Sb sublattice in the IBM (left) and BM (right) phases: open dashed
  circles mark the ideal kagome sites and the grey dashed network the
  corresponding bonds, filled pink circles the inverse-breathing and breathing
  Sb positions, and arrows the displacements, magnified for visibility. Only the
  first-layer carbon atoms below the Sb$_3$ triangles are shown, as filled grey circles.
  \textbf{c},~Phonon dispersion along $\Gamma$--M--K--$\Gamma$ of the
  undistorted Sb kagome phase (left panel) and of the breathing-distorted
  structure (right panel), on a common frequency scale. Each branch is coloured
  by the atomic species that carries most of the vibration at that wavevector
  (legend). Imaginary frequencies are plotted as negative; the shaded region
  marks $\omega<0$ and the dashed line $\omega=0$.
  \textbf{d},~Displacement pattern of the $\Gamma$-point soft mode of the
  undistorted phase, shown in top view (left panel) and in a cross-sectional view along the blue dashed line in the top view (right panel). Arrows show how each atom moves, with lengths in
  proportion to the true displacements. Shading marks the Sb$_3$ triangles that
  contract (red) and expand (blue), and the dashed rhombus is the
  $(2\times2)$ cell relative to the SiC(0001) lattice. Substrate atoms are drawn
  faded, since they carry a negligible share of the motion.
    }
  \label{fig:S9}
\end{figure*}

{As shown in Fig.~\ref{fig:S9}a, our total-energy calculations confirm that the optimized
breathing mode (BM) represents the globally stable phase. To further explore the energy
landscape we performed nudged elastic band (NEB) calculations~\cite{henkelman2000climbing}
to examine the transition pathway between the normal breathing mode and the inverse
breathing mode (IBM), in which the topmost and second-sublayer carbon atoms of the
substrate preferentially stabilize the larger and smaller kagome triangles, respectively,
as illustrated in the left panel of Fig.~\ref{fig:S9}b. The artificial, orbital-filtered, freestanding Sb
kagome lattice would undergo spontaneous symmetry breaking into one of two energetically
equivalent breathing configurations. The presence of the substrate lifts this degeneracy by
stabilizing one specific configuration over the other, thereby providing an effective route
to engineer the breathing mode, as evident from Fig.~\ref{fig:S9}a.  This behaviour can be understood from the bonding to the substrate: the topmost carbon atoms of the substrate preferentially bind to the Sb atoms, thereby forming smaller triangles around them (right panel of Fig.~\ref{fig:S9}b). Consequently the breathing mode emerges as the substrate-favoured distortion, making it the globally stable phase, while the IBM remains only a local minimum. Hence, the electronically driven instability of orbital-filtered kagome antimonene is effectively guided by the SiC substrate, whose two inequivalent carbon sublayers beneath the kagome triangles selectively stabilize the breathing mode observed in experiment.

This picture is supported by our phonon calculations. With the three Sb atoms at the ideal kagome sites and a uniform Sb--Sb separation, the overlayer is
dynamically unstable. Several phonon branches exhibit imaginary frequencies, with the strongest instability occurring at the zone center (Fig.~\ref{fig:S9}c, left). The leading instability is therefore commensurate with the periodicity of the system and can be accommodated within the current unit cell, without requiring the further enlargement that a zone-boundary instability would entail.

The eigenvector of this mode identifies the corresponding distortion (Fig.~\ref{fig:S9}d). The mode is almost entirely Sb in character, with the three Sb atoms accounting for essentially the entire in-plane breathing-mode displacement. Each Sb atom moves directly inward or outward along the line connecting it to the center of its Sb$_3$ triangle, resulting in a radial displacement pattern. Consequently, corner-sharing triangles of the kagome net alternately contract and expand in antiphase.

Displacing the overlayer along this eigenvector therefore provides a direct route to stabilizing the structure. Relaxation of the Sb layer in the presence of the substrate potential allows the Sb atoms to shift in-plane in close correspondence with the eigenvector, together with a small inward relaxation of the Sb plane. This transforms the uniform network into one with alternating short and long Sb--Sb bonds, corresponding to a trimerized kagome lattice (right panel of Fig.~\ref{fig:S9}b). The resulting distortion coincides with the breathing minimum of the total-energy profile shown in Fig.~\ref{fig:S9}a, demonstrating that the two independent routes to the ground state yield the same structure.

As expected, the distorted breathing structure is dynamically stable
(Fig.~\ref{fig:S9}c, right). The soft mode hardens into the
lowest optical branch at a real frequency, the zone-centre acoustic
modes are numerically zero, and the
high-frequency spectrum is left untouched. This shows that the
instability is confined to the Sb sublattice and is fully removed by
the breathing distortion.  A shallow residual excursion of the flexural
branch survives in two narrow windows about $\Gamma$; species
projection identifies it as substrate-derived rather than Sb, and its
quadratic dispersion makes it the branch most sensitive to slab
thickness, so it does not reflect an instability of the distorted
phase.

Taken together, the undistorted kagome geometry lies at the barrier between the BM
and IBM minima (Fig.~\ref{fig:S9}a), with negative curvature along the breathing
distortion, and the trimerized structure is the dynamically stable
ground state of the Sb overlayer.}

\begin{figure}[hbt]
\centering
  \includegraphics[width=1.0\columnwidth]{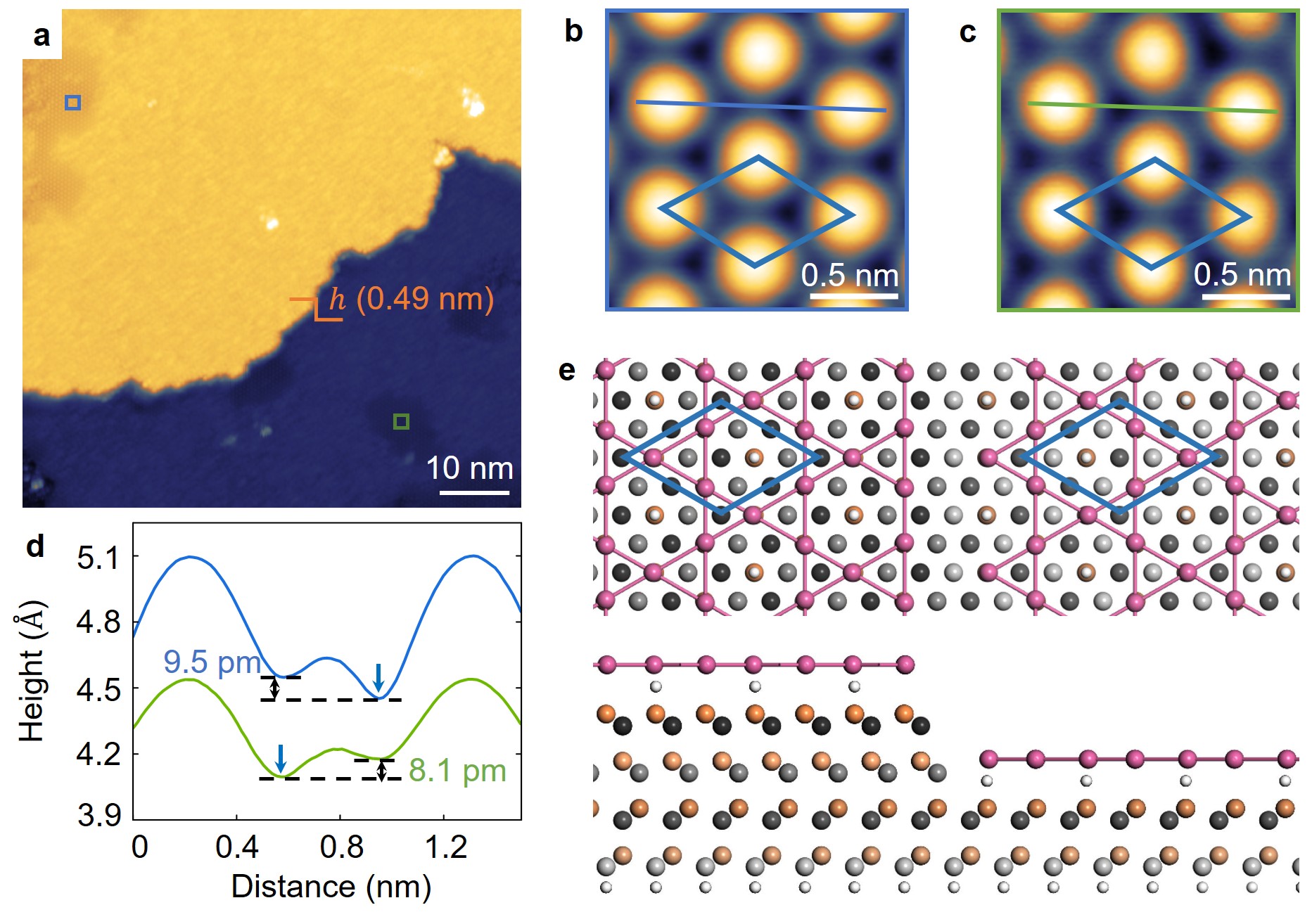}
  \caption{\textbf{Breathing kagome antimonene on adjacent SiC(0001) terraces.} \textbf{a,} Large-scale STM image (-3.0\,V, 10\,pA) acquired on an area
including two adjacent terraces with a height difference of 0.49~nm, i.e.\ half a SiC unit
cell. \textbf{b and c,} Zoomed-in STM images (-2.0\,V, 30\,pA) acquired on the top and bottom terraces in \textbf{a}, marked by blue and green squares, respectively, showing the 180$^\circ$ rotational offset between the kagome
lattices on the two terraces.
\textbf{d,} Line profiles extracted along the blue and green lines in \textbf{b} and
\textbf{c}, respectively. Blue arrows indicate the lowest positions in the topography.
\textbf{e,} Atomic model of the Sb breathing kagome lattice on two terraces separated by a
half SiC(0001) step, illustrating the lattice geometry in \textbf{b} and \textbf{c}.}
  \label{fig:S10}
\end{figure}

This picture makes a clear structural prediction: a single breathing sense of kagome antimonene on each SiC(0001) terrace. Because the C atoms in the topmost and
second sublayers of the substrate trap the smaller and larger triangles respectively, two
adjacent SiC(0001) terraces separated by half a unit cell (0.49~nm) along the $z$ direction
should host kagome domains with a 180$^\circ$ rotational shift between them, while
maintaining $C_3$ symmetry, as illustrated in Fig.~\ref{fig:S10}e. Consistent with this
model, kagome antimonene on two adjacent
terraces (Fig.~\ref{fig:S10}a) shows such a 180$^\circ$ rotational shift in the
STM topography (Figs.~\ref{fig:S10}b and \ref{fig:S10}c). The line profiles in
Fig.~\ref{fig:S10}d make this quantitative: the lowest region in the topography, indicated by blue arrows, shifts from the right side on
the top terrace to the left side on the bottom terrace.

\clearpage

\section{Projector, Wilson-loop and Wannier diagnosis of the orbital-resolved obstruction}
\label{sec:wilson}

This section provides the technical diagnosis supporting the band-resolved obstructed
atomic limit discussed in the main text. The goal is to distinguish the Wannier
representation of the isolated one-band and two-band sectors from that of the complete
half-filled occupied manifold. The analysis combines smooth band projectors, embedded
Wilson-loop Wannier centres, $C_3$ symmetry characters, maximally localized Wannier functions and finite-geometry corner-charge markers, and closes with the adiabatic
interpolation that establishes the site-centred character of the full valence manifold.

For this analysis it is convenient to place the origin on a kagome site, so that the
breathing lattice is described by the plane group $p3m1$ with the sublattices at
$\boldsymbol{\tau}_A=(0,0)$, $\boldsymbol{\tau}_B=(0,1/2)$ and
$\boldsymbol{\tau}_C=(1/2,1/2)$ in fractional coordinates with respect to
$(\mathbf{a}_1,\mathbf{a}_2)$. Within this embedding the strong-bond ($t_2$) and weak-bond
($t_1$) triangle centres and the summed site position are represented by
\begin{equation}
\boldsymbol{\tau}_{t_2}=(\boldsymbol{\tau}_A+\boldsymbol{\tau}_B+\boldsymbol{\tau}_C)/3
=\left(1/6,1/3\right),
\qquad
\boldsymbol{\tau}_{t_1}=\left(5/6,2/3\right),
\qquad
\sum_\alpha\boldsymbol{\tau}_\alpha\equiv\left(1/2,0\right),
\label{eq:supp_embedding_reference}
\end{equation}
the last modulo lattice vectors. These positions serve as the reference points for
identifying triangle-centred and kagome-site Wannier representations throughout. This embedding follows from the unit cell of Sec.~\ref{sec:tb} by moving the
origin to site $A$. Hybrid Wannier centres are defined only modulo a lattice
vector, so the meaningful statement is never the absolute value of a polarization but its
position relative to the reference points of
Eq.~\eqref{eq:supp_embedding_reference}. All numerical values quoted below refer to this
convention. Since all hopping amplitudes of the model are real, spinless time-reversal
symmetry is preserved and the Chern numbers of the isolated subspaces analysed below vanish; exponentially localized Wannier functions are therefore permitted, and the
atomic-limit classification is controlled by the positions and symmetry
representations of the Wannier centres.

For an isolated set of bands $\mathcal{S}$ we define the Bloch projector
\begin{equation}
\mathcal{P}_{\mathcal{S}}(\mathbf{k})
=\sum_{n\in\mathcal{S}}\ket{u_{n\mathbf{k}}}\bra{u_{n\mathbf{k}}},
\label{eq:supp_projector_general}
\end{equation}
where $\ket{u_{n\mathbf{k}}}$ are the cell-periodic Bloch eigenstates. The Hamiltonian is
written in the two-orbital kagome basis $(A1,A2,B1,B2,C1,C2)$, where $A,B,C$ denote the
sublattices and $1,2$ the two local orbital flavours. The two orbitals on a given
sublattice share the same real-space position, so the orbital embedding entering the
Wilson-loop calculation is $\boldsymbol{\tau}=(\boldsymbol{\tau}_A,\boldsymbol{\tau}_A,
\boldsymbol{\tau}_B,\boldsymbol{\tau}_B,\boldsymbol{\tau}_C,\boldsymbol{\tau}_C)$. In the
periodic gauge of Sec.~\ref{sec:tb} the physical overlap matrix between neighbouring
momenta must therefore carry the embedding phase explicitly,
\begin{equation}
M_j=V^\dagger(\mathbf{k}_j)\,
e^{-i\Delta\mathbf{k}\cdot\boldsymbol{\tau}}\,
V(\mathbf{k}_{j+1}),
\label{eq:supp_wilson_overlap}
\end{equation}
where $V(\mathbf{k})$ contains the occupied eigenvectors of the chosen subspace. We take
the unitary part of $M_j$ by singular-value decomposition and form the non-Abelian Wilson
loop $W_i=\prod_j M_j$, with $i=1,2$ labelling a loop along the reciprocal vector
$\mathbf{b}_i$. The determinant Wilson phase, averaged over the transverse momentum $k_\perp$, gives the total Wannier centre,
\begin{equation}
P_i=-\bigl\langle\arg\det W_i(k_\perp)\bigr\rangle_{k_\perp}/(2\pi) \quad (\mathrm{mod}\;1),
\label{eq:supp_det_wilson}
\end{equation}
which for a single occupied band is the Wannier centre of that band and for a multi-band
subspace the sum of the occupied Wannier centres modulo a lattice vector. For the parameter set of Fig.~1f of the main text both relevant direct gaps,
$\Delta_{1/6}=\min_{\mathbf{k}}[E_2(\mathbf{k})-E_1(\mathbf{k})]$ and
$\Delta_{1/2}=\min_{\mathbf{k}}[E_4(\mathbf{k})-E_3(\mathbf{k})]$, remain finite
throughout the Brillouin zone, so the isolated one-band sector and the three-band occupied
sector define smooth projectors.

At $1/6$ filling the occupied subspace consists of the isolated, predominantly
$B_{2u}$-derived $\gamma$ band, and the determinant Wilson loop gives
\begin{equation}
(P_1,P_2)_{\gamma}=\left(1/6,1/3\right)=\boldsymbol{\tau}_{t_2},
\label{eq:supp_polarization_one_band}
\end{equation}
which coincides with the strong-bond triangle centre of
Eq.~\eqref{eq:supp_embedding_reference}. The maximally localized Wannier function
constructed for this isolated band returns the same centre and displays a triangle-centred
molecular character, and an independent real-space Wannier-centre analysis of the
\textit{ab initio}-derived model band structure of Fig.~4c of the main text reproduces the
same position. The assignment is corroborated by the $C_3$ eigenvalues: when the threefold
rotation is defined about the strong-bond triangle centre, the $\gamma$ band has eigenvalue
$+1$ at $\Gamma$, $K$ and $K'$, as required for a Wannier state centred at that position.
It also becomes transparent in the decoupled-trimer limit $t_1\rightarrow0$, which is
reached without closing the intravalence gap $\Delta_{1/6}$ (Fig.~\ref{fig:S4}): there the
$\gamma$ band evolves continuously into an exactly flat band of isolated, predominantly
$B_{2u}$-derived $t_2$-trimer eigenstates centred on the strong triangles. Because this
symmetry-allowed Wannier centre does not coincide with a physical kagome site, the isolated
$\gamma$ band realizes a triangle-centred obstructed atomic limit relative to the Sb sites.

In the same limit the $\alpha$ and $\beta$ bands evolve into the complementary $C_3$ doublet
of the strong triangle. These molecular states illustrate the trimer decomposition; the
Wannier centres follow independently from the Wilson loops and $C_3$ characters.

The $\alpha$ and $\beta$ bands are degenerate at $\Gamma$, as required by the
two-dimensional irrep listed in Table~\ref{tab:degeneracies}, and form a connected two-band
manifold. The gauge-invariant object associated with this sector is therefore the full
two-band projector
$\mathcal{P}_{\alpha\beta}(\mathbf{k})=\ket{u_{1\mathbf{k}}}\bra{u_{1\mathbf{k}}}
+\ket{u_{2\mathbf{k}}}\bra{u_{2\mathbf{k}}}$ rather than either individual eigenvector,
since the two states can be mixed by smooth unitary transformations within the isolated
two-dimensional subspace. The determinant Wilson loops give the total polarization
\begin{equation}
(P_1,P_2)_{\alpha\beta}=\left(1/3,2/3\right)=2\,\boldsymbol{\tau}_{t_2}
\ (\mathrm{mod}\ 1),
\label{eq:supp_polarization_two_band}
\end{equation}
and the $C_3$ eigenvalues at the high-symmetry momenta are consistent with two Wannier
states centred on the strong-bond triangle. As an isolated subspace, this sector is therefore obstructed relative to a site-centred
kagome limit.

At half filling the full occupied projector is
$\mathcal{P}_{\mathrm{occ}}(\mathbf{k})=\sum_{n=1}^{3}
\ket{u_{n\mathbf{k}}}\bra{u_{n\mathbf{k}}}$, and the determinant Wilson loops give
\begin{equation}
(P_1,P_2)_{1/2}=\left(1/2,0\right),
\label{eq:supp_polarization_half}
\end{equation}
which equals the summed kagome-site position of Eq.~\eqref{eq:supp_embedding_reference}.
Since $3\boldsymbol{\tau}_{t_2}$ coincides with the same point modulo a lattice vector, the
polarization alone does not distinguish the two limits; the $C_3$ content and the adiabatic
path below do.
With the threefold rotation taken about the strong-bond triangle centre, the $C_3$
representation of this three-band subspace is
\begin{equation}
\Gamma:\{0,1,2\},\qquad K:\{0,1,2\},\qquad K':\{0,2,1\},
\label{eq:supp_c3_half}
\end{equation}
where $m$ labels the eigenvalue $\lambda_m=e^{2\pi im/3}$, so that the half-filled manifold
carries the full regular set $\{1,\omega,\omega^2\}$ at the $C_3$-invariant momenta,
with the $K$ and $K'$ assignments interchanged by time-reversal symmetry.

With the rotation taken about the kagome hexagon centre instead, the bands $\gamma$, $\beta$
and $\alpha$ carry $m=0,1,2$ at $\Gamma$, $m=1,2,0$ at $K$ and $m=2,1,0$ at $K'$, with or
without the small $\pi$ hopping. The labels of individual bands depend on the rotation
centre, since displacing it by $\mathbf{d}$ multiplies every eigenvalue by
$e^{i\mathbf{k}\cdot\mathbf{d}}$; the occupied manifold as a whole, however, carries each of
$1$, $\omega$ and $\omega^2$ exactly once at every $C_3$-invariant momentum. This content is
compatible with the kagome-site representation. The symmetry-preserving interpolation below
establishes the adiabatic connection to the corresponding site-centred atomic limit.

The same band-resolved structure is reflected in the boundary charge of finite triangular
flakes with a charge-neutral, $C_3$-symmetric termination and the ionic reference adopted
in our calculations. The isolated $\gamma$ sector contributes a fractional corner marker of
approximately $e/3$ modulo $e$, and the complementary $\alpha\beta$ sector approximately
$2e/3$ modulo $e$, for the same termination. The fractional value arises because the
electronic Wannier centres sit on the triangle centres while the compensating ionic charge
sits on the kagome sites; this is a termination-dependent filling anomaly and does not, by
itself, require protected in-gap corner states. When the complete three-band manifold is
occupied, the two contributions add to $e/3+2e/3=e\equiv0$ (mod $e$) per corner, consistent with the kagome-site atomic representation
obtained above. The corner-charge interpretation is tied to the chosen
symmetric finite geometry and boundary termination, whereas the bulk diagnosis is provided
by the projectors, Wilson centres and $C_3$ characters.

Finally, the site-centred character of the full manifold can be established explicitly by
adiabatic continuation, using the on-site orbital splitting $\mathcal{H}_\Delta$ of
Eq.~\eqref{eq:H_Delta} between the two local orbital channels, which carry opposite
eigenvalues under the site-preserving vertical mirror. This perturbation preserves the full plane-group symmetry (Sec.~\ref{sec:tb}) and therefore
defines a symmetry-preserving deformation. As $\Delta>0$ is increased, lowering the
$B_{2u}$ orbital, the
$B_{2u}$- and $B_{3u}$-derived contributions become progressively separated between the
valence and conduction sectors, while the gap separating the occupied and unoccupied
manifolds remains open throughout the interpolation. For sufficiently large $\Delta$ the
occupied three-band subspace evolves into a completely filled kagome manifold derived
predominantly from a single local orbital channel, which admits one Wannier centre on each
kagome site and therefore realizes a site-centred atomic limit without a Wannier
obstruction.

The analysis above is spinless. With spin--orbit coupling each band splits into two
spin-split branches that are Kramers degenerate at $\Gamma$ and $M$.
For the parameters of Table~\ref{tab:params}, the in-plane manifold keeps both direct gaps
open, $\Delta_{1/6}\approx0.49$~eV and $\Delta_{1/2}\approx0.86$~eV (0.80 and 1.07~eV without
spin--orbit coupling). The band-resolved atomic-limit structure carries over to the
spinful case provided these gaps remain open along the whole path from
$\lambda_{\mathrm{SO}}=\lambda_R=\lambda'_R=0$ and the $\mathbb{Z}_2$ invariants of the separated
spinful sectors vanish.

It remains to justify the model reduction used for the real-space maps of the main text. To
isolate the essential kagome-layer physics, the calculations shown in Figs.~4c--f of the
main text use a simplified freestanding hydrogenated Sb monolayer that preserves the
relevant symmetry and orbital character of the full system while omitting the SiC
substrate; the two valence bands closest to the Fermi level are predominantly
$B_{3u}$-derived and the isolated $\gamma$ band carries the complementary $B_{2u}$-derived
texture, reproducing the band-resolved structure of the minimal two-orbital model. In the
realistic material, three $p_z$-derived bands lie energetically between the $\alpha,\beta$
and $\gamma$ in-plane bands and are comparatively weakly dispersive in the absence of
spin--orbit coupling (Fig.~3b of the main text and Fig.~\ref{fig:S7}c). Although they
modify the energy ordering of the spectrum, the in-plane kagome subspace remains
identifiable through its orbital character and through a smooth, symmetry-preserving
Wannier projection (Fig.~\ref{fig:S8}). Since the in-plane sectors are then not
isolated in energy, their Wannier centres are defined through this projection, and the
band-resolved atomic-limit character refers to this projected in-plane subspace. Within it,
the three-band manifold can be deformed adiabatically to Sb-site-centred Wannier
functions without closing the separating direct gap.

Together, these results show the band-resolved nature of the obstruction: the
internally separated one-band and two-band sectors are naturally described by
triangle-centred Wannier representations when treated separately, whereas the complete half-filled subspace is compatible with, and adiabatically connected
to, a kagome-site atomic limit.

\clearpage

\section{More evidence for the orbital-resolved atomic obstruction}
\label{sec:more_evidence}

\begin{figure}[hbt]
\centering
  \includegraphics[width=1.0\columnwidth]{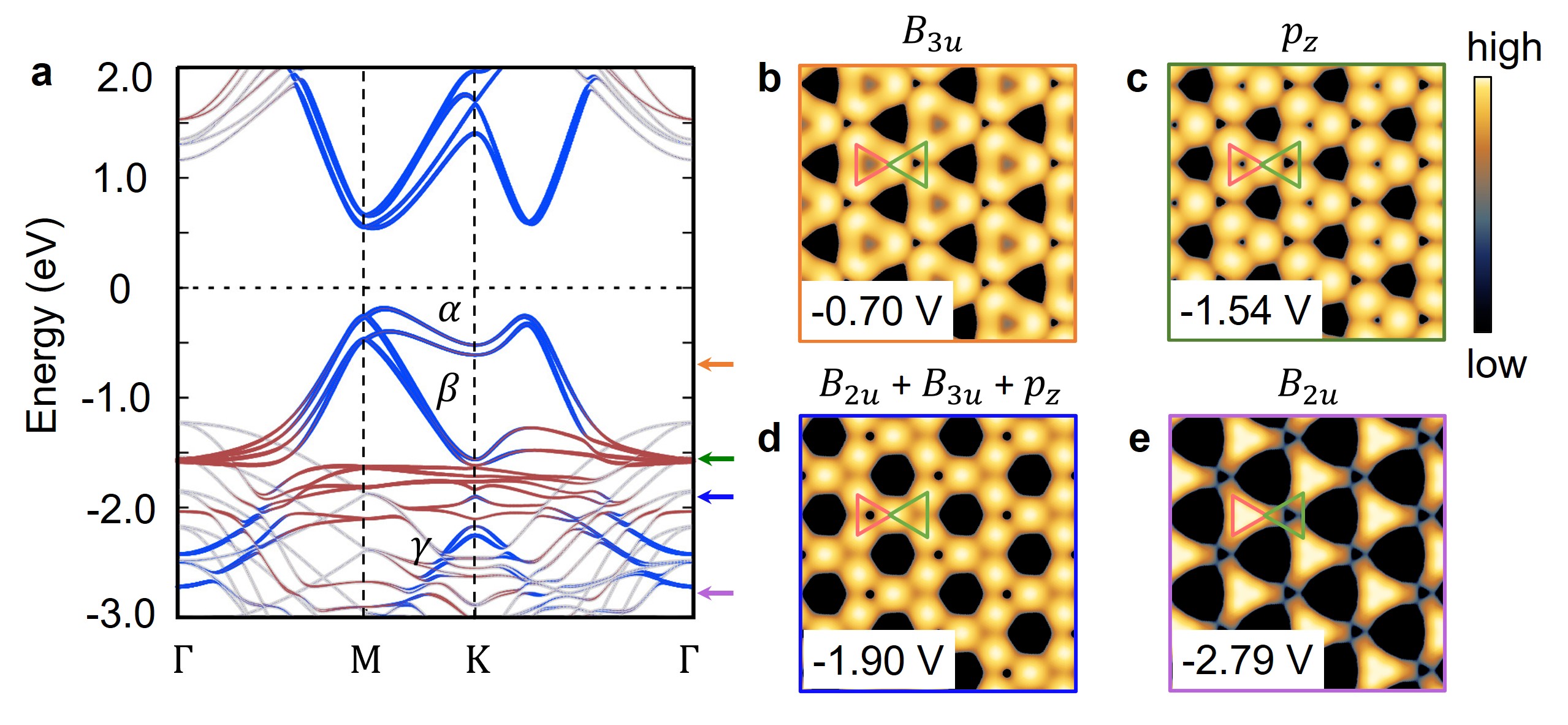}
  \caption{\textbf{DFT-calculated energy-dependent LDOS for the realistic model of breathing
kagome antimonene on SiC(0001).}
\textbf{a,} Band structure along the high-symmetry path, including substrate and SOC. The
coloured arrows indicate the energy positions at which the LDOS maps are plotted.
\textbf{b--e,} In-plane-projected LDOS of the breathing kagome antimonene layer at
$-0.70$~eV (\textbf{b}, dominated by the $B_{3u}$ orbital), $-1.54$~eV (\textbf{c},
dominated by the $p_z$ orbital), $-1.90$~eV (\textbf{d}, jointly contributed by the
$B_{2u}$, $B_{3u}$ and $p_z$ orbitals), and $-2.79$~eV (\textbf{e}, dominated by the
$B_{2u}$ orbital). The colours of the frames correspond to those of the arrows in
\textbf{a}. The smaller and larger triangles of the breathing kagome lattice are indicated
in pink and green, respectively, and are overlaid on top of the LDOS maps.}
  \label{fig:S11}
\end{figure}

To provide further evidence for the band-resolved atomic obstruction in breathing kagome
antimonene supported by the SiC(0001) substrate, we calculated the energy-dependent local
density of states (LDOS) based on the realistic structural model, as shown in
Fig.~\ref{fig:S11}. The DFT-calculated band structure, including both the substrate and
spin--orbit coupling, is shown in Fig.~\ref{fig:S11}a, and the LDOS maps were extracted at
the energies marked by the coloured arrows.

At $-0.70$~eV, marked by the orange arrow, the corresponding band is dominated by the
$B_{3u}$ orbital character. The real-space LDOS shown in Fig.~\ref{fig:S11}b exhibits a
localization pattern closely resembling that obtained from the simplified breathing kagome
model in Figs.~4a, 4d and 4g of the main text, with the charge density accumulated at the
centres of the smaller triangles. Figure~\ref{fig:S11}c shows the LDOS calculated at
$-1.54$~eV, marked by the green arrow. This state is dominated by the $p_z$ orbital and is
primarily localized at the Sb atomic sites; owing to the spatial extension of the
$p_z$-derived states and the inequivalent sizes of the neighbouring triangles, the LDOS
nevertheless appears more concentrated within the smaller triangles than within the larger
ones. Consequently, if the STM tip does not provide sufficient atomic resolution, the
corresponding scanning tunnelling spectroscopy (STS) image may appear as enhanced intensity
over the smaller triangles rather than as clearly resolved atomic-site features. At the
energy marked by the blue arrow ($-1.90$~eV) the electronic states arise from joint
contributions of the $B_{3u}$, $p_z$ and $B_{2u}$ orbitals, and this hybrid orbital
contribution shifts the charge localization from the smaller to the larger triangles, as
shown in Fig.~\ref{fig:S11}d. At still higher binding energy, marked by the purple arrow,
the states are dominated by the $B_{2u}$ orbital and the corresponding LDOS is shown in
Fig.~\ref{fig:S11}e; it exhibits the characteristic spatial pattern associated with the
$B_{2u}$-derived band, similar to that obtained from the simplified model in Figs.~4b, 4e
and 4h of the main text. Compared with Fig.~\ref{fig:S11}d, the charge localization shifts
back to the smaller triangles.

\begin{figure}[hbt]
\centering
  \includegraphics[width=0.78\columnwidth]{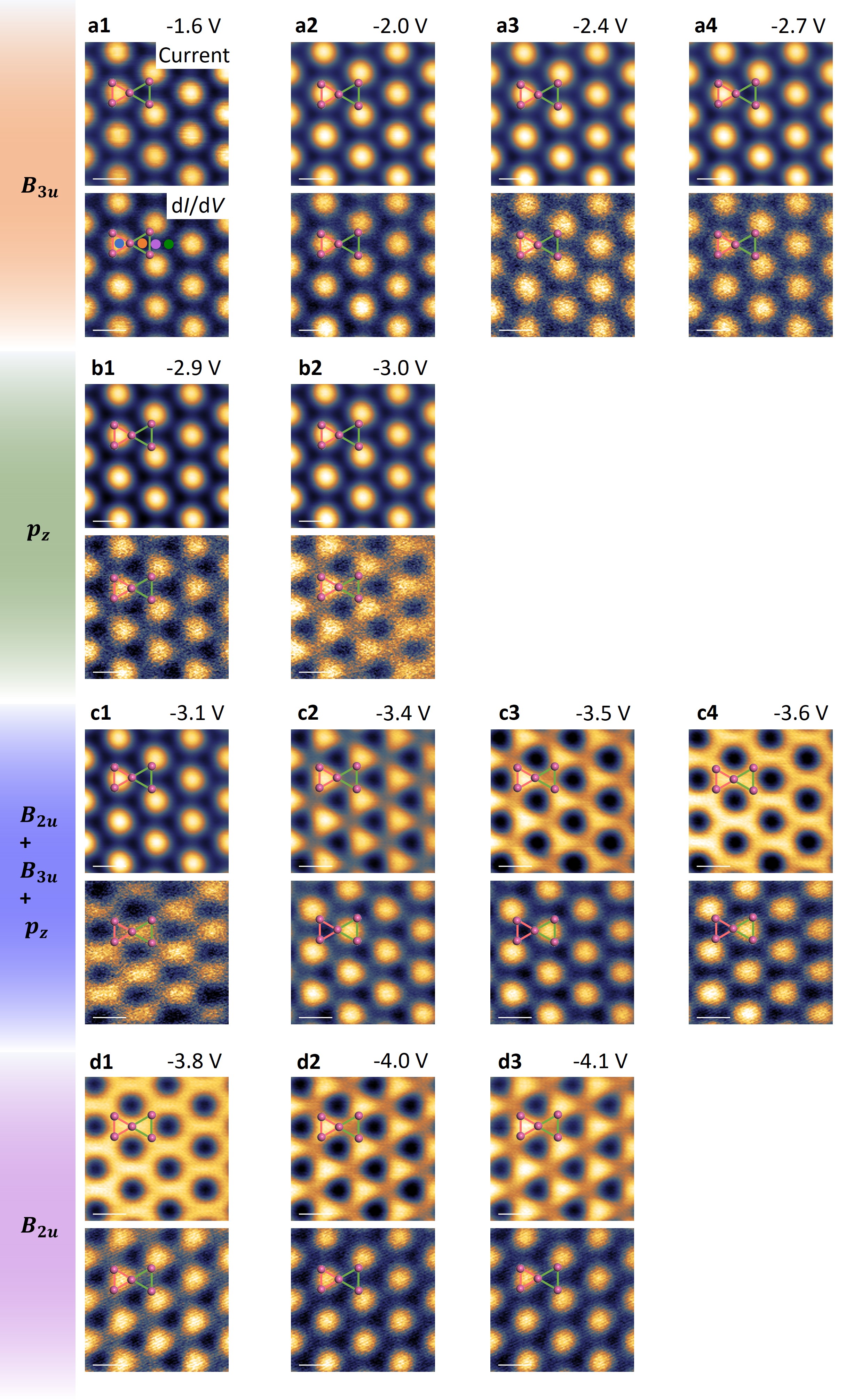}
  \caption{\textbf{STM-detected current images (upper panels) and $dI/dV$ maps (lower
panels) of breathing kagome antimonene}, acquired at the same position and various bias
voltages in constant-height mode. At low binding energies (\textbf{a1}--\textbf{a4}) the
electronic states are dominated by the $B_{3u}$ orbital. At increased binding energies of
$-2.9$ and $-3.0$~V (\textbf{b1} and \textbf{b2}) the states are dominated by the $p_z$
orbital. At still higher binding energies (\textbf{c1}--\textbf{c4}) the $B_{2u}$, $B_{3u}$
and $p_z$ orbitals jointly contribute to the electronic states. Finally, starting from
$-3.8$~V (\textbf{d1}--\textbf{d3}), the states are dominated by the $B_{2u}$ orbital. The
scale bar applies to all images and corresponds to 0.5~nm. The smaller and larger triangles
of the breathing kagome lattice are marked in pink and green, respectively, and both the
representative triangles and the atomic lattice are overlaid on the images. The colours used
to mark the orbital characters are consistent with those in Fig.~\ref{fig:S11}.}
  \label{fig:S12}
\end{figure}

Experimentally, we further acquired energy-dependent tunnelling-current images and
differential-conductance ($dI/dV$) maps in constant-height mode, as shown in
Fig.~\ref{fig:S12}, for comparison with the theoretical LDOS calculations in
Fig.~\ref{fig:S11} and with the charge distributions predicted for the band-resolved obstructed and non-obstructed atomic
limits. The $dI/dV$ maps and tunnelling-current images reflect the energy-resolved and
energy-integrated LDOS, respectively.

At relatively low binding energies, as shown in Figs.~\ref{fig:S12}a1--a4 ($-1.6$ to
$-2.7$~V), the relevant bands are expected to be dominated by the $B_{3u}$ orbital
according to the band structure. In this energy range the $dI/dV$ maps show that the LDOS
is mainly accumulated at the centres of the smaller triangles, consistent with the
$B_{3u}$-dominated states predicted in Fig.~\ref{fig:S11} and Fig.~4 of the main text, and
the corresponding tunnelling-current images also show more pronounced intensity over the
smaller triangles than over the larger ones. At $-2.9$ to $-3.0$~V
(Figs.~\ref{fig:S12}b1--b2) the $dI/dV$ maps still show localization around the smaller
triangles, but the pattern evolves from a circular feature to a triangular shape. This
evolution suggests the increasing contribution of the $p_z$ orbital; however, owing to the
limited real-space resolution of the STM tip, the individual Sb atomic sites cannot be
clearly resolved. The tunnelling-current images acquired in this energy range still show
enhanced intensity at the smaller triangles, with a predominantly circular appearance.

From $-3.1$ to $-3.6$~V, the LDOS localization in the $dI/dV$ maps shifts from the smaller
to the larger triangles, as shown in Figs.~\ref{fig:S12}c1--c4. This behaviour reproduces
the theoretical prediction for states jointly contributed by the $B_{3u}$, $p_z$ and
$B_{2u}$ orbitals. In contrast, the energy-integrated tunnelling-current images do not show
the same abrupt shift as the energy-resolved LDOS; instead the current pattern gradually
evolves from a circular feature to a triangular shape at $-3.4$~V, then to a three-lobed
pattern at $-3.5$~V, and finally to a more atomically resolved kagome lattice at $-3.6$~V.

At still higher binding energies, below approximately $-3.8$~V, the states are dominated by
the $B_{2u}$ orbital, as shown in Figs.~\ref{fig:S12}d1--d3. In this regime the
energy-resolved LDOS shifts back to the smaller triangles, in good agreement with the
theoretical LDOS maps in Fig.~\ref{fig:S11} and Fig.~4 of the main text. The
tunnelling-current image at $-3.8$~V still exhibits a relatively atomically resolved kagome
pattern, whereas those at $-4.0$ and $-4.1$~V show a more triangular intensity
distribution. Since the tunnelling-current image reflects the energy-integrated state
localization, the inclusion of the $B_{2u}$ orbital together with the $B_{3u}$ orbital
leads to a broader spatial distribution of the electronic states towards the Sb atomic
sites compared with the case dominated only by the $B_{3u}$ orbital. This is consistent with the non-obstructed atomic limit predicted at half filling, when both the
$B_{3u}$ and $B_{2u}$ orbitals are included, although the integrated current also contains
$p_z$-derived contributions.

\begin{figure}[hbt]
\centering
  \includegraphics[width=0.8\columnwidth]{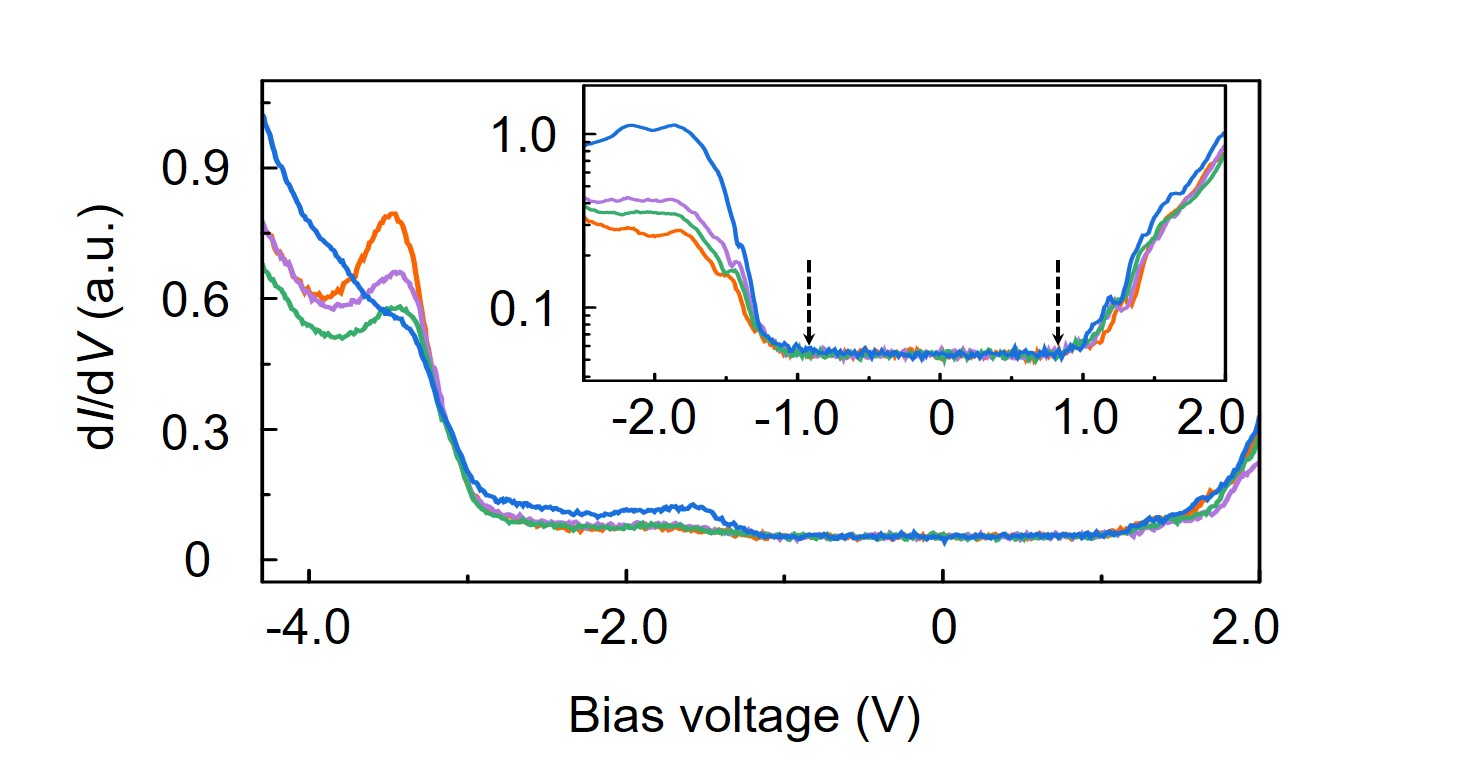}
  \caption{\textbf{STM-measured differential conductance spectra acquired at different
positions within a unit cell of breathing kagome antimonene.} The measurement positions are
indicated by the corresponding colours in the lower panel of Fig.~\ref{fig:S12}a1, and all
spectra were acquired with the tip held at the same height. The inset shows zoomed-in
spectra plotted with a logarithmic $y$-axis. The valence-band maximum and conduction-band
minimum are marked by dotted black arrows.}
  \label{fig:S13}
\end{figure}

The energy-dependent evolution of the LDOS is also consistent with the position-dependent
$dI/dV$ spectra, as shown in Fig.~\ref{fig:S13}. From $-1.20$ to $-3.00$~V, the spectrum
acquired at the centre of the smaller triangle (blue curve) exhibits a stronger intensity
than that acquired at the centre of the larger triangle (orange curve). From $-3.00$ to
$-3.73$~V, however, the spectral intensity at the centre of the larger triangle becomes
stronger than that at the smaller triangle, and below $-3.73$~V the intensity shifts back
to the smaller triangle. This position-dependent spectral-intensity shift is in good
agreement with the evolution of charge localization observed in Fig.~\ref{fig:S12}.

In conclusion, by combining the STM and STS measurements of Figs.~\ref{fig:S12} and
\ref{fig:S13} with the theoretical calculations of Fig.~\ref{fig:S11}, we find that the
electronic states at low binding energies (biases above approximately $-2.9$~V) are dominated by
the $B_{3u}$ orbital, whereas those at higher binding energies (below approximately
$-3.8$~V) are dominated by the $B_{2u}$ orbital. Both orbital-dominated states show
charge localization centred at the smaller triangles of
the breathing kagome lattice, consistent with the predicted obstructed atomic limits. In contrast, when the $B_{2u}$ and $B_{3u}$ orbitals are
considered together at half filling, the observed redistribution is consistent with the
non-obstructed atomic limit obtained from the projector, Wilson-loop and $C_3$ analysis of
Sec.~\ref{sec:wilson}.

\bigskip
\clearpage

\bibliography{Sb_kagome_SiC}